\documentclass{article}

\usepackage{arxiv}

\usepackage[utf8]{inputenc} 
\usepackage[T1]{fontenc}    
\usepackage{hyperref}       
\usepackage{booktabs}       
\usepackage{amsmath}
\usepackage{amssymb}
\usepackage{amsfonts}       
\usepackage{nicefrac}       
\usepackage{microtype}      
\usepackage{graphicx}
\usepackage{natbib}
\usepackage{doi}
\usepackage{algorithm, algpseudocode}
\usepackage{subcaption}

\title{\emph{hyperFastRL}: Hypernetwork-Based Reinforcement Learning for Unified Control of Parametric Chaotic PDEs}

\author{ \href{https://orcid.org/0000-0002-3780-3929}{\includegraphics[scale=0.06]{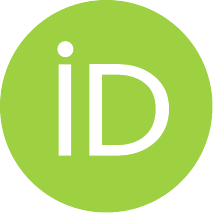}\hspace{1mm}Anil Sapkota} \\
	Department of Mechanical and Aerospace Engineering\\
	University of Tennessee\\
	Knoxville, Tennessee \\
	\texttt{asapkot1@vols.utk.edu} \\
	\And
	\href{https://orcid.org/0000-0000-0000-0000}{\includegraphics[scale=0.06]{orcid.pdf}\hspace{1mm}Omer San} \\
	Department of Mechanical and Aerospace Engineering\\
	University of Tennessee\\
	Knoxville, Tennessee \\
	\texttt{osan@utk.edu} \\
}

\renewcommand{\shorttitle}{hyperFastRL}

\hypersetup{
pdftitle={hyperFastRL: Hypernetwork-Based Reinforcement Learning for Unified Control of Parametric Chaotic PDEs},
pdfsubject={Machine Learning, Reinforcement Learning, PDE Control},
pdfauthor={Anil Sapkota, Omer San},
pdfkeywords={PDE control, reinforcement learning, hypernetworks, Kuramoto-Sivashinsky},
}

\begin{document}
\maketitle


\begin{abstract}
Spatiotemporal chaos in fluid systems exhibits severe parametric sensitivity, rendering classical adjoint-based optimal control intractable because each operating regime requires recomputing the control law. We address this bottleneck with \emph{hyperFastRL}, a parameter-conditioned reinforcement learning framework that leverages \textbf{Hypernetworks} to shift from tuning isolated controllers per-regime to learning a unified parametric control manifold. By mapping a physical forcing parameter $\mu$ directly to the weights of a spatial feedback policy, the architecture cleanly decouples parametric adaptation from spatial boundary stabilization. To overcome the extreme variance inherent to chaotic reward landscapes, we deploy a pessimistic distributional value estimation over a massively parallel environment ensemble. We evaluate three Hypernetwork functional forms, ranging from residual MLPs to periodic Fourier and Kolmogorov-Arnold (KAN) representations, on the Kuramoto-Sivashinsky equation under varying spatial forcing. All forms achieve robust stabilization. KAN yields the most consistent energy-cascade suppression and tracking across unseen parametrizations, while Fourier networks exhibit worse extrapolation variability. Furthermore, leveraging high-throughput parallelization allows us to intentionally trade a fraction of peak asymptotic reward for a 37\% reduction in training wall-clock time, identifying an optimal operating regime for practical deployment in complex, parameter-varying chaotic PDEs.
\end{abstract}


\section{Introduction}\label{sec:intro}

The active control of fluid flows is a foundational challenge in engineering because many relevant regimes are strongly nonlinear and chaotic. In such systems, small perturbations can produce large trajectory divergence, making robust feedback essential. Foundational chaos-control results, such as the OGY method, established that unstable chaotic dynamics can be steered with targeted interventions \cite{ott1990}. In fluid mechanics, the Kuramoto-Sivashinsky (KS) equation remains a canonical benchmark for spatiotemporal chaos and turbulence-like behavior \cite{bucci2019, garnier2021, zhu2020, wang2020}. Across applications, control strategies span open-loop forcing, model-based closed-loop control, and learning-based adaptation, each with different trade-offs in model fidelity, robustness, and computational cost \cite{bewley2001, kim2007}.

Classical flow-control methods remain essential and have delivered major advances, including linear systems approaches, adjoint-based optimization, and model predictive control variants \cite{bewley2001, kim2007}. Typical targets include transition delay and disturbance suppression in boundary layers, turbulence reduction in wall-bounded flows, and wake stabilization/drag reduction in bluff-body configurations. Representative examples include input--output model reduction and $H_2$ feedback design for flat-plate boundary layers \cite{bagheri2009}, MEMS-based feedback concepts for turbulent skin-friction reduction \cite{kasagi2009}, gain-scheduled relaminarization control in channel flow \cite{hogberg2003}, and broader linear closed-loop frameworks for transitional and unstable flows \cite{sipp2013, sipp2016amr}. In applied aerodynamics, fluidic oscillator development and sweeping-jet actuation studies also provide important classical AFC design guidance for practical forcing architectures \cite{gregory2013}. Complementary studies developed robust model-based feedback design \cite{jones2015}, localized estimation/control in shear flows \cite{tol2017}, iterative closed-loop control of quasiperiodic flows \cite{leclercq2019}, and ERA-based direct modelling for unstable-flow feedback control \cite{flinois2016}. The review in \cite{garnier2021} also highlights adjoint-based drag-optimization benchmarks around bluff-body geometries, which remain strong references for model-based optimal control in fluids.

However, these methods are typically tailored to a nominal model and parameter regime. In parameter-dependent chaotic PDEs, changing the physical parameter (e.g., Reynolds number, forcing amplitude, viscosity-related quantities) generally requires recomputation or retuning of reduced models, gradients, and controllers. This weak interpolation capability across a continuous parameter axis limits real-time adaptive deployment. These limitations motivate a complementary paradigm: instead of re-deriving controllers for each operating condition, one can learn a feedback policy from data that directly maps observed flow states to control actions. In this context, DRL becomes attractive for nonlinear, high-dimensional, and parameter-varying flow systems.

Deep reinforcement learning (DRL) provides a complementary data-driven paradigm for control by learning feedback policies directly from interaction and shifting heavy computation to training, after which inference is fast. DRL methods are commonly grouped into value-based approaches such as DQN \cite{mnih2015dqn}, policy-gradient/actor-critic approaches such as A3C, DDPG, PPO, and TD3 \cite{mnih2016a3c,lillicrap2016ddpg,schulman2017ppo,fujimoto2018td3,schulman2022unified,li2025sgaoffpolicy}, and distributional/conservative variants for improved value estimation and robustness \cite{kuznetsov2020tqc,kumar2019bear,kumar2020cql,wu2019brac}. Historically, RL-based chaos control predates deep RL, with early optimal-chaos-control results using reinforcement learning \cite{gadaleta1999optimal,gadaleta2001ijcnnchaos}, followed by deep-RL studies showing restoration of chaotic dynamics \cite{vashishtha2020restoring}, model-free continuous deep-Q approaches \cite{ikemoto2019continuousDqnChaos}, and recent spatiotemporal-chaos modulation studies \cite{han2025modulatingChaos,bhatia2022scientificControlPde,han2021,froehlich2021complexDynamicsRl,weissenbacher2025chaosflowcontrol}. DQN demonstrated that a single agent can learn directly from pixels and reach human-competitive Atari performance \cite{mnih2015dqn}. AlphaGo showed that deep RL combined with search can solve long-horizon strategic planning at superhuman level in Go \cite{silver2016alphago}. In robotics and humanoid control, recent high-throughput actor-critic pipelines have produced agile and robust locomotion behaviors \cite{seo2025fasttd3}. In autonomous-driving decision stacks, deep RL has been used for tactical control tasks such as lane-change and merge decision-making under dynamic multi-agent traffic interactions \cite{chen2022jiotautono}. Closely related learning-based advances include deep-network methods for high-dimensional PDE computation \cite{han2018pnas,e2017cms} and reinforcement-learning-based controller design for hybrid UAV flight \cite{xu2019uav}.

In fluid and flow control, RL/DRL strategies are often grouped by control structure: direct closed-loop actuation, low-dimensional design/placement optimization, and chaotic-dynamics stabilization \cite{vignon2023pof,garnier2021,lampton2008,foo2023drl,peitz2023flowControlSml}. A key point emphasized by Vignon et al. is that classical RL formulations (tabular or weakly approximated value methods) become difficult to scale in realistic AFC settings because observation spaces are high-dimensional, action spaces are often continuous, and sample budgets are dominated by expensive CFD rollouts \cite{vignon2023pof}. In the same context, DQN-style methods can be effective when actions are discretized, but action discretization itself can become restrictive for fine-grained actuation and may require extensive tuning to remain stable in non-stationary flow environments \cite{mnih2015dqn,vignon2023pof}. This is one reason policy-based/actor-critic families (A3C, PPO, DDPG, TD3) are frequently preferred in AFC: they naturally handle continuous controls and are more flexible for real-time feedback parameterizations \cite{mnih2016a3c,schulman2017ppo,lillicrap2016ddpg,fujimoto2018td3,vignon2023pof}.

Classical and DRL methods are most informative when compared on the same target tasks. For wake stabilization and drag reduction, classical approaches rely on linearized models, reduced-order dynamics, and adjoint/model-based synthesis \cite{bewley2001,kim2007}. DRL reaches the same objective through end-to-end feedback policies learned from interaction, with demonstrated gains on cylinder/bluff-body configurations, weakly turbulent active flow-control settings, and turbulent channel drag-reduction cases \cite{rabault2019,fan2020,ren2021weaklyTurbulentAfc,guastoni2023turbulentDragReduction,vignon2023pof,wang2019rlflowcontrol,li2024nn,liu2025}. For transitional and unstable shear flows, classical pipelines use input--output model reduction and robust/$H_2$ feedback design with stronger interpretability near design conditions \cite{bagheri2009,jones2015,sipp2016amr}. DRL relaxes explicit model requirements and can discover nonlinear policies directly, but typically with heavier data requirements and weaker formal robustness guarantees \cite{garnier2021,vignon2023pof}.

At this stage, the dominant practical bottleneck is computational throughput: full-order CFD is expensive, so data generation for RL is also expensive. Multiple implementation papers report this constraint explicitly and show that training speed depends strongly on how aggressively rollouts are parallelized \cite{rabault2019multiEnv,kurz2022jocs,wang2022drlinfluids}. In particular, the DRLinFluids framework demonstrates a practical coupling of deep RL with OpenFOAM for CFD-based training workflows, highlighting both usability gains and persistent runtime pressure in high-fidelity settings \cite{wang2022drlinfluids}. This is particularly important in chaotic PDE control, where policy quality depends not only on sample count but also on diverse trajectory coverage. To mitigate this bottleneck, one line of work uses reduced or surrogate models instead of full-order CFD during policy optimization. Examples include reduced-order neural-ODE models for spatiotemporal-chaos control \cite{zeng2022spatiotemporalChaos}, symmetry-reduction-enhanced DRL for active control of chaotic spatiotemporal dynamics \cite{zeng2021symmetryReduction}, and model-based RL perspectives that report better sample efficiency than model-free baselines in PDE-control settings \cite{werner2024ecc,mayfrank2025}. Closely related data-driven modeling work has also advanced reduced-order and partial-observation forecasting of chaotic dynamics, including neural-ODE reduced models and inertial-manifold-based constructions \cite{linot2022ddrom,ozalp2023reconstruction,liu2024inertialmanifold,sitzmann2020siren}. A second line addresses complexity by control architecture through multi-agent systems and distributed control formulations: multi-agent RL decomposes large control domains into coordinated local agents, improving scalability of sensing/actuation and enabling effective control in high-dimensional 2D convection settings \cite{vignon2023rbcMultiAgent}, while distributed convolutional RL has also been demonstrated for PDE control \cite{peitz2024distributedPdeRl}. Additional application-focused studies in aerodynamics (e.g., airfoil AFC) show practical deployment potential, but also reinforce that training cost and generalization remain central constraints \cite{portalPorras2023airfoilAfc}.

\begin{figure}[!htb]
    \centering
    \includegraphics[width=0.6\columnwidth]{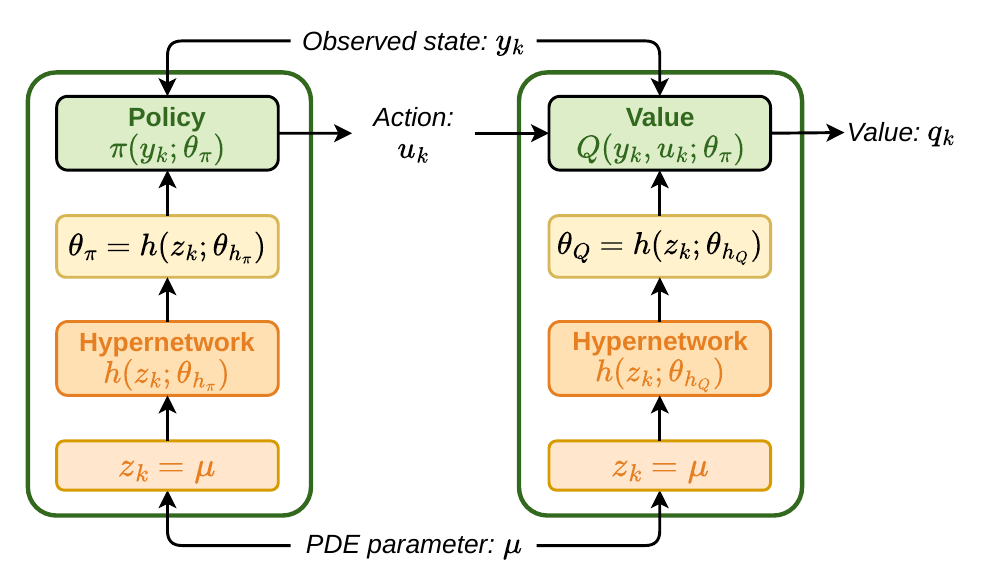}
    \caption{Architecture of the parameter-conditioned hypernetwork topology (Adapted from \cite{keynan2021, botteghi2025hyperl}). The hypernetwork cleanly disentangles parametric adaptation from the spatial feedback control problem by conditioning both actor and critic weights on the continuous physical parameter, enabling cross-regime generalization without per-regime retraining.}
    \label{fig:concept}
\end{figure}

Taken together, the literature still leaves three central gaps: (i) parameter-general control instead of per-regime retraining, (ii) stable value learning under chaotic rewards and overestimation-sensitive updates, and (iii) high-throughput rollout pipelines that scale without degrading control quality or generalization \cite{vignon2023pof,botteghi2025hyperl,werner2023pdeRl}. These gaps motivate combining parameter-conditioned policies with scalable off-policy learning and conservative/distributional critics. In this work, we study this combination through \emph{hyperFastRL}, a parameter-conditioned framework for control of parametric chaotic PDEs. Building on HypeRL \cite{botteghi2025hyperl}, we use \textbf{Hypernetworks} to generate actor and critic weights from the conditioning parameter $\mu$, separating contextual adaptation from spatial feedback control \cite{ha2016hypernetworks,keynan2021}. Figure~\ref{fig:concept} illustrates this conditioning mechanism.

At a high level, \emph{hyperFastRL} is used here as a unified parameter-conditioned control framework for chaotic PDEs, with emphasis on cross-regime behavior and practical training throughput. Specifically, we make three contributions that map directly to the empirical study: (i) a parameter-conditioned policy/value construction via hypernetworks for cross-regime control in KS (evaluated through seen-parameter, interpolation, and mild extrapolation tests) \cite{botteghi2025hyperl,ha2016hypernetworks}; (ii) a conservative distributional critic design based on TQC to reduce overestimation-driven instability in chaotic-return training (evaluated with stabilization and variance-oriented metrics) \cite{kuznetsov2020tqc}; and (iii) a scalable parallel off-policy training pipeline following FastTD3-style updates (evaluated with wall-clock and speed--performance trade-off analyses) \cite{seo2025fasttd3}. We evaluate this combined design on KS control across multiple seeds and operating conditions. Detailed algorithmic mechanics are deferred to subsequent sections. The remainder of this paper is organized as follows: Section~\ref{sec:problem_formulation} presents the problem formulation, theoretical foundations, and methods; Section~\ref{sec:results} reports empirical evaluation and comparative analysis; and the final sections summarize conclusions and supporting material.

\section{Problem Formulation and Theoretical Foundations}\label{sec:problem_formulation}

This section establishes a single through-line from control objective to implementation choices. We first define the KS control problem and its RL form, then justify the critic and parameter-conditioning design decisions, and finally describe the high-throughput training system that motivates the protocol choices in Section~\ref{sec:exp_setup}.

\subsection{KS Control Problem, Rewards, and Core Setting}\label{sec:ks_control}
The stabilization of the parametric Kuramoto--Sivashinsky (KS) equation is used as our primary benchmark for feedback control in turbulent-like regimes. KS is widely used as a reduced yet dynamically rich setting for spatiotemporal chaos: it exhibits nonlinear mode coupling, broadband energy transfer, and sensitive dependence on perturbations while remaining computationally tractable in one spatial dimension. This makes it suitable for systematically studying the trade-off between control quality, robustness, and computational throughput.

Let $\Omega=[0,L]$ be a periodic spatial domain, $t\in[0,T]$ the time interval, and $y(x,t;\mu)$ the scalar state for a regime parameter $\mu\in\mathcal{P}\subset\mathbb{R}$. In abstract form, we write the controlled parametric dynamics as
\begin{equation}
\partial_t y = \mathcal{F}_{\mu}(y) + \mathcal{B}u,
\end{equation}
with boundary and initial conditions
\begin{align}
    y(\cdot,0;\mu) &= y_0(\cdot;\mu), \nonumber \\
    y(x+L,t;\mu) &= y(x,t;\mu),
\end{align}
where $u(t)\in\mathbb{R}^{N_a}$ is the actuator vector and $\mathcal{B}:\mathbb{R}^{N_a}\to L^2(\Omega)$ maps actuator amplitudes to a distributed forcing field. A convenient decomposition is
\begin{equation}
\mathcal{F}_{\mu}(y)=\mathcal{A}y+\mathcal{N}(y) + f_{\mu},
\end{equation}
with an intrinsic linear operator $\mathcal{A}$ (instability/dissipation balance), a quadratic nonlinear convection term $\mathcal{N}(y)$ capturing nonlinear energy transfer, and a parameter-conditioned external spatial forcing field $f_{\mu}$.

In concrete KS implementations, this corresponds to a fourth-order dissipative PDE with quadratic advection and a parameter-varying spatial forcing term, for example
\begin{equation}
\partial_t y + y\,\partial_x y + \nu_2\,\partial_{xx}y + \nu_4\,\partial_{xxxx}y = f_{\mu}(x) + \sum_{i=1}^{N_a} b_i(x)u_i(t),
\end{equation}
where $b_i(x)$ denotes the spatial profile of actuator $i$, $\nu_2$ and $\nu_4$ define the intrinsic instability and dissipation scales, and $f_{\mu}(x)$ introduces the parameter-dependent external continuous forcing. We consider admissible controls
\begin{equation}
\begin{aligned}
    \mathcal{U}=\{u \in L^2(0,T;\mathbb{R}^{N_a}) : |u_i(t)| \le 1, 
    i=1,\dots,N_a\},
\end{aligned}
\end{equation}
which encode actuator saturation and finite control authority.

For each parameter value $\mu$, the finite-horizon objective is a quadratic tracking-effort trade-off,
\begin{equation}
\begin{aligned}
    J_{\mu}(u)=\int_0^T\!\Bigl( &\|y(\cdot,t;\mu)-y_{\mathrm{ref}}(\cdot,t)\|_{L^2(\Omega)}^2  
    + \alpha\|u(t)\|_2^2 \Bigr)dt,
\end{aligned}
\end{equation}
where $\alpha>0$ is a penalty parameter,
where $y_{\mathrm{ref}}$ is the target field (case-dependent in our experiments: zero reference or prescribed multi-mode cosine profile).
and the single-regime optimal-control problem is
\begin{equation}
u_{\mu}^{\star}=\arg\min_{u\in\mathcal{U}} J_{\mu}(u).
\end{equation}
In the parametric setting of interest, however, the practical target is not one optimizer per regime but a unified policy that performs well over a continuum of $\mu$. This motivates the policy-level objective
\begin{equation}
\pi^{\star}=\arg\min_{\pi}\ \mathbb{E}_{\mu\sim\rho(\mathcal{P})}\big[J_{\mu}(\pi)\big],
\end{equation}
with $u(t)=\pi(y(\cdot,t),\mu)$ and sampling measure $\rho$ over operating conditions. Equivalently, in value-function form,
\begin{equation}
\begin{aligned}
    V^{\pi}(y_0,\mu) = \mathbb{E}\biggl[ \int_0^T \Bigl( &\|y(\cdot,t;\mu)-y_{\mathrm{ref}}(\cdot,t)\|_{L^2(\Omega)}^2 
    + \alpha\|u(t)\|_2^2 \Bigr) dt \biggr],
\end{aligned}
\end{equation}
and the goal is to minimize $V^{\pi}$ jointly across initial conditions and parameters.

The core challenge is handling nonlinear chaos, actuator constraints, and parameter variability without per-regime retraining. Adjoint/model-based methods can work for a fixed point, but recomputation across dense parameter continuums is costly \cite{bewley2001, kim2007, botteghi2025hyperl}; this motivates the parameter-conditioned RL pipeline developed next. This supports our parameter-conditioned policy architecture (Section~\ref{sec:hypernetwork}).

\subsubsection{From Controlled KS PDE to Optimal Control and RL}
For numerical control experiments, we instantiate the above formulation using the 1D forced KS equation on a periodic domain with $L=22$,
\begin{equation}
\begin{aligned}
    y_t + y y_x + y_{xx} + y_{xxxx} = &\mu \cos\!\left(\frac{4\pi x}{L}\right) + \sum_{i=1}^{N_a} u_i(t)\, g_i(x),
\end{aligned}
\end{equation}
with $\mu\in[-0.225,0.225]$, $N_a=8$ Gaussian actuators, and bounded amplitudes $u_i(t)\in[-1,1]$. The Gaussian actuator kernels use periodic distance and fixed width,
\begin{equation}
    g_i(x)=A\exp\!\left(-\left(\frac{\mathrm{dist}(x,c_i)}{\sigma}\right)^2\right),
\end{equation}
with $A=1.0$ and $\sigma=0.8$.

This controlled PDE is cast as a finite-horizon constrained optimal-control problem on admissible controls $\mathcal{U}$:
\begin{equation}
    \min_{u\in\mathcal{U}}\; J_\mu(u)=\int_0^T \ell\big(y(\cdot,t;\mu),u(t)\big)\,dt,
\end{equation}
with stage cost $\ell(y,u)=\|y-y_{\mathrm{ref}}\|_{L^2(\Omega)}^2+\alpha\|u\|_2^2$ and $\alpha>0$. This directly exposes the trade-off between stabilization quality and control energy. In continuous time, the associated value function is
\begin{equation}
    V(y,t;\mu)=\inf_{u\in\mathcal{U}}\int_t^T \ell\big(y(\tau),u(\tau)\big)\,d\tau,
\end{equation}
which leads formally to the Hamilton--Jacobi--Bellman framework for optimal feedback. For turbulent-like KS regimes with parametric uncertainty, solving that PDE directly at every $\mu$ is computationally prohibitive.

After temporal discretization with control interval $\Delta t_{\mathrm{ctrl}}$, the same problem is written as an MDP $\mathcal{M}=(\mathcal{S},\mathcal{A},\mathcal{P},\mathcal{R},\gamma)$ with state $s_k=[y(x,t_k),\mu]$, bounded continuous action $u_k\in[-1,1]^{N_a}$, and transitions induced by the KS CFD solver. RL then seeks
\begin{equation}
    \pi^* = \arg\max_{\pi}\; \mathbb{E}_\pi\!\left[\sum_{k=0}^{K-1}\gamma^k r_k\right],
\end{equation}
with Bellman optimality
\begin{equation}
    Q^*(s,u)=\mathbb{E}\left[r+\gamma\sup_{u'\in\mathcal{A}}Q^*(s',u')\mid s,u\right].
\end{equation}

To keep optimization consistent with the continuous objective, we define reward from tracking error and control effort:
\begin{equation}
    r_k = -\frac{1}{2T_{\max}}\left(\|e_k\|_{L^2(\Omega)}^2 + \alpha \frac{L}{N} \|u_k\|_2^2\right),
\end{equation}
where $e_k=y_k-y_{\mathrm{ref}}$, $\|f\|_{L^2(\Omega)}^2\approx\tfrac{L}{N}\sum_{i=1}^N f_i^2$, $\alpha=0.1$, and $T_{\max}=250$. Note that the spatial integral normalization ($\frac{L}{N}=\Delta x$) is applied to both the state tracking error and the squared Euclidean norm of the discrete control vector, ensuring dimensional consistency between the physical space and the actuator amplitudes. This normalization keeps per-step reward magnitudes comparable across trajectories while preserving the intended stabilization-effort trade-off.
With this sign convention, maximizing return is equivalent to minimizing a discounted version of the tracking-effort objective; in practice we use $\gamma\approx 1$ to retain a long effective horizon while keeping temporal-difference targets stable.

In our setting, this classical formulation is conceptual: high-dimensional states/actions and expensive PDE transitions require function approximation, motivating the specific DRL realization developed in Section~\ref{sec:rl_to_drl} and validated experimentally in Section~\ref{sec:exp_setup}.

\subsubsection{CFD Process}
The CFD pipeline is designed to preserve stiff KS dynamics while supporting high-throughput rollout generation on GPU. Spatial derivatives are computed spectrally on a periodic grid and time advancement uses ETDRK4 \cite{kassam2005}. Following the Kassam--Trefethen contour-integral construction, ETDRK4 coefficients are precomputed with 32 complex roots in high precision (CPU float64/complex128) and then reused in GPU training (float32) to avoid runtime instability. For the quadratic nonlinearity, we apply the standard $3/2$-rule de-aliasing (pad in Fourier space, compute $y^2$ in real space, then truncate), which reduces aliasing artifacts during long chaotic rollouts.

To scale rollout generation we implement a zero-copy, GPU-native environment and massively parallel ensemble of KS instances, following prior multi-environment and HPC-focused efforts in flow-control RL \cite{rabault2019multiEnv, kurz2022jocs, wang2022drlinfluids}. This parallelization strategy trades per-step latency for sustained wall-clock throughput and is essential for our off-policy training loop that reuses large replay buffers \cite{seo2025fasttd3,kurz2022relexi}.

Solver settings (solver substep $\Delta t=0.1$ combined with time substepping and frameskip) were chosen to balance numerical stability and control cadence. Time substepping stabilizes stiff gradients while frameskip reduces the effective control frequency to match actuator bandwidth and amortize compute, a pragmatic choice consistent with prior KS/CFD-RL work \cite{rabault2019multiEnv,kassam2005}. In our default setup each control action is held across four solver substeps, yielding an effective control cadence $\Delta t_{\mathrm{ctrl}}=0.2$ in the RL loop. The controlled forcing parameterization, actuator layout, and training-time $\mu$ range are defined in Section~\ref{sec:ks_control} and are used unchanged in the CFD rollout engine.

Initial states are generated from randomized multi-mode sine superpositions (8 modes), normalized to fixed energy, and then evolved through an uncontrolled burn-in phase to reach attractor-like chaotic patterns before logging transitions. This initialization-plus-burn-in protocol increases trajectory diversity and reduces synchronized transients across parallel environments, consistent with earlier RL-for-flow studies \cite{bucci2019,rabault2019multiEnv}. Episodes are terminated early on numerical instability (e.g., NaN or large-amplitude blow-up) to prevent corrupted samples from entering the replay buffer \cite{wang2022drlinfluids}.

To prevent unphysical long-time drift, the solver explicitly controls the $k=0$ Fourier coefficient. In the experiments reported here (see Section~\ref{sec:results} for definitions), we enforce mean-zero (zero the $k=0$ mode) for Case 1 (zero-reference stabilization) and Case 2 (four-mode cosine tracking, which has zero spatial mean). For Case 3 (four-mode cosine tracking with a non-zero mean) we instead pin the $k=0$ mode to the non-zero value, enabling offset tracking without drift.

\subsection{RL to DRL}\label{sec:rl_to_drl}
Section~\ref{sec:ks_control} defines the KS control objective, MDP, and reward. We now realize that formulation with deep function approximation using a deterministic actor and distributional critics. The policy is parameterized as
\begin{align}
    u_k &= \pi_{\theta}(s_k), \quad
    s_k = [y(x,t_k),\mu], \quad u_k \in [-1,1]^{N_a}, \nonumber
\end{align}
and is trained off-policy from replayed transitions $(s_k,u_k,r_k,s_{k+1})\sim\mathcal{D}$.

Off-policy actor-critic families such as TD3 are commonly preferred in continuous-action flow-control problems because they balance sample efficiency and stability under function approximation \cite{fujimoto2018td3,seo2025fasttd3,suttonbarto2018}.

As baseline, TD3 uses twin scalar critics and a deterministic actor. With smoothed target action
\begin{equation}
    \tilde{u}_{k+1}=\mathrm{clip}\!\left(\pi_{\theta^-}(s_{k+1})+\epsilon,\,-1,1\right),
\end{equation}
where $\epsilon\sim\mathrm{clip}(\mathcal{N}(0,\sigma_n^2),-c,c)$ is target policy noise,
the TD3 target is
\begin{equation}
    y_k^{\mathrm{TD3}} = r_k + \gamma\, \min_{i\in\{1,2\}} Q_{\phi_i^-}(s_{k+1},\tilde{u}_{k+1}),
\end{equation}
and critic fitting minimizes
\begin{equation}
    \mathcal{L}_{\mathrm{TD3}}(\phi_i)=\mathbb{E}_{\mathcal{D}}\left[\left(Q_{\phi_i}(s_k,u_k)-y_k^{\mathrm{TD3}}\right)^2\right].
\end{equation}
This baseline is useful, but it approximates only a point estimate of return.

Our final critic design uses \textbf{Truncated Quantile Critics (TQC)} \cite{kuznetsov2020tqc} on top of this TD3 backbone \cite{fujimoto2018td3}. Rather than regressing a scalar estimate, we adopt a distributional RL perspective \cite{bellemare2017distributional} in which each critic predicts a return distribution via quantile atoms \cite{dabney2018distributional}. Target construction discards the highest quantiles to obtain conservative Bellman targets in chaotic regimes. Let each critic output $M$ quantiles and let $d$ denote the number of truncated top atoms after pooling/sorting target quantiles. The resulting critic objective is quantile Huber regression:
\begin{equation}
    \mathcal{L}_Q(\phi)=\frac{1}{B}\sum_{k=1}^{B}\sum_{m=1}^{M}\sum_{j=1}^{2M-d}
    \rho_{\hat{\tau}_m}\!\left(Y_j-q_m(s_k,u_k;\phi)\right),
\end{equation}
where $Y_j$ are truncated quantile targets. Relative to TD3's scalar target, this provides a richer approximation of the return law and is intended to improve target robustness under heavy-tailed or intermittent returns, which is relevant in chaotic PDE control where rare high-disturbance scenarios can dominate learning.

This choice is grounded in recent applications reporting that quantile-based distributional methods can improve robustness under noisy or heterogeneous reward signals across diverse domains \cite{foo2023drl} such as active flow control \cite{xia2024activeflowcontrol}, including reward-model robustness settings \cite{dorka2024qrm}.

Actor updates use deterministic policy gradients with delayed target-network updates, as in TD3. The next subsection introduces parameter-conditioned function approximation, and Section~\ref{sec:fasttd3} then describes the high-throughput optimization schedule used to train that combined design.

\subsection{Hypernetwork and its Variants}\label{sec:hypernetwork}
Standard DRL architectures for parametric control often rely on a single fixed set of weights to represent feedback laws across all physical regimes. In chaotic PDE settings, this forces the same parameters to encode both the spatial control map and the regime-dependent adaptation, which can induce interference between tasks and degrade generalization \cite{keynan2021}. Hypernetworks address this limitation by letting the conditioning variable determine the policy weights themselves, rather than asking one static controller to cover the entire parameter family \cite{ha2016hypernetworks, botteghi2025hyperl}. Naive concatenation of semantically distinct inputs (e.g., state and action in Q-functions, or state and context in policies) can lead to poor gradient approximation in actor-critic algorithms and high learning-step variance; conditioning on a low-dimensional context via a primary network that generates the dynamic weights of actor and critic has been shown to improve gradient quality and reduce variance \cite{keynan2021}.

To strictly decouple the contextual parameter from the high-frequency spatial observation, we employ \textbf{Hypernetworks} \cite{ha2016hypernetworks}. A Hypernetwork encoder $H_\phi$, parameterized by $\phi$, serves as a primary neural network that ingests only the scalar $\mu$ and outputs the complete set of weights for both the actor and critic networks:
\begin{equation}
    \theta_\pi, \theta_Q = H_\phi(\mu)
\end{equation}
Consequently, both the policy and value functions operate entirely on the spatial manifold, while their functional topologies and filter strengths are dynamically instantiated by the Hypernetwork based on the physical regime. This separation of parametric adaptation (Hypernetwork) from spatial feedback control (conditioned networks) is used to reduce cross-regime interference without retraining a separate controller per parameter value. Prior work has shown that hypernetwork conditioning can improve cross-regime generalization in parametric control tasks \cite{ha2016hypernetworks,keynan2021,botteghi2025hyperl}.

\paragraph{Architectural refinements for parametric embeddings.}
Mapping the low-dimensional scalar $\mu$ into a massive, expressive space of policy weights requires overcoming the neural \emph{spectral bias}: the extensively documented phenomenon where standard MLPs struggle to learn high-frequency mappings from low-dimensional inputs \cite{rahaman2019}. In our setting, this is naturally a function-space approximation problem: the encoder must represent both smooth global trends and sharper regime-dependent structure in the map $\mu \mapsto \theta$ while remaining stable under high-throughput optimization.

\begin{figure}[!htb]
    \centering
    \begin{subfigure}{0.49\linewidth}
        \centering\includegraphics[width=0.8\textwidth]{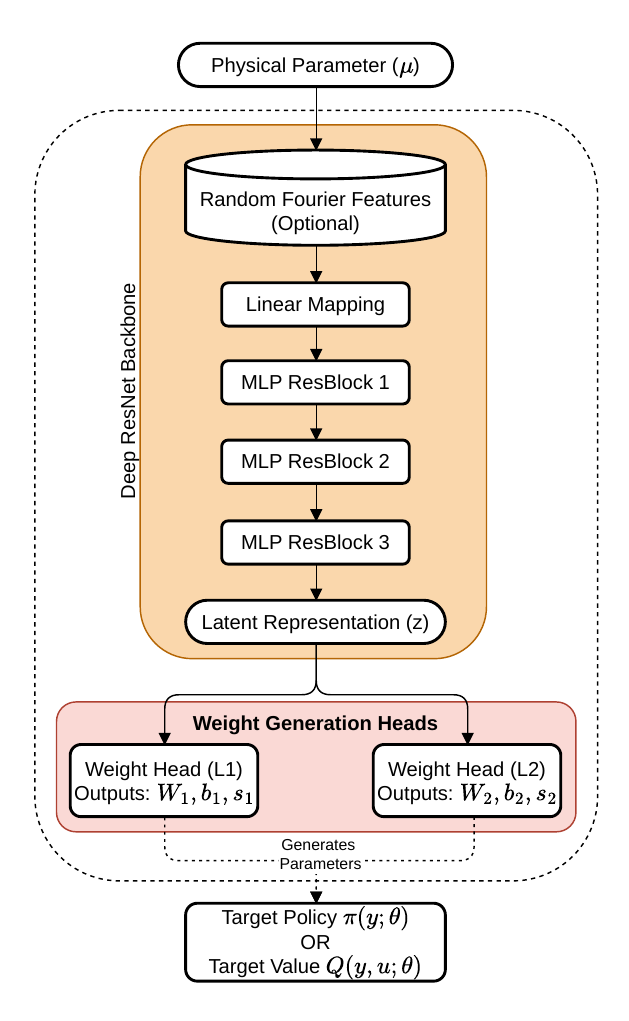}
    \end{subfigure}
    \begin{subfigure}{0.49\linewidth}
        \centering\includegraphics[width=0.8\textwidth]{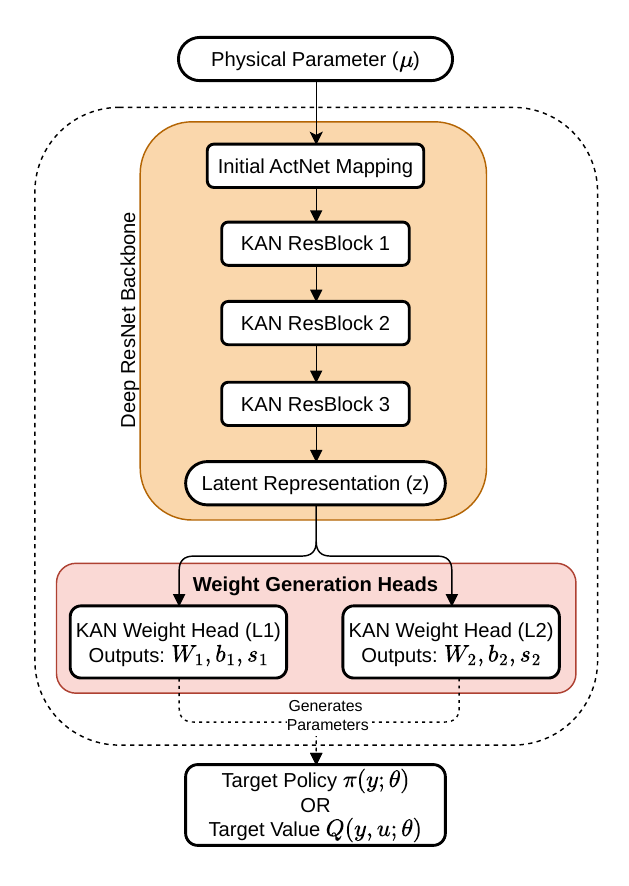}
    \end{subfigure}
    
    \caption{Detailed internal structure of the Hypernetwork \cite{keynan2021} encoders. \textbf{Left}: The ResNet backbone processes the parameter latent $z$, which is then split into respective heads to generate the exact topological weights, biases, and scales for the target network. \textbf{Right}: The ActNet backbone maps the physical parameter $\mu$ into a latent space, while parallel ActNet weight heads dynamically generate the full target network weights.}
    \label{fig:hypernet_structures}
\end{figure}

We explore two advanced primitives to supersede the standard MLP backbone in the Hypernetwork (see Figure~\ref{fig:hypernet_structures} for the internal topologies). This is motivated by a practical question used later in Section~\ref{sec:exp_setup}: whether richer parameter embeddings improve cross-regime behavior under a fixed training protocol. First, we employ \textbf{Random Fourier Features (RFF)} \cite{tancik2020fourierFeatures}. The original RFF approach expands $\mu$ into a periodic space via sine/cosine projections; we extend this by also concatenating the original scalar:
\begin{equation}
    \gamma(\mu) = \bigl[\mu,\; \sin(2\pi \sigma\, \mathbf{B} \mu),\; \cos(2\pi \sigma\, \mathbf{B} \mu)\bigr]
\end{equation}
where $\mathbf{B}$ is a frozen matrix with i.i.d.\ $\mathcal{N}(0,1)$ entries and $\sigma$ is a frequency scale. This concatenation of the original state $\mu$ (prepended identity skip) supplies a non-periodic global coordinate that can stabilize behavior outside the strict training grid. Second, we integrate the \textbf{Kolmogorov-Arnold Network (KAN)} architecture \cite{liu2024kan}, utilizing the computationally efficient \textbf{ActNet} \cite{guilhoto2024actnet} formulation. In ActNet, a hidden feature $x$ is first projected onto a shared sinusoidal basis with learnable frequencies and phases,
\begin{align}
    \psi_k(x) &= \sin(\omega_k^{\mathrm{eff}} x + \phi_k), \quad
    \omega_{k,\ell}^{\mathrm{eff}} = \omega_k w_{0,\ell}, \nonumber
\end{align}
where the original ActNet uses a fixed global scaling constant $w_0$, but our implementation uses a \emph{learnable per-layer} scaling parameter $w_{0,\ell}$ for layer $\ell$. To stabilize optimization, each basis response is analytically normalized using its closed-form Gaussian mean and variance computed with the effective frequencies (assuming normalized pre-activation inputs $x \sim \mathcal{N}(0,1)$, which is structurally enforced via standard LayerNorm in our network backbone):
\begin{align}
    \mathbb{E}[\psi_{k,\ell}] &= e^{-(\omega_{k,\ell}^{\mathrm{eff}})^{2}/2}\sin(\phi_k), \quad
    \mathrm{Var}[\psi_{k,\ell}] = \frac{1}{2} - \frac{1}{2}e^{-2(\omega_{k,\ell}^{\mathrm{eff}})^{2}}\cos(2\phi_k) - \mathbb{E}[\psi_{k,\ell}]^2, \nonumber
\end{align}

before being combined through learnable edge coefficients. In compact form, one ActNet layer can be written as
\begin{equation}
    h' = \sum_{k=1}^{K} \beta_k \Bigl( \widehat{\psi}_k(h) \Lambda \Bigr) + \mathbf{W}_{\mathrm{lin}} h + b,
\end{equation}
where $\widehat{\psi}_k$ denotes the normalized sinusoidal basis, $\Lambda$ and $\beta_k$ are learnable mixing weights, and $\mathbf{W}_{\mathrm{lin}} h$ is a linear residual branch. This construction preserves the expressivity of periodic basis expansions while remaining fully differentiable and computationally compatible with high-throughput backpropagation in PDE control environments.

Related approaches that learn parametric solution operators for PDEs, such as the Fourier Neural Operator and DeepONet, offer an alternative route for handling parametric families of PDEs by directly mapping parameters or initial/boundary data to solution fields \cite{li2021fno, lu2019deeponet}. These neural-operator methods are complementary to hypernetwork-based control: they can accelerate forward prediction or provide surrogate rollouts, while hypernetwork methods focus on producing parameter-conditioned controller weights for closed-loop feedback.

\paragraph{Hypernetwork Weight Generation}
The central design idea of HyperFastRL is to condition both the policy and value functions entirely on the physical parameter, without burdening the state-dependent backbones with multi-task interference (see also the foundational analysis in \cite{keynan2021}).
Given $\tilde{\mu}$, a Hypernetwork $H_\phi$ generates the complete weight tensors of both the target-policy and target-critic networks in a single forward pass:
\begin{align}
    H_\phi(\tilde{\mu}) &\;\longrightarrow\; \bigl\{\,(\mathbf{W}_\ell,\, \mathbf{b}_\ell,\,
    \mathbf{s}_\ell)\,\bigr\}_{\ell=1}^{L}
    \label{eq:hypernet}
\end{align}
where $L$ is the number of target-network layers, and $\mathbf{s}_\ell \in \mathbb{R}^{d_\ell}$
is a learned per-neuron scale initialized near unity,
    $\mathbf{s}_\ell = \mathbf{1} + W_s\,\mathbf{z}$,
    acting as an adaptive feature-wise gain on each generated layer. The target-network forward pass for layer $\ell$ is then:
\begin{equation}
    h_\ell = \text{ReLU}\!\left( \mathbf{s}_\ell \odot (\mathbf{W}_\ell\, h_{\ell-1})
    + \mathbf{b}_\ell \right)
    \label{eq:target_fwd}
\end{equation}
with $\text{softsign}(\cdot)$ replacing ReLU at the final actor layer, eliminating the
gradient saturation of $\tanh$ while bounding actions to $[-1, 1]$.

In vectorized training, many samples can share the same conditioning parameter. The
Hypernetwork therefore needs to be evaluated only on the unique parameter values in a
mini-batch, and the resulting weights are reused across all matching samples. This
\emph{unique-weight optimization} is formalized as:
\begin{equation}
    \theta_b = H_\phi(\tilde{\mu}_{\sigma(b)}), \quad
    \sigma(b) = \text{UniqueIndex}(\tilde{\mu}_b)
    \label{eq:unique_weight}
\end{equation}
which reduces redundant Hypernetwork evaluations; since each unique parameter value
    must materialise a full weight tensor in GPU memory, this deduplication yields
    substantial VRAM savings when many batch samples share the same conditioning parameter.

The Hypernetwork backbone is a three-stage ResNet with width-doubling stages
$(256 \!\to\! 512 \!\to\! 1024)$ \cite{ha2016hypernetworks}, each stage containing two stacked residual blocks followed by LayerNorm.
Spectral normalisation is applied to every linear layer in the backbone and output heads to
control the Lipschitz constant of $H_\phi$ and improve optimization stability \cite{miyato2018}.
Concrete architectural widths and implementation hyperparameters are deferred to Section~\ref{sec:exp_setup} and Appendix~\ref{sec:app_hyper}, where the comparison protocol is defined.

\subsection{Training with FastTD3: High-Throughput Implementation}\label{sec:fasttd3}
Here we extend on previous sections and focus on unique training implementation details for the high-throughput FastTD3/TQC pipeline. 

The control problem is the MDP $(\mathcal{S}, \mathcal{A}, \mathcal{P}, \mathcal{R}, \gamma)$ from Section~\ref{sec:ks_control}.
To avoid ambiguity in later figures, we distinguish between two time horizons used in this work. In the RL objective above, $T_{\max}=250$ refers to the reward-normalization control horizon (250 control steps at $\Delta t_{\mathrm{ctrl}}=0.2\,$s, i.e., 50 s). For qualitative spacetime heatmaps, we intentionally use a longer visualization rollout of $T_{\mathrm{heat}}=1000$ control steps (200 s), with control activated after step 500, so the plots show both pre-control and post-control behavior in one panel.

\begin{figure*}[!htb]
    \centering
    \includegraphics[width=0.9\textwidth]{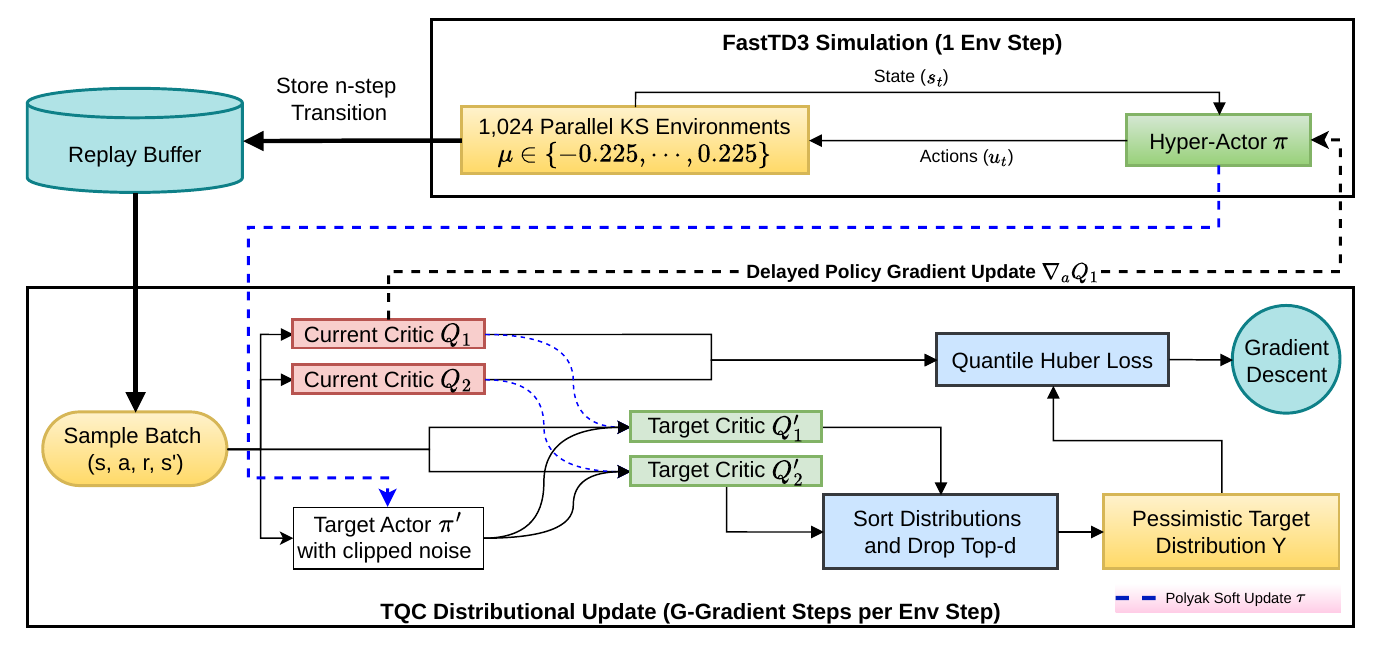}
    \caption{FastTD3 and TQC Optimization Workflow. 1,024 KS environments are simulated in parallel with $\mu \in \{-0.225, \dots, 0.225\}$. Rollouts populate a replay buffer which is sampled to compute a conservative target distribution via Truncated Quantile Critics by sorting pooled atoms from twin critics and dropping the top-$d$ values (e.g., $d=5$ out of $2M=50$, i.e., 10\%), achieving robust continuous control under chaotic forcing.}
    \label{fig:tqc_parallel_rl}
\end{figure*}

This formulation corresponds directly to the KS-RL setting already introduced in Section~\ref{sec:ks_control}, with the same actor-critic specification (TD3/FastTD3).
Training uses $N_\text{env} = 1024$ parallel independent KS instances with staggered initial
conditions, whose transitions are stored in an N-step replay buffer
($n = 3$, buffer size $= 4 \times 10^6$) \cite{suttonbarto2018,stableBaselines2024vecEnv,stableBaselines2025rlTips}.
Observations are z-score normalized online via running Welford statistics (mean and variance) \cite{welford1962,ji2022robustRlNonlinear,liu2021}; rewards are scaled by their running standard deviation without mean-centering, preserving the sign of the episodic return while stabilizing critic training.
The full training procedure which couples the parallel environment rollouts with the gradient-update pipeline is summarized in Figure~\ref{fig:tqc_parallel_rl} and detailed in Algorithm~\ref{alg:hyperfastrl}.

\begin{algorithm}[htb]
\caption{HyperFastRL Training Loop}
\label{alg:hyperfastrl}
\begin{algorithmic}[1]
    \Require Parallel KS environments $\{\mathcal{E}_k\}_{k=1}^{N_\text{env}}$,
          Hypernetwork $H_\phi$, Actor $\pi$, Critic $Q^{(1)}, Q^{(2)}$,
          Target networks $H_{\phi^-}\!, \pi^-, Q^{(1)-}\!, Q^{(2)-}$,
          Replay buffer $\mathcal{D}$, gradient steps $GS$, batch size $B$, $\tau$
    \State Initialize all networks; populate $\mathcal{D}$ with random rollouts
    \For{each environment step}
     \State Observe $s_t = [y_t, \tilde{\mu}]$ from all $N_\text{env}$ environments
     \State $\theta_\pi \leftarrow H_\phi(\tilde{\mu})$ \Comment{Generate actor weights}
     \State $u_t \leftarrow \text{softsign}(\pi(y_t;\,\theta_\pi)) + \epsilon,\quad
         \epsilon \sim \mathcal{N}(0,\, 0.05^2)$
     \State Step environments; store $n$-step transitions in $\mathcal{D}$
     \State Update obs.~normalizer with $s_t$; update reward normalizer with $r_t$
    \For{$g = 1,\ldots, GS$}
         \State Sample mini-batch $\{(s, u, r_n, s', \gamma^n)\} \sim \mathcal{D}$ of size $B$
         \State $(\theta^-_\pi,\,\theta^-_Q) \leftarrow H_{\phi^-}(\tilde{\mu}')$
         \Comment{Generate target networks for next state}
         \State $u' \leftarrow \pi^-(y';\,\theta^-_\pi)
             + \text{clip}(\mathcal{N}(0,0.2^2),\,-0.5,\,0.5)$
         \State Compute TQC target $Y$ by pooling, sorting, and dropping top-$d$ quantiles
             from target atoms $\{Z^{(j)-}(s', u';\, \theta^-_Q)\}_{j=1}^{2}$
         \State $(\theta_\pi, \theta_Q) \leftarrow H_\phi(\tilde{\mu})$
         \Comment{Generate current networks}
         \State Update hypernetwork critic parameters $\phi_Q$ by minimizing Quantile Huber Loss 
             between $Y$ and $\{Z^{(j)}(s, u;\, \theta_Q)\}_{j=1}^{2}$
         \If{$g \bmod 2 = 0$} \Comment{Delayed actor update}
          \State Update hypernetwork actor parameters $\phi_\pi$ by maximizing 
              $\frac{1}{B}\sum Q^{(1)}(s,\pi(y;\,\theta_\pi);\, \theta_Q)$
          \State Polyak update: $\phi^- \leftarrow (1-\tau)\phi^- + \tau\phi$
              for all target networks
         \EndIf
     \EndFor
    \EndFor
\end{algorithmic}
\end{algorithm}

HyperFastRL adopts the \textbf{FastTD3} training protocol \cite{seo2025fasttd3}, where experience collection is decoupled from optimization and multiple critic/actor updates can be performed per environment interaction. This is motivated by the fundamental efficiency trade-off between gradient updates and environment interactions in off-policy RL. While reusing buffer experience improves wall-clock sample efficiency, over-aggressive regimes risk policy mismatch: Liu et al. \cite{liu2025_rl_collapse} analyze collapse modes under high update frequency, Goodall et al. \cite{goodall2025_behaviour_policy_opt,goodall2024bpo} bound variance in behavior-policy estimation, and unified analyses (e.g., \cite{luo2024ompo, kallus2020_statistically}) motivate an explicitly controlled reuse ratio. This update-to-data mechanism is summarized by
\begin{equation}
    \text{Reuse Ratio} = \frac{GS \cdot B}{N_\text{env}}
\end{equation}
where $GS$ is the number of gradient updates per environment step, $B$ is the mini-batch
size, and $N_\text{env}$ is the number of parallel environments. Specific values are
reported in the Experimental Setup section.

Rather than the standard twin-critic minimum, we use \textbf{Truncated Quantile Critics
(TQC)} \cite{kuznetsov2020tqc} to produce conservative value targets. This explicitly targets gap (ii) from Section~\ref{sec:intro}, where chaotic reward distributions can induce severe overestimation bias in scalar critics; related flow-control studies have also reported practical robustness benefits from distributional quantile critics \cite{xia2024activeflowcontrol}. Each critic predicts a set of
quantile atoms; these atoms are pooled across target critics, sorted, and the largest
tail is truncated before constructing Bellman targets:
\begin{align}
    Y_j &= R^{(n)}_t + \gamma^n \cdot Z^{-}_{(j)}, \\
    j &= 1,\ldots, 2M - d, \nonumber
\end{align}
where $R^{(n)}_t = \sum_{i=0}^{n-1} \gamma^i r_{t+i}$ is the $n$-step discounted return, $Z^{-}_{(j)}$ denotes the $j$-th sorted quantile from the pooled target set at step $t+n$,
$M$ is the number of quantiles per critic, and $d$ is the number of truncated top atoms.
Each critic is then updated by minimizing the Quantile Huber loss:
\begin{equation}
    \mathcal{L}_Q(\phi_Q) = \frac{1}{B} \sum_{i=1}^{B} \sum_{m=1}^{M}
    \sum_{j=1}^{2M - d}
    \rho_{\hat{\tau}_m}\!\left( Y_j - q_m(s_i, u_i;\, \phi_Q) \right)
\end{equation}
where $\hat{\tau}_m = (m - 0.5)/M$ are uniform quantile midpoints and $\rho_\tau$ is the
asymmetric Huber loss:
\begin{equation}
\begin{aligned}
    \rho_\tau(\delta) &= \bigl|\tau - \mathbf{1}[\delta < 0]\bigr| \cdot \mathcal{L}_\delta(\delta), \\
    \mathcal{L}_\delta(\delta) &=
    \begin{cases} 
        \frac{1}{2}\delta^2 & |\delta| \leq 1 \\ 
        |\delta| - \frac{1}{2} & \text{otherwise} 
    \end{cases}
\end{aligned}
\end{equation}
This distributional treatment biases value estimates downward in highly chaotic
environments, mitigating the overestimation-driven policy collapses common when applying
standard TD3 to the KS equation.

\subsection{Experimental Setup}\label{sec:exp_setup}
The present study is intentionally scoped to a controlled 1D KS benchmark with a scalar forcing parameter, as HypeRL has shown parameter-conditioning to be advantageous in 1D and 2D flow applications \cite{botteghi2025hyperl}. Thus, we interpret results as benchmark-level evidence for training stability, parametric adaptation,
and practical control performance in this setting, rather than as universal claims across PDE classes.
Interpolation and mild extrapolation tests are treated as structured distribution-shift checks within a
narrow one-dimensional parameter family \cite{tobin2017,pinto2017}. For fairness, all encoder variants share the same RL pipeline, target-network topology, optimizer schedule,
evaluation protocol, and seed set; only the Hypernetwork encoder is changed. Because backbone sizes are close
but not perfectly matched (Table~\ref{tab:param_counts}), architecture comparisons are interpreted as practical
protocol-controlled comparisons rather than strict capacity-controlled causal attribution.
Finally, uncertainty estimates are based on five seeds, which are sufficient for trend-level confidence but not for definitive significance claims. Reported runtimes in the ablation and single-seed experiments (e.g., Section~3.2) are single-seed wall-clock times; runtime values given for the architecture-comparison (Section~3.3) are cumulative across the five seeds and reported as aggregate wall-clock time. In our setup, online reward normalization is used only inside the critic-update pipeline during training; all train/eval/test rewards reported in this section are raw episodic returns from the environment, so values remain directly comparable across architectures. The full shared hyperparameter table is provided in Appendix A (Table~\ref{tab:hyperparams}). All reported runtime measurements in this section are training wall-clock times recorded on the UTK ISAAC HPC cluster (H100 GPUs).

\noindent\textbf{Setup summary.}
\begin{itemize}
    \item \textbf{Training parameter sweep:} forcing parameters are sampled from the 19-point grid
    \[
    \mu \in \{-0.225 + k\Delta\mu\}_{k=0}^{18},\qquad \Delta\mu=0.025.
    \]
    \item \textbf{Post-training test set:} seven representative seen values from the training grid
    ($\mu \in \{-0.225, -0.15, -0.075, 0.0, 0.075, 0.15, 0.225\}$), plus one unseen interpolation point ($\mu=0.1125$)
    and one mild extrapolation point ($\mu=-0.25$).
    \item \textbf{Exploration phase:} the first 5\% of total timesteps are collected with purely random actions
    (no learned policy control), corresponding to approximately 375,000 steps in the 7.5M-step budget.
    \item \textbf{Reset protocol:} each environment reset uses staggered initialization across parallel workers and
    applies a burn-in of 100 solver steps before control rollouts are logged, improving trajectory decorrelation and
    reducing near-identical initial transients.
\end{itemize}

\section{Results}\label{sec:results}

We evaluate \emph{hyperFastRL} on the parametric KS control task described in
Section~\ref{sec:exp_setup}. The core contribution tested in this section is the coupled
\textbf{parameter-conditioned Hypernetwork + FastTD3/TQC} training framework,
with three encoder instantiations for the Hypernetwork. The residual
\textbf{MLP} serves as the baseline encoder, implemented as the HypeRL-style
parameter-conditioned MLP backbone \cite{botteghi2025hyperl} trained with the
FastTD3+TQC \cite{seo2025fasttd3, kuznetsov2020tqc} optimization stack introduced in this work; the \textbf{Fourier Feature}
and \textbf{ActNet-KAN} encoders are the two novel architectures introduced here.
All experiments are run for $7.5 \times 10^6$ environment steps.
Multi-seed results are reported as mean with 95\% confidence intervals over five
independent seeds, providing an uncertainty quantification of performance across random initialisations. All three encoders share the same target-network topology, optimizer settings, rollout budget, evaluation protocol, and seed schedule; only the Hypernetwork encoder is changed, isolating the contribution of each encoder architecture.

\subsection{Computational Efficiency: Gradient Steps Ablation}\label{sec:efficiency}

Following the theoretical formulation in Section~\ref{sec:fasttd3}, we quantify this efficiency using the Reuse Ratio. For our specific high-throughput configuration ($B=32,768$, $N_{\text{env}}=1024$) detailed in Appendix~\ref{sec:app_hyper}, this relationship simplifies to:
\[
    \text{Reuse Ratio} = 32\,GS.
\]

\begin{figure*}[!htb]
    \centering
    \begin{subfigure}{0.32\textwidth}
        \centering
        \includegraphics[width=\linewidth]{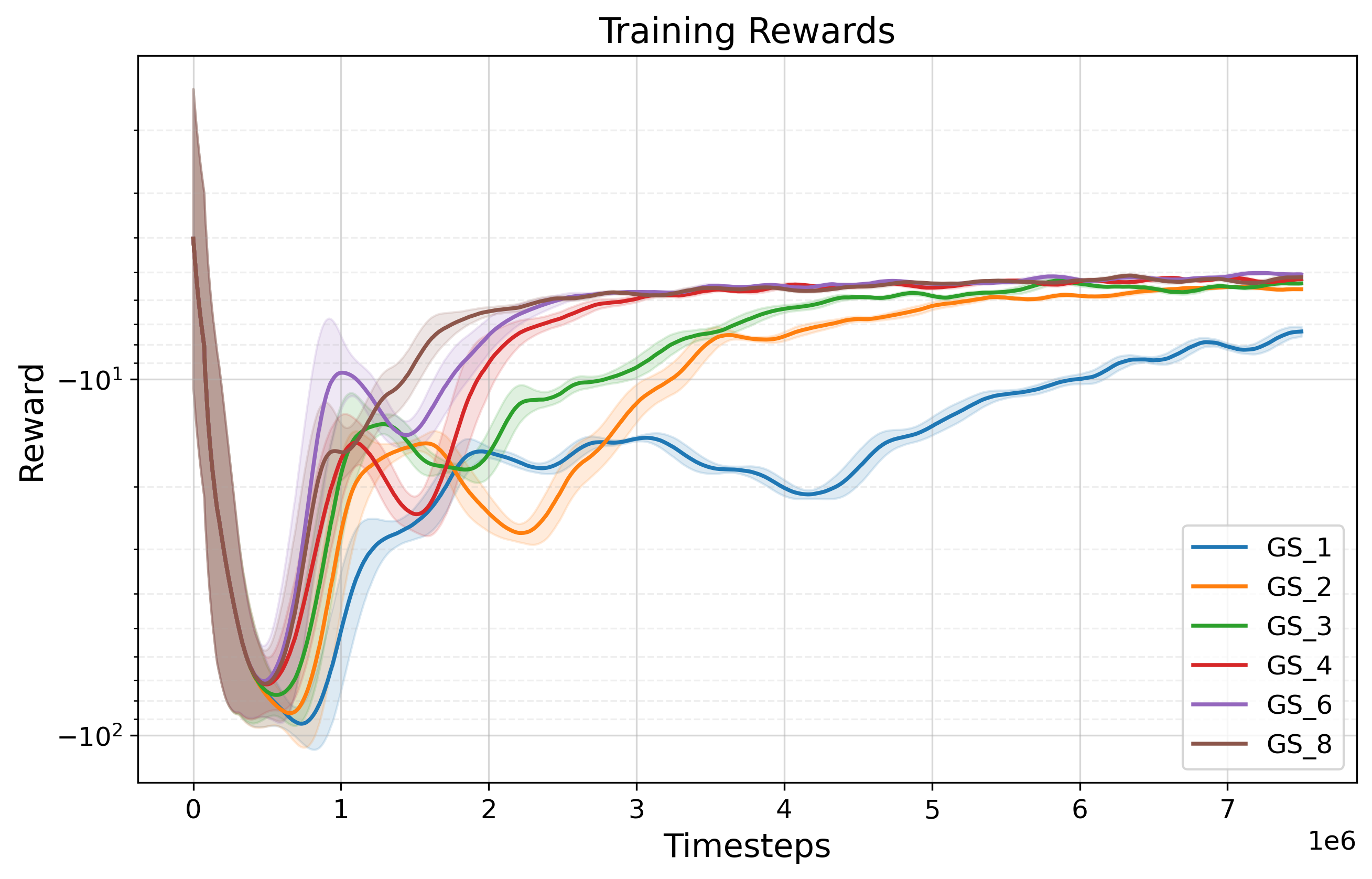}
        \caption{Training reward.}
    \end{subfigure}
    \hfill
    \begin{subfigure}{0.32\textwidth}
        \centering
        \includegraphics[width=\linewidth]{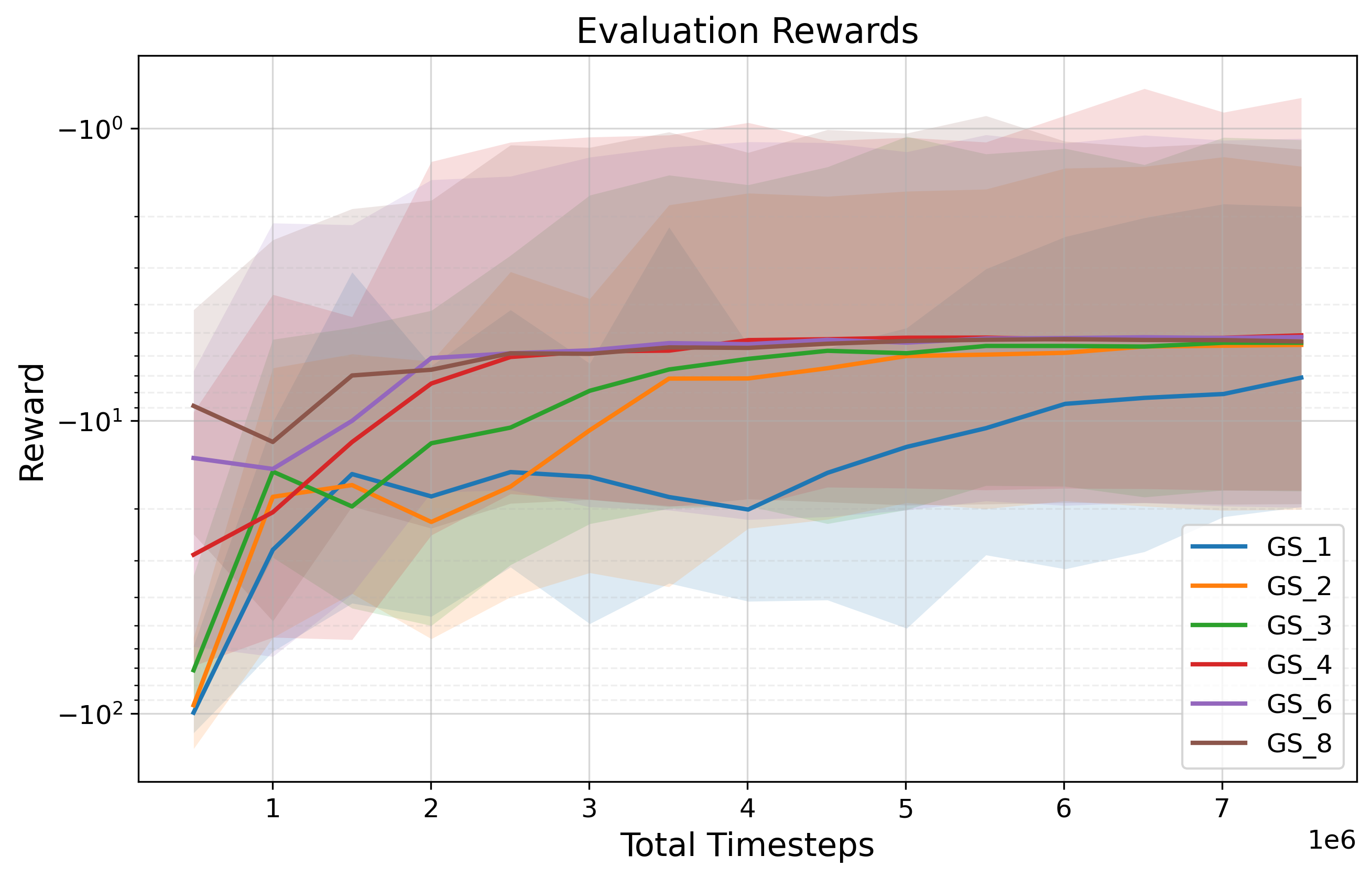}
        \caption{Evaluation reward.}
    \end{subfigure}
    \hfill
    \begin{subfigure}{0.32\textwidth}
        \centering
        \includegraphics[width=\linewidth]{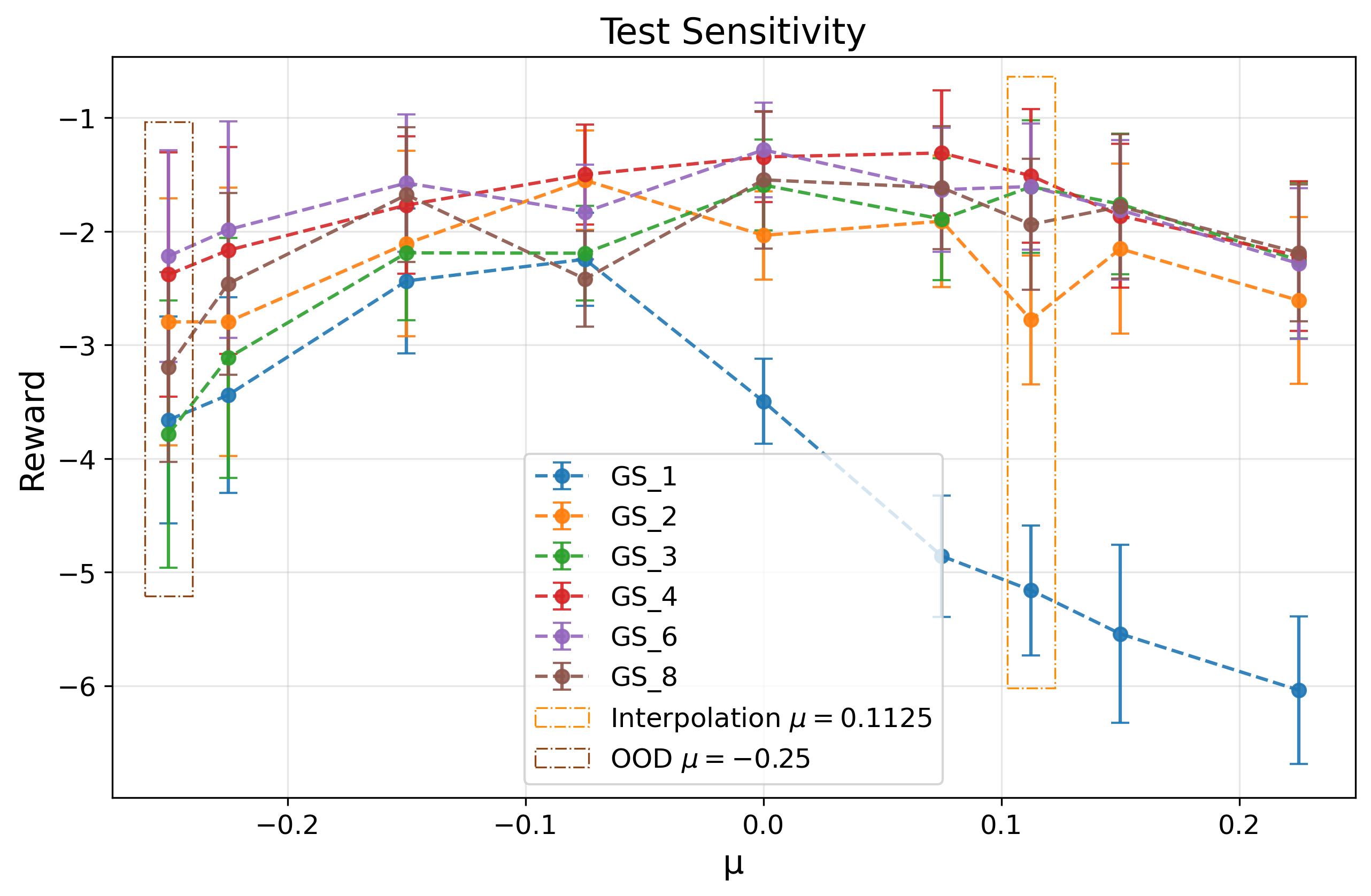}
        \caption{Test reward.}
    \end{subfigure}
    \caption{Training, evaluation, and test reward domains across varying gradient-step (GS) configurations for the baseline MLP encoder. (a) Training curves illustrate sample efficiency, with higher GS yielding faster initial ascent. (b) Evaluation returns over the training period, showing mean and min--max ranges across 10 evaluation episodes per checkpoint. (c) Post-training generalization mapping the final policy across 9 discrete forcing configurations (7 seen training parameters $\mu \in [-0.225, 0.225]$ alongside 2 unseen interpolation/extrapolation targets). While higher $GS$ configurations (e.g., $GS=4, 6$) maximize early data reuse, they exhibit diminishing asymptotic returns and drastically increased wall-clock costs, motivating a moderate $GS=2$ trade-off.}
    \label{fig:gs_curves}
\end{figure*}

Here, we use GS and 'Reuse Ratio' interchangeably. We ablated $GS \in \{1, 2, 3, 4, 6, 8\}$ using the MLP encoder to characterize the throughput/accuracy trade-off of the proposed \emph{hyperFastRL} architecture (Figure~\ref{fig:gs_curves}, Table~\ref{tab:gs_ablation_summary}).

\begin{table*}[!htb]
    \centering
    \caption{Quantitative summary of the gradient-step ablation study (Section~\ref{sec:efficiency}) measuring the continuous-time control throughput against asymptotic accuracy for the MLP baseline. 'Runtime' denotes total wall-clock hours for $7.5 \times 10^6$ environment steps. The test block reports the full generalization range across evaluated $\mu$ configurations. Note how $GS=2$ provides the most balanced compromise between minimizing runtime (39m 11s) and closing the performance gap with higher-reuse regimes.}
    \label{tab:gs_ablation_summary}
    \begin{tabular}{lccccc}
        \toprule
        \textbf{GS} & \textbf{Runtime} & \textbf{Final Train ($\pm\sigma$)} & \textbf{Final Eval ($\pm\sigma$)} & \textbf{Test Range [min, max]} & \textbf{Test $\sigma$} \\
        \midrule
        1 & 29m 32s & -7.32 $\pm$ 0.30 & -7.12 $\pm$ 2.87 & [-6.04, -2.25] & 1.28 \\
        2 & 39m 11s & -5.61 $\pm$ 0.04 & -5.50 $\pm$ 2.57 & [-2.80, -1.55] & 0.43 \\
        3 & 55m 05s & -5.38 $\pm$ 0.08 & -5.41 $\pm$ 2.50 & [-3.78, -1.59] & 0.69 \\
        4 & 1h 02m & -5.20 $\pm$ 0.06 & -5.10 $\pm$ 2.49 & [-2.38, -1.31] & 0.38 \\
        6 & 1h 28m & -5.11 $\pm$ 0.07 & -5.17 $\pm$ 2.46 & [-2.28, -1.28] & 0.30 \\
        8 & 1h 49m & -5.18 $\pm$ 0.07 & -5.36 $\pm$ 2.48 & [-3.20, -1.55] & 0.50 \\
        \bottomrule
    \end{tabular}
\end{table*}

The massively parallel architecture allows us to intentionally navigate the performance--throughput Pareto front. While $GS = 4$ achieves fractionally better peak evaluation scores, transitioning to $GS = 2$ intentionally trades a statistically minor reduction in asymptotic reward for a massive 37\% reduction in training wall-clock time. Because architecture-comparison runs include heavier encoders (especially KAN), where extra gradient updates exponentially amplify computational cost, this efficiency gain is critical. Accordingly, for the main architecture-comparison campaign we adopt a data reuse ratio of 64 ($GS = 2$) as the optimal practical operating point. This configuration balances robust control fidelity with compute-resource tractability as established in Table~\ref{tab:gs_ablation_summary}, and is used consistently throughout Sections~\ref{sec:arch_comparison} and \ref{sec:heatmaps}.

\subsection{Performance Overview: Architecture Encoder Comparison}\label{sec:arch_comparison}

To test whether the GS choice from Section~\ref{sec:efficiency} changes encoder ranking, we compare MLP, Fourier, and KAN across five independent seeds at both GS=2 and GS=4 under the same protocol. This two-setting check is important because Section~\ref{sec:efficiency} ablates GS with the MLP encoder only, while the full architecture study includes heavier encoders.

\begin{figure*}[!htb]
    \centering
    \begin{subfigure}[t]{0.32\textwidth}
        \centering
        \includegraphics[width=\linewidth]{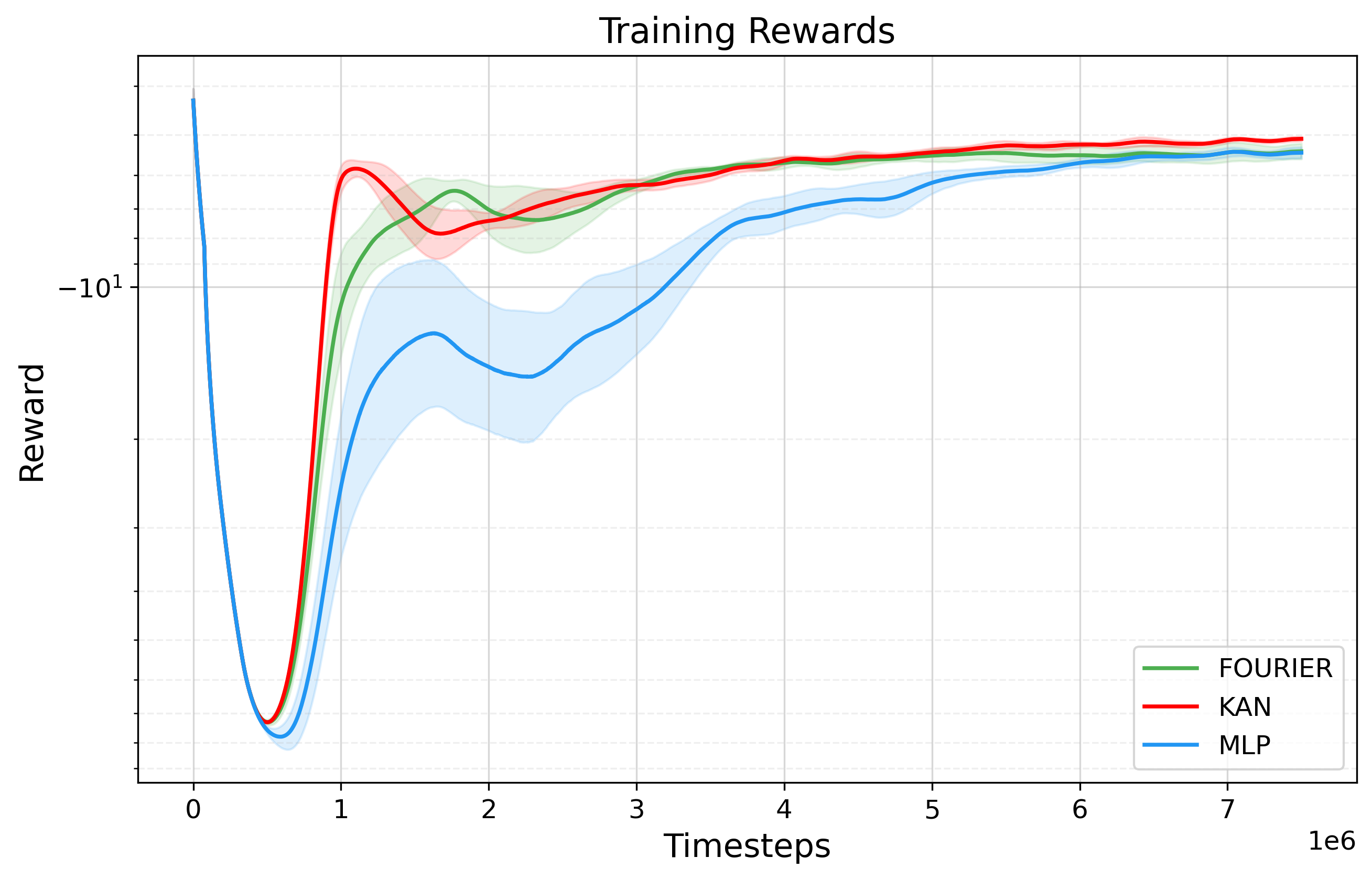}
    \end{subfigure}
    \hfill
    \begin{subfigure}[t]{0.32\textwidth}
        \centering
        \includegraphics[width=\linewidth]{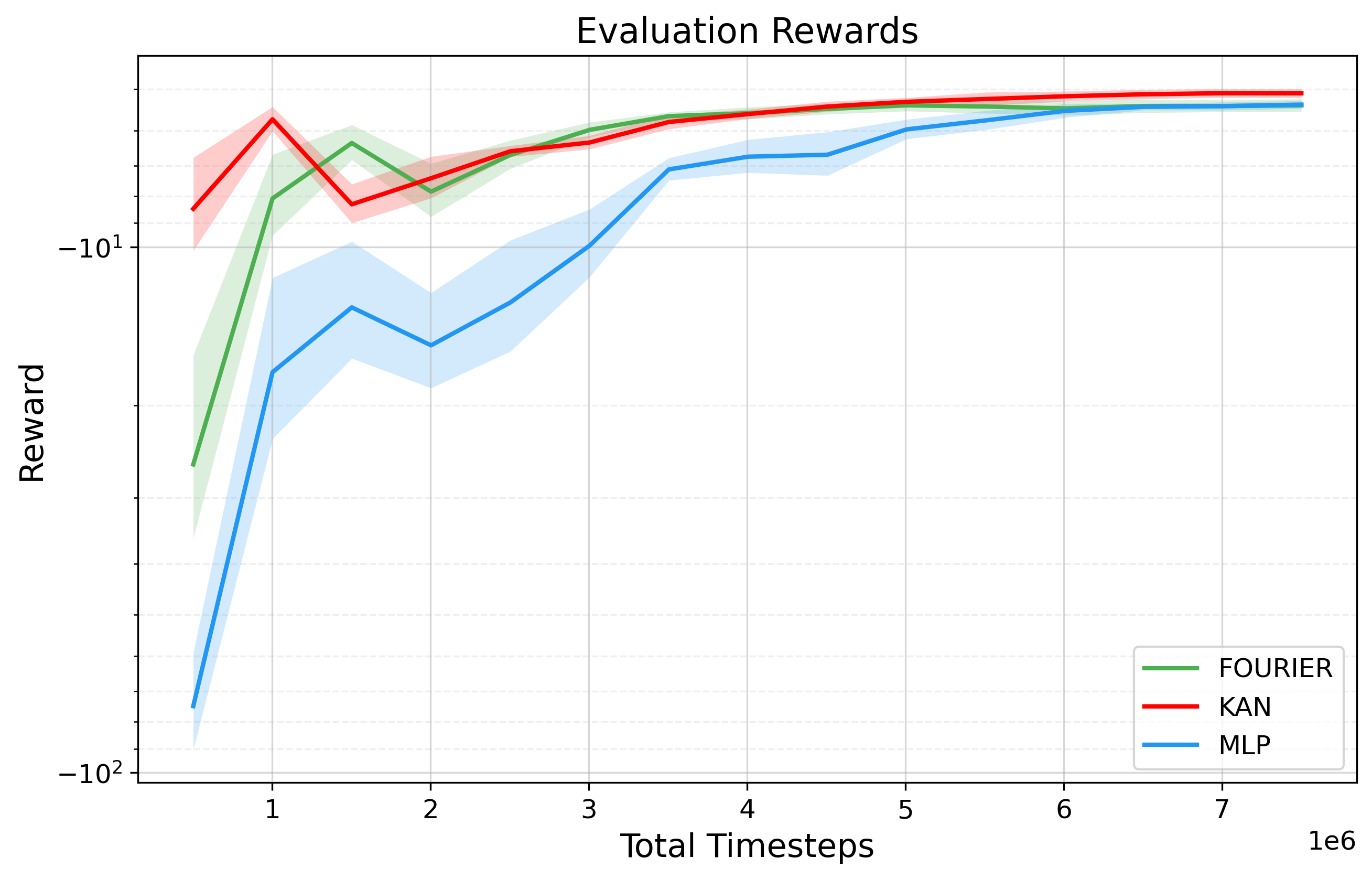}
    \end{subfigure}
    \hfill
    \begin{subfigure}[t]{0.32\textwidth}
        \centering
        \includegraphics[width=\linewidth]{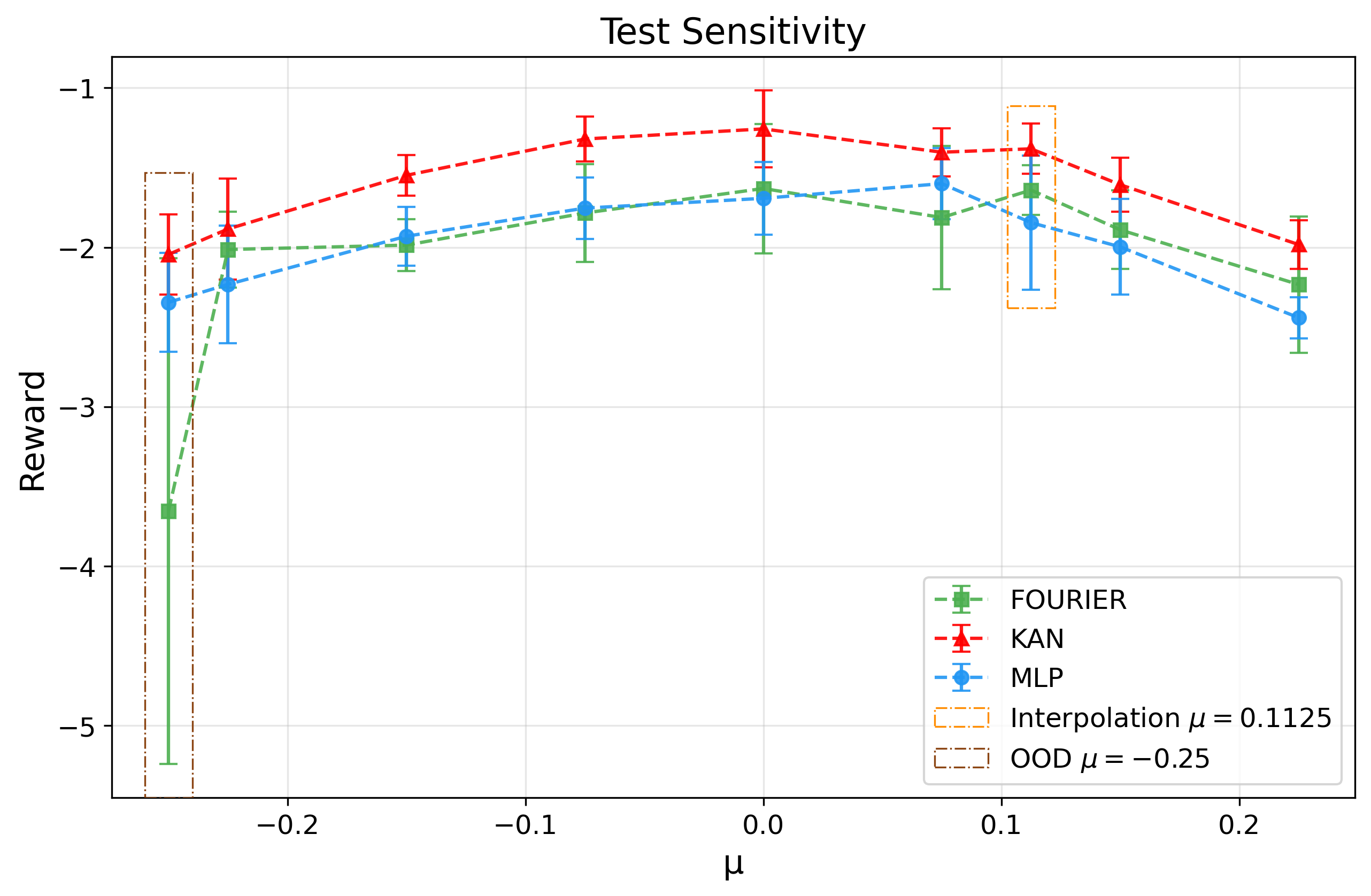}
    \end{subfigure}

    \vspace{0.5em}
    \begin{subfigure}[t]{0.32\textwidth}
        \centering
        \includegraphics[width=\linewidth]{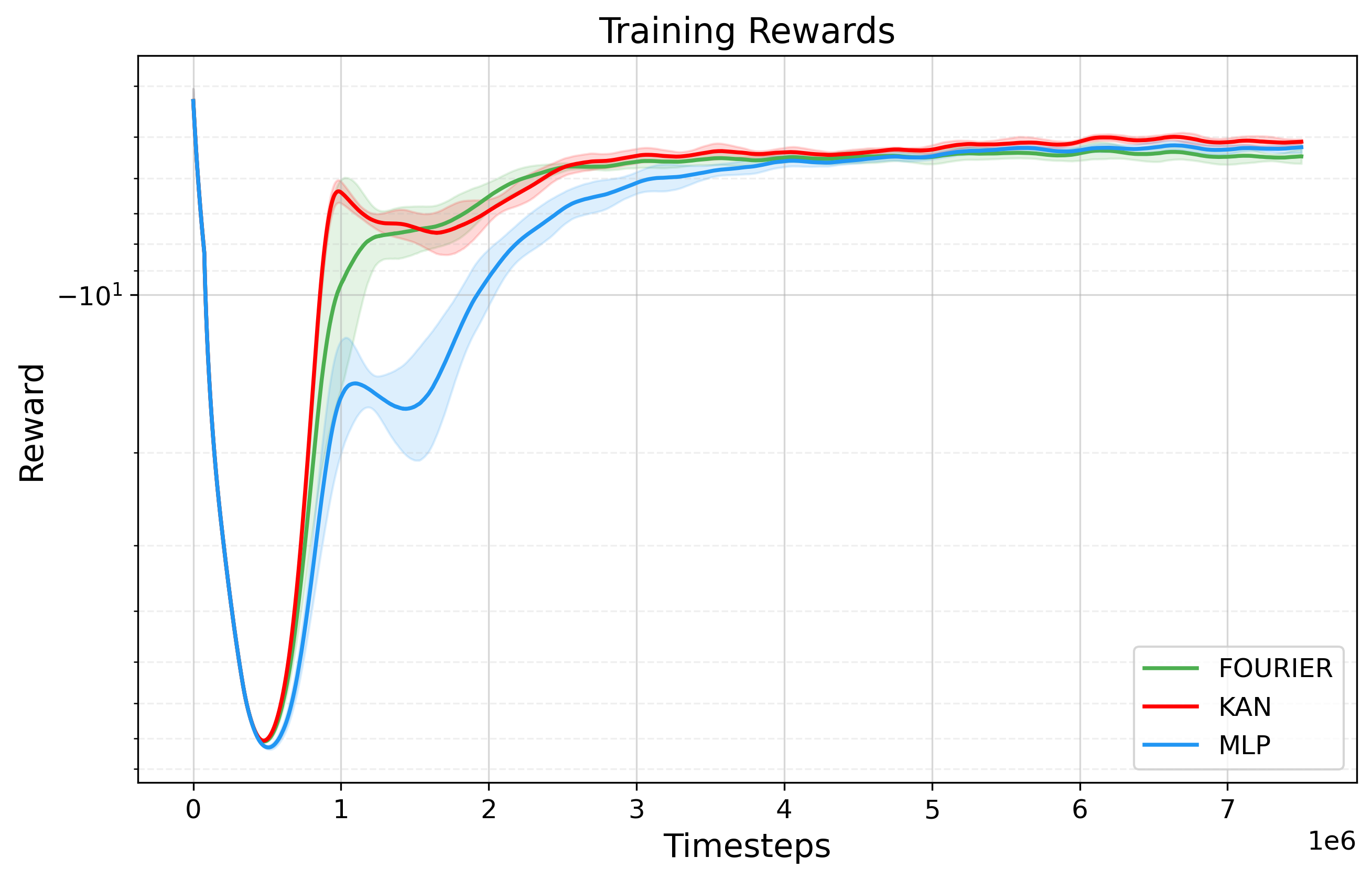}
    \end{subfigure}
    \hfill
    \begin{subfigure}[t]{0.32\textwidth}
        \centering
        \includegraphics[width=\linewidth]{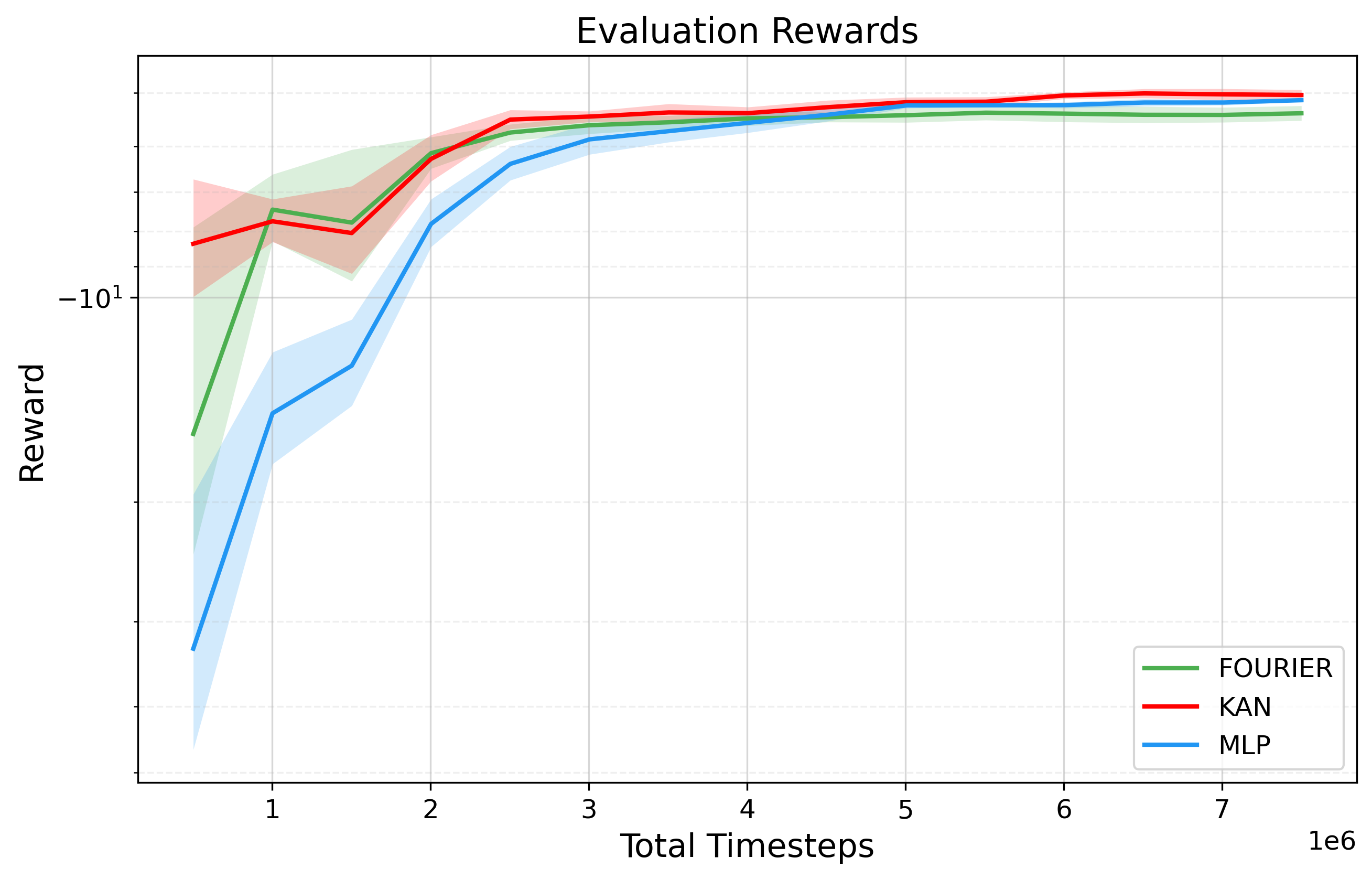}
    \end{subfigure}
    \hfill
    \begin{subfigure}[t]{0.32\textwidth}
        \centering
        \includegraphics[width=\linewidth]{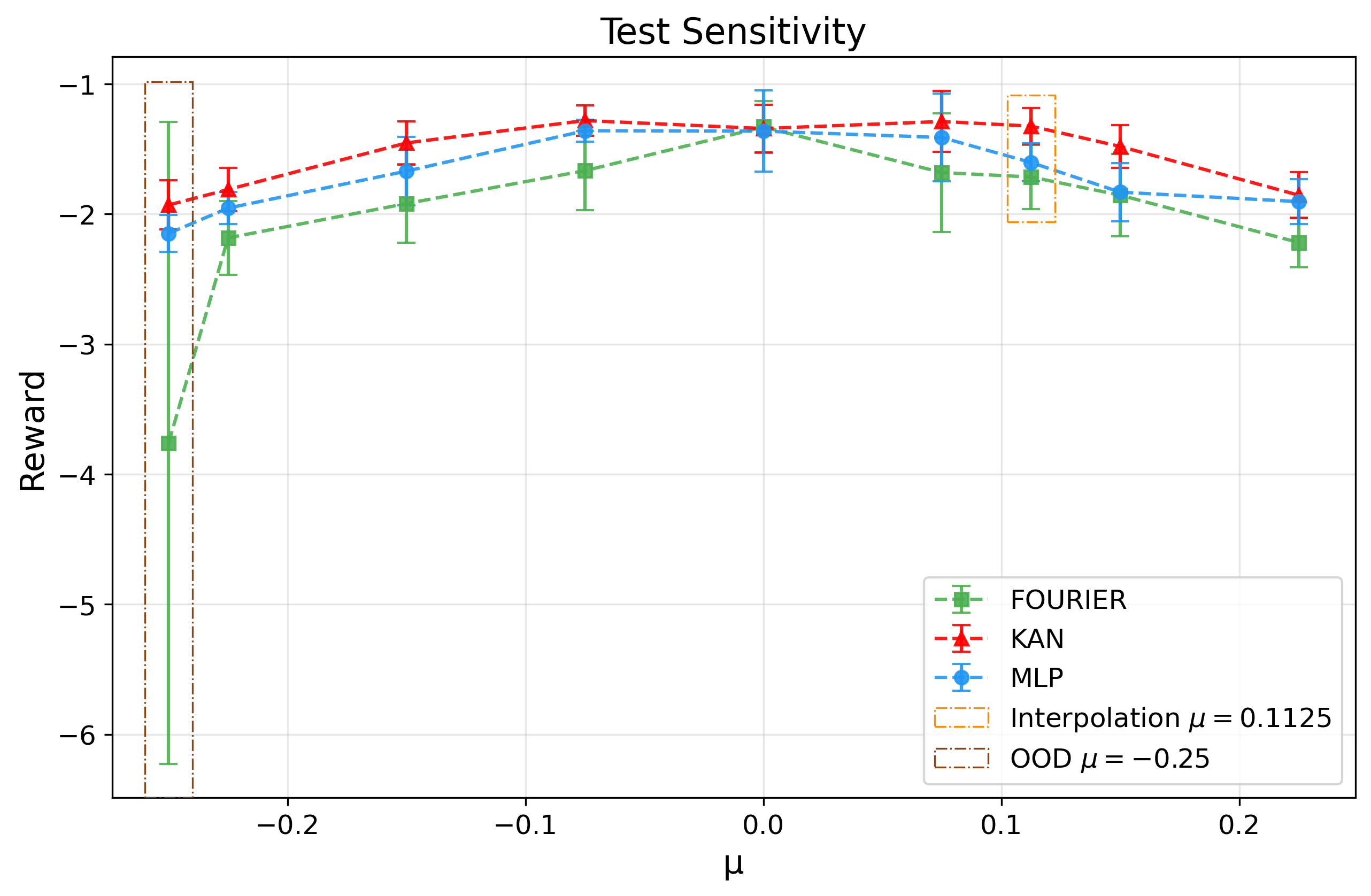}
    \end{subfigure}
    \caption{Direct multi-seed performance comparison of MLP, Fourier, and KAN encoders navigating the baseline $GS=2$ (top row) and high-reuse $GS=4$ (bottom row) training protocols. The three columns display rolling training rewards, periodic cross-domain evaluation means (bounded by 95\% confidence intervals across 5 orthogonal random seeds), and final generalization profiles evaluated across 9 discrete forcing amplitudes (7 seen training values, 2 unseen interpolation/extrapolation values). Across both update settings, actuator architectures show overlapping asymptotic limits due to extreme chaotic sensitivity, but KAN consistently secures tighter lower-bound robustness (variance reduction) under out-of-distribution testing.}
    \label{fig:gs2_vs_gs4}
\end{figure*}

\begin{table*}[!htb]
    \centering
    \small
    \caption{Consolidated architecture comparison corresponding to Section~\ref{sec:arch_comparison}. Metrics reflect the mean performance and standard deviation across 5 independent seeds for both $GS=2$ and $GS=4$ settings. Here, the 'Test Range' encapsulates the absolute minimum and maximum generalization rewards achieved across all seeds. The exponential run-cost penalty of $GS=4$ is particularly severe for the ActNet-KAN encoder, making the $GS=2$ regime mandatory for scalable experimental iteration.}
    \label{tab:gs_runtime_compare}
    \begin{tabular}{llcccc}
        \toprule
        \textbf{Encoder} & \textbf{GS} & \textbf{Time} & \textbf{Final Train ($\pm\sigma$)} & \textbf{Final Eval ($\pm\sigma$)} & \textbf{Test Range [min, max], $\sigma$} \\
        \midrule
        MLP     & 2 & 3h 17m & -5.42 $\pm$ 0.14 & -5.36 $\pm$ 0.10 & [-2.80, -1.16], 0.42 \\
        MLP     & 4 & 5h 21m & -5.20 $\pm$ 0.15 & -5.13 $\pm$ 0.03 & [-2.38, -0.93], 0.37 \\
        \midrule
        Fourier & 2 & 3h 24m & -5.39 $\pm$ 0.22 & -5.37 $\pm$ 0.17 & [-7.15, -0.99], 0.91 \\
        Fourier & 4 & 5h 21m & -5.42 $\pm$ 0.22 & -5.36 $\pm$ 0.15 & [-9.33, -1.01], 1.19 \\
        \midrule
        KAN     & 2 & 4h 36m & -5.08 $\pm$ 0.05 & -5.10 $\pm$ 0.12 & [-2.29, -0.89], 0.36 \\
        KAN     & 4 & 7h 56m & -5.08 $\pm$ 0.06 & -5.04 $\pm$ 0.10 & [-2.23, -0.96], 0.32 \\
        \bottomrule
    \end{tabular}
\end{table*}

Across both update settings, KAN is the most consistent encoder under this protocol, while Fourier is mixed: its mean train/eval rewards are competitive with MLP, but its test-range tails and variance are notably worse (Figure~\ref{fig:gs2_vs_gs4}, Table~\ref{tab:gs_runtime_compare}). The gain from GS=4 is present but small relative to added wall-clock cost. These results support a practical default of GS=2 for the main campaign, with GS=4 treated as a higher-cost option when small peak-reward gains are worth the extra runtime. 

It is worth noting that the overlapping reward distributions across the five seeds (Figure~\ref{fig:gs2_vs_gs4}) reflect the extreme sensitivity of the chaotic KS reward landscape to initial conditions. Rather than strictly dominating the mean asymptotic performance, KAN's structural advantage manifests primarily as variance reduction and worst-case scenario mitigation during out-of-distribution tracking (evidenced by the tight Test Range standard deviation). This suggests that the decoupled sinusoidal basis of ActNet-KAN is better suited for dynamically capturing the modal spatial responses of the PDE than the generalized approximation of the densely connected MLP.

\subsection{Qualitative Stabilization: Heatmaps}\label{sec:heatmaps}

To qualitatively assess control effectiveness across reward settings, we split the
stabilization analysis into three cases:
\begin{itemize}
    \item \textbf{Case 1:} stabilization control to the zero reference.
    \item \textbf{Case 2:} four-mode cosine tracking.
    \item \textbf{Case 3:} four-mode cosine tracking with a non-zero mean.
\end{itemize}
In each case, we compare MLP, Fourier, and KAN policies at
two representative unseen parameter values: $\mu = 0.1125$ (in-range interpolation) and
$\mu = -0.25$ (mild extrapolation outside the training grid). In Cases 2 and 3, the
controller is asked to follow a
four-mode cosine reference built from spatial modes $k=1,\ldots,4$; Case 3 adds a non-zero spatial mean.

\begin{figure}[!htb]
    \centering
    \begin{subfigure}{\textwidth}
        \centering
        \begin{subfigure}{0.32\textwidth}
            \centering
            \includegraphics[width=\linewidth]{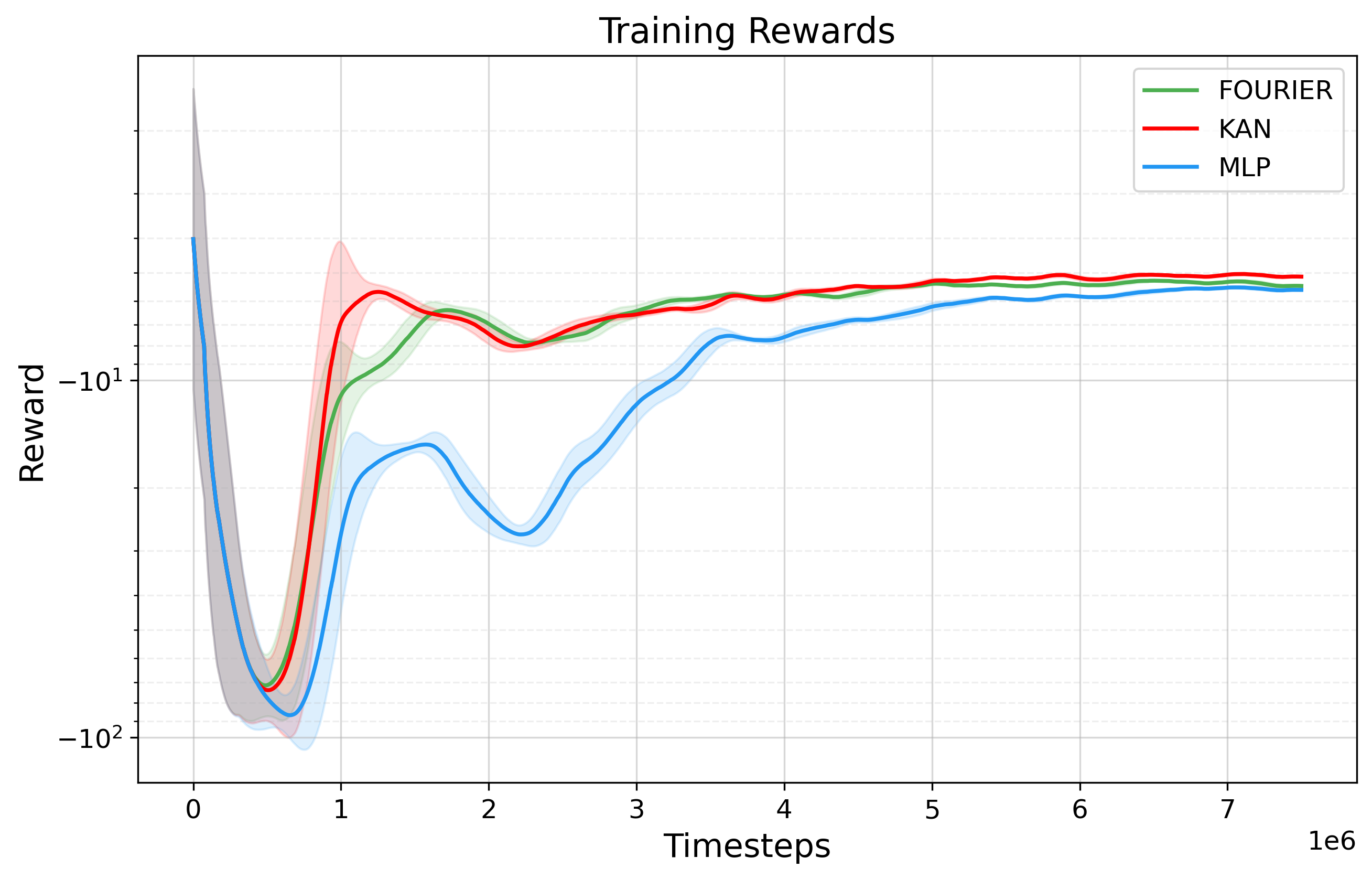}
        \end{subfigure}\hfill
        \begin{subfigure}{0.32\textwidth}
            \centering
            \includegraphics[width=\linewidth]{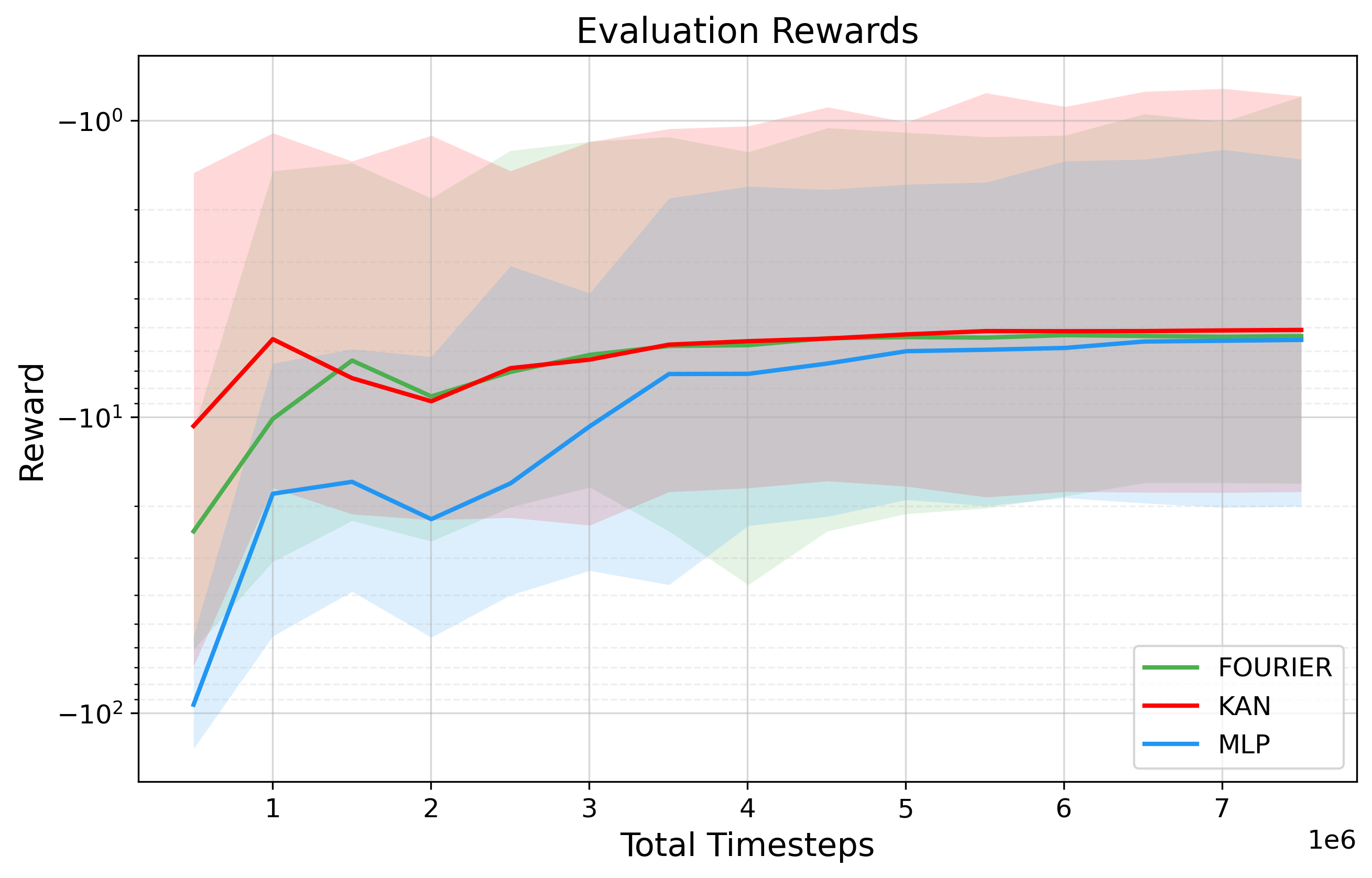}
        \end{subfigure}\hfill
        \begin{subfigure}{0.32\textwidth}
            \centering
            \includegraphics[width=\linewidth]{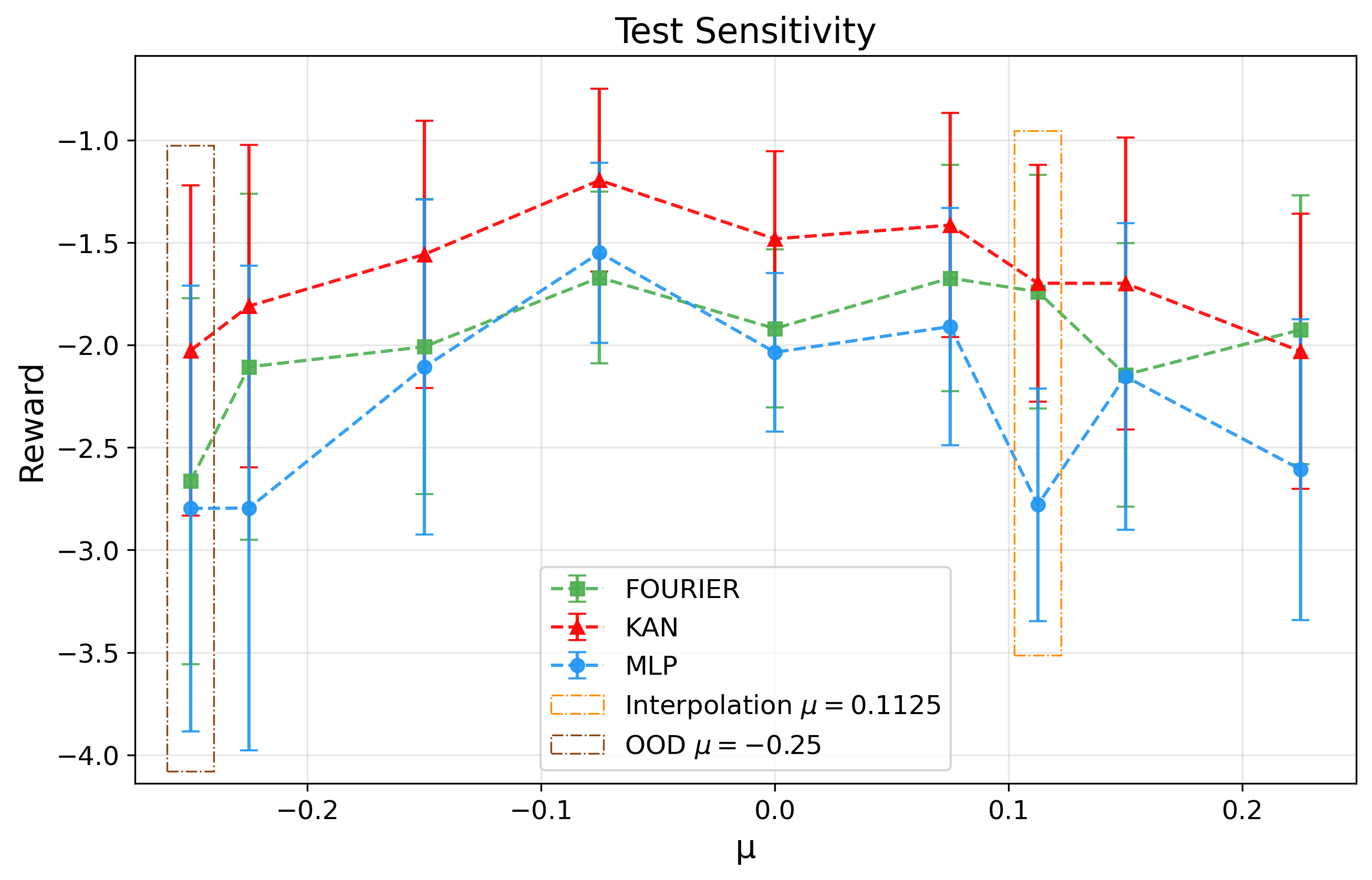}
        \end{subfigure}
        \subcaption{Case 1 (zero-reference stabilization control)}
    \end{subfigure}
    \vspace{0.5em}

    \begin{subfigure}{\textwidth}
        \centering
        \begin{subfigure}{0.32\textwidth}
            \centering
            \includegraphics[width=\linewidth]{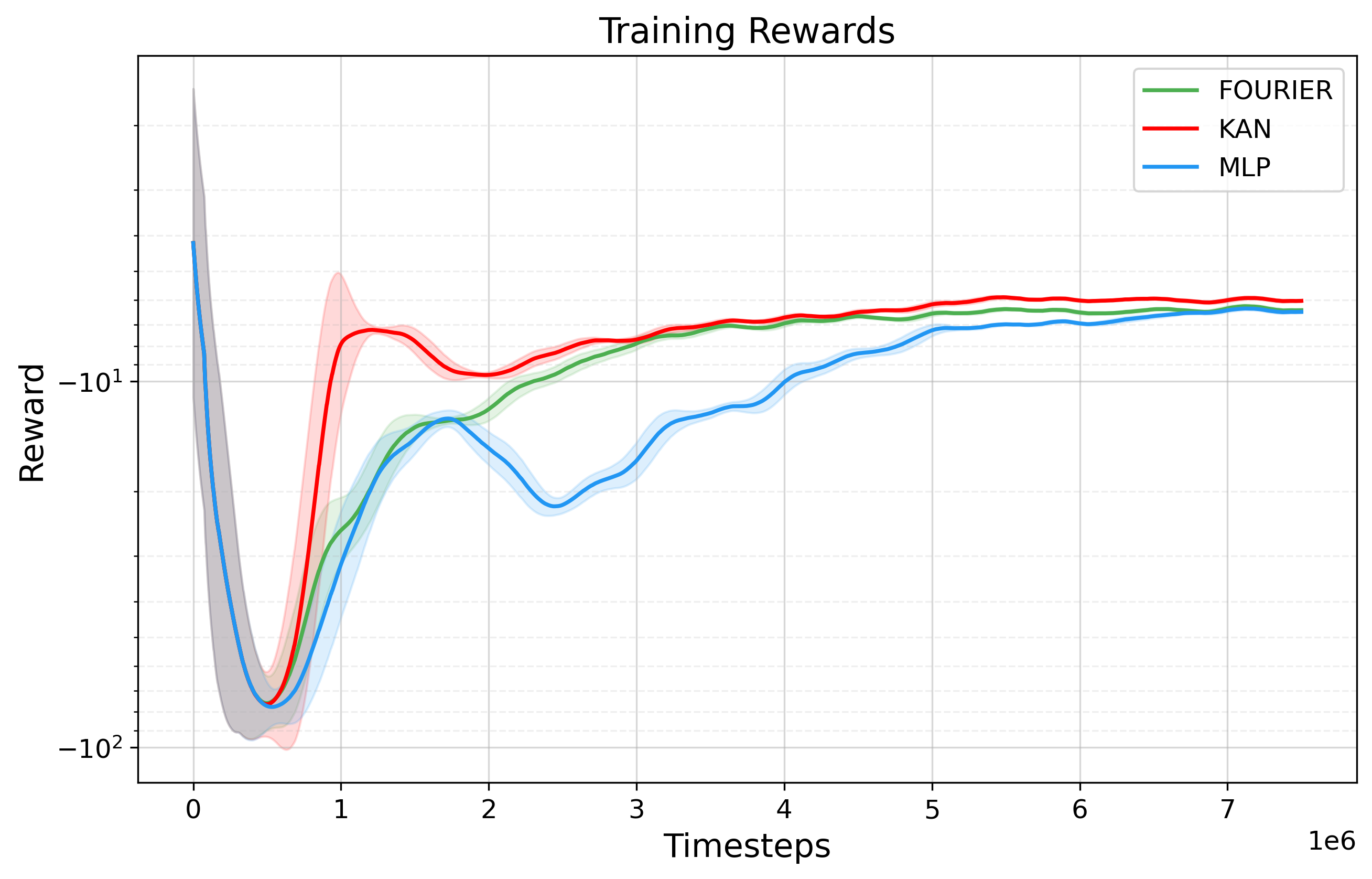}
        \end{subfigure}\hfill
        \begin{subfigure}{0.32\textwidth}
            \centering
            \includegraphics[width=\linewidth]{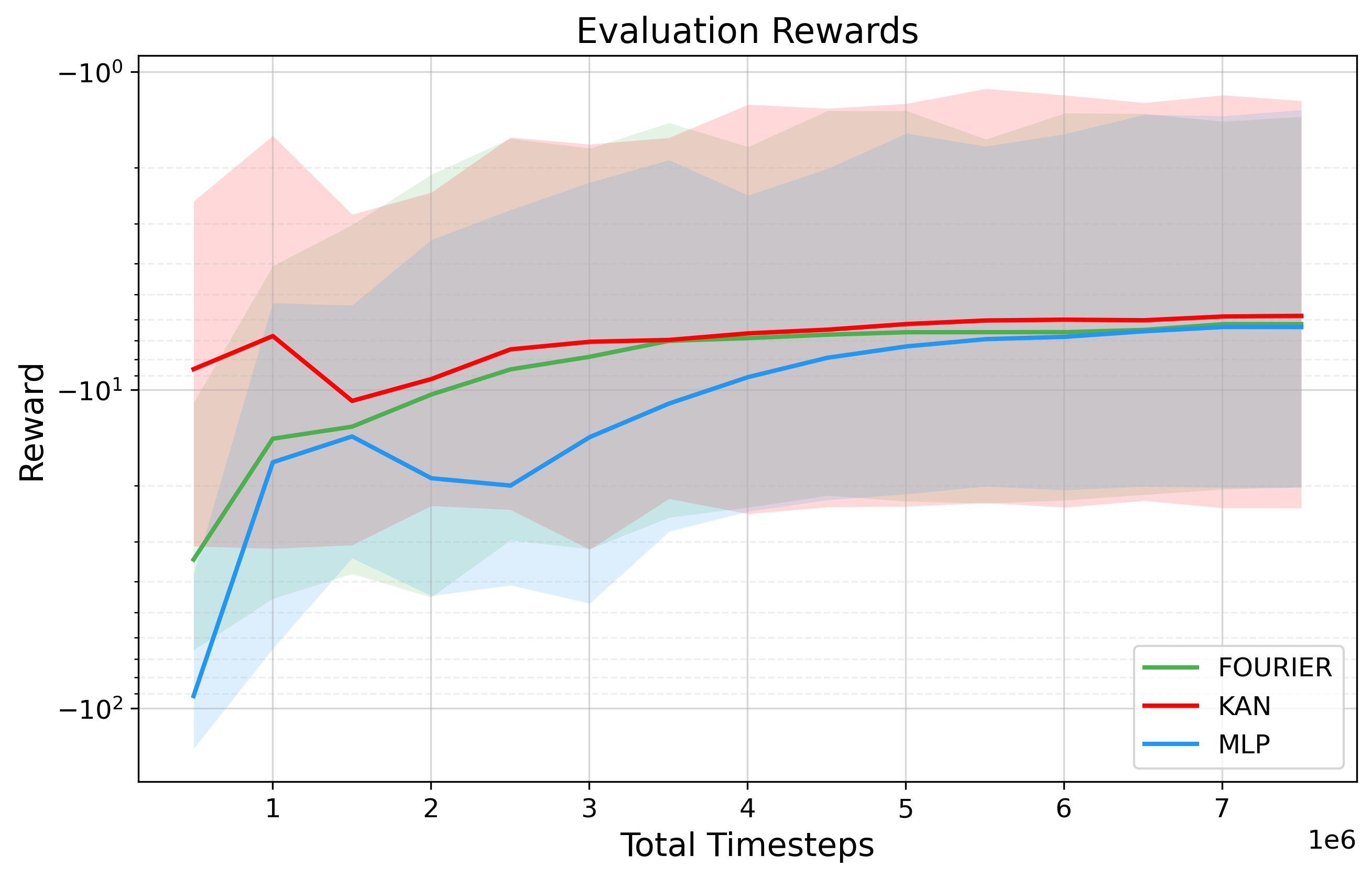}
        \end{subfigure}\hfill
        \begin{subfigure}{0.32\textwidth}
            \centering
            \includegraphics[width=\linewidth]{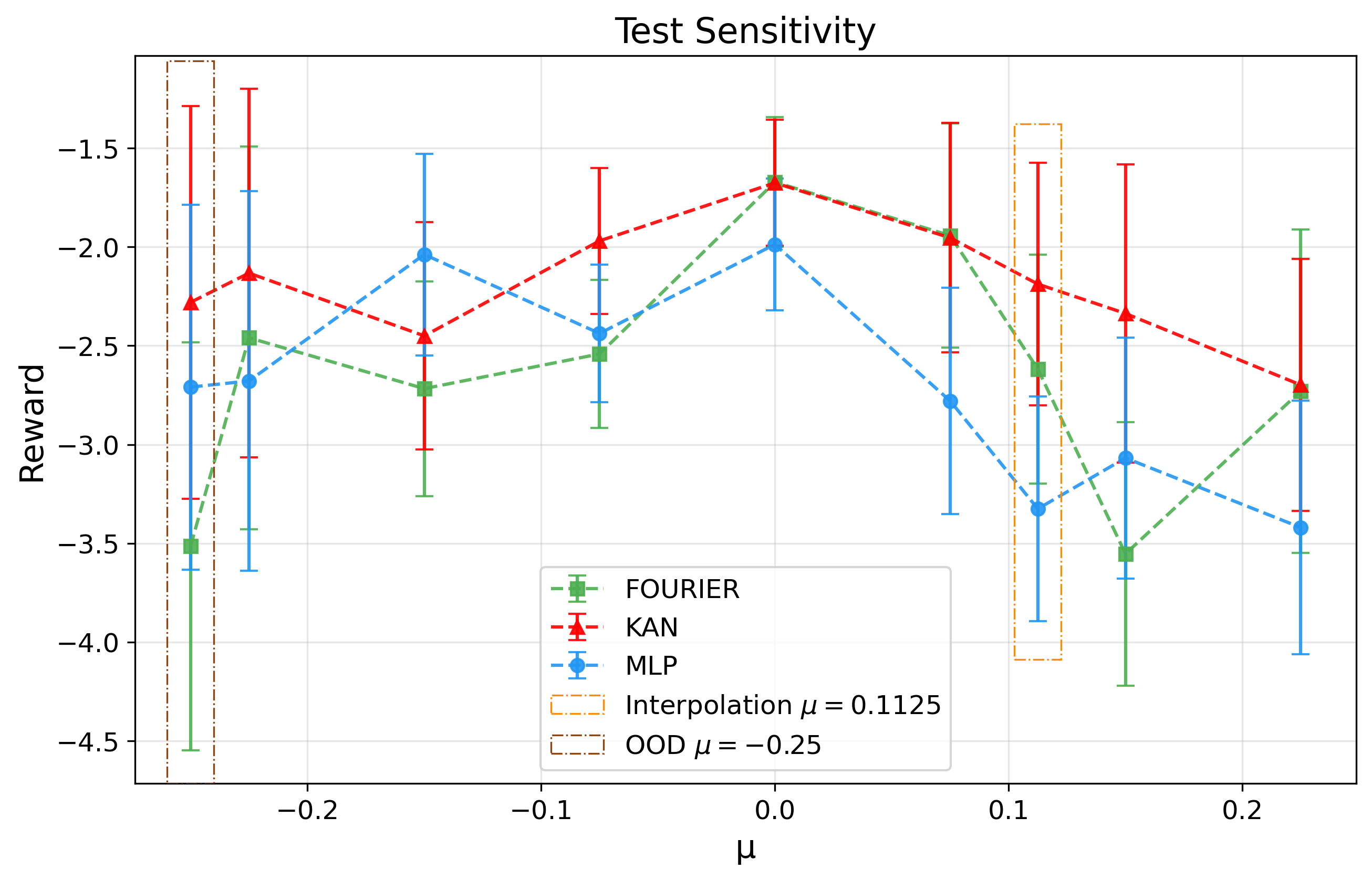}
        \end{subfigure}
        \subcaption{Case 2 (four-mode cosine tracking)}
    \end{subfigure}
    \vspace{0.5em}

    \begin{subfigure}{\textwidth}
        \centering
        \begin{subfigure}{0.32\textwidth}
            \centering
            \includegraphics[width=\linewidth]{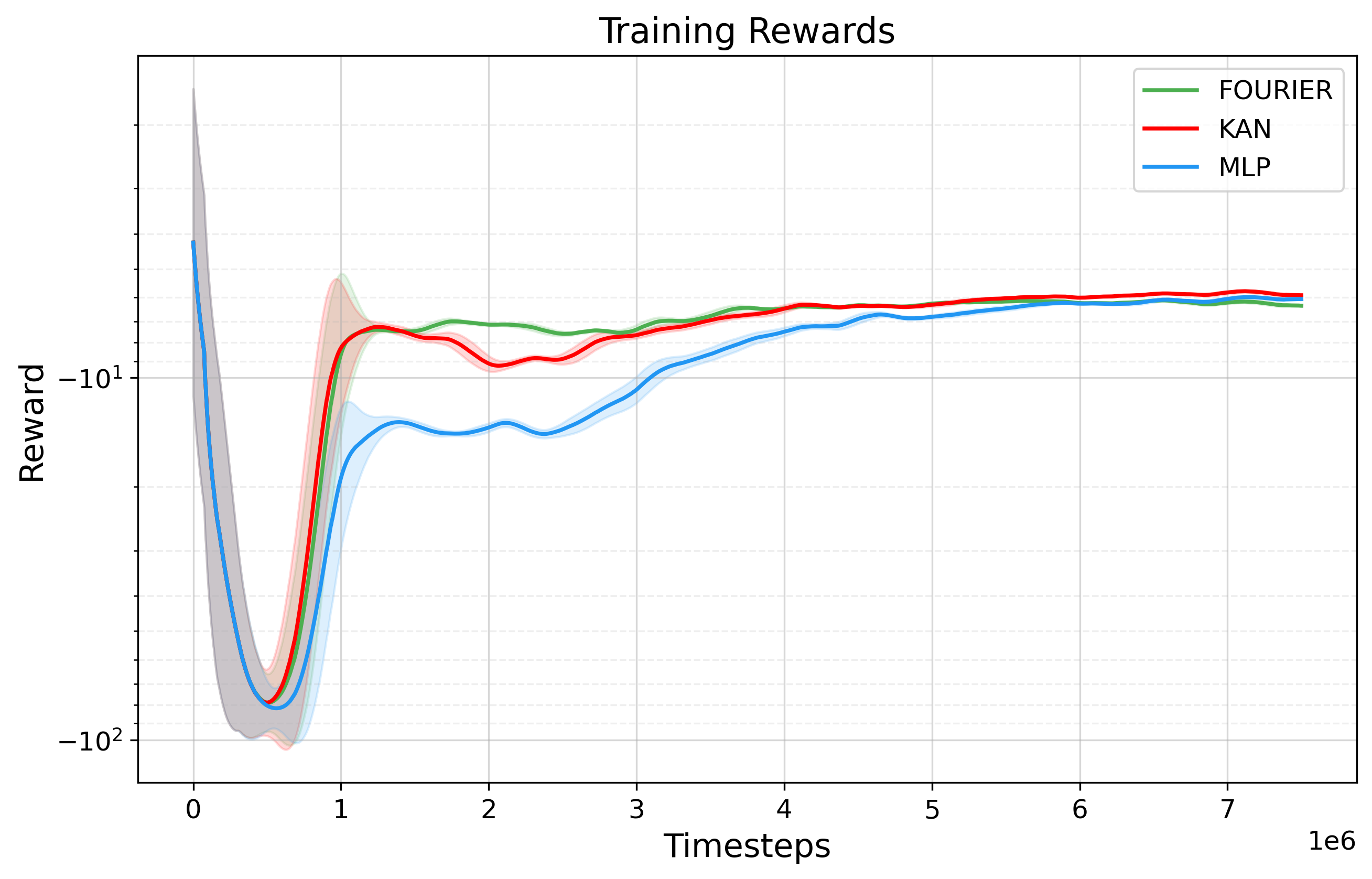}
        \end{subfigure}\hfill
        \begin{subfigure}{0.32\textwidth}
            \centering
            \includegraphics[width=\linewidth]{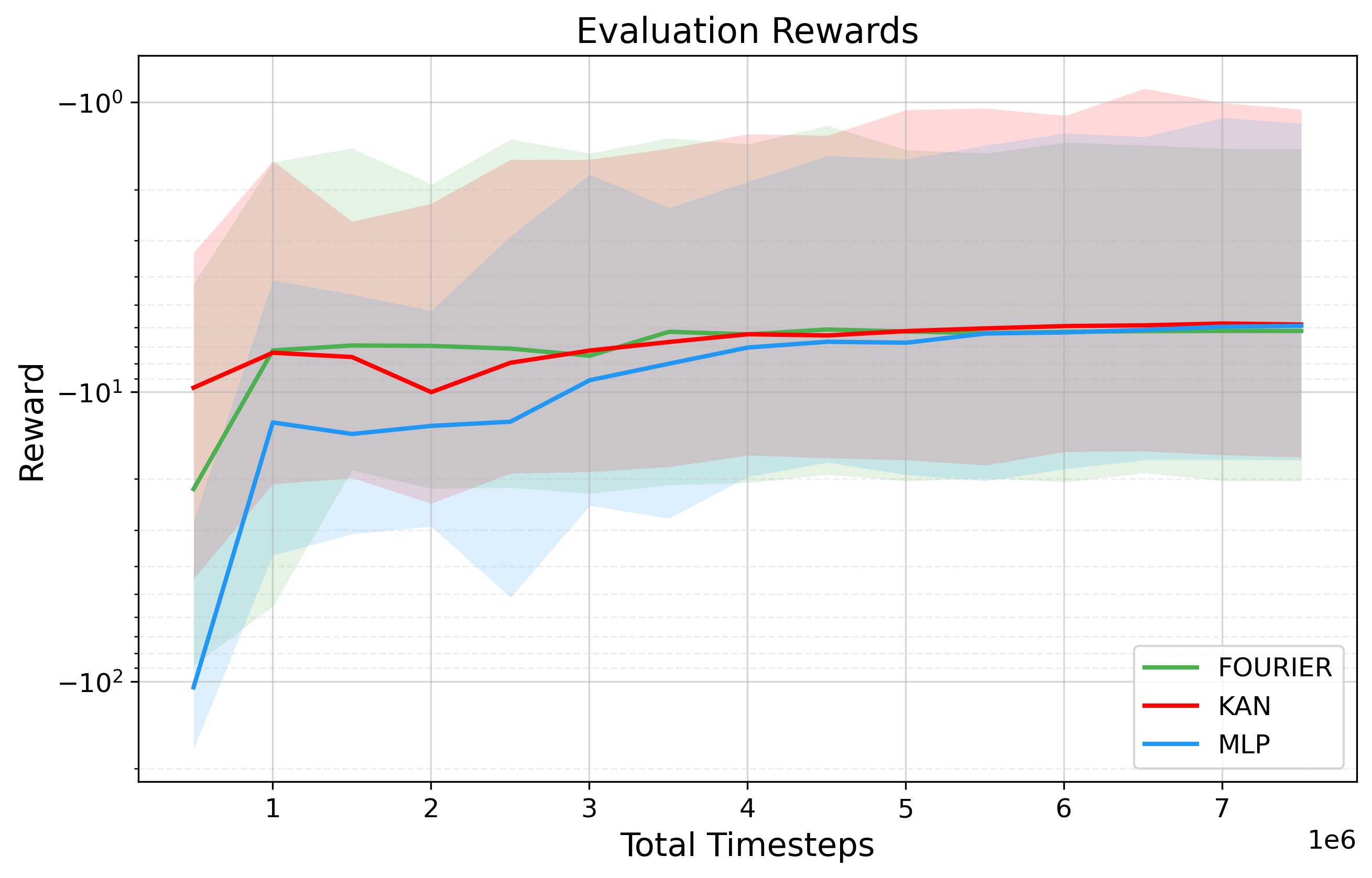}
        \end{subfigure}\hfill
        \begin{subfigure}{0.32\textwidth}
            \centering
            \includegraphics[width=\linewidth]{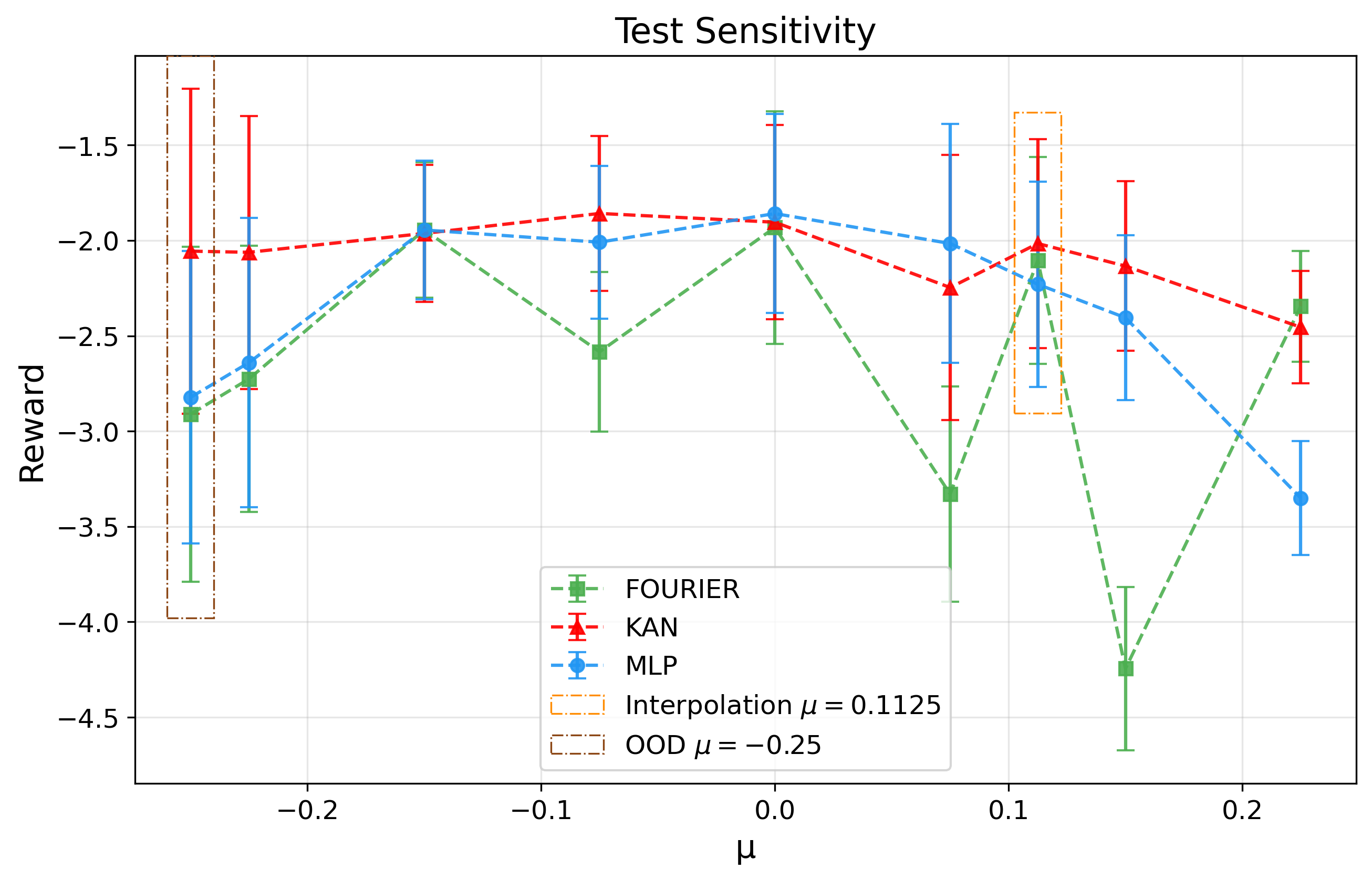}
        \end{subfigure}
        \subcaption{Case 3 (four-mode cosine tracking with a non-zero mean)}
    \end{subfigure}

    \caption{Comparative RL learning dynamics across three core physical tasks: (a) zero-reference stabilization, (b) four-mode cosine target tracking, and (c) mean-shifted offset cosine target tracking. Each subpanel illustrates the training convergence, periodic evaluation distribution across the validation domain, and post-training generalization capabilities. While all three parameter-conditioned encoders effectively constrain the KS system and stabilize chaotic transients, ActNet-KAN and Fourier explicitly dominate the MLP backbone in worst-case out-of-distribution tracking performance (far right columns).}
    \label{fig:rewards_all}
\end{figure}

\begin{figure*}[p]
    \centering
    \begin{subfigure}{0.49\textwidth}
        \centering
        \includegraphics[width=0.9\linewidth]{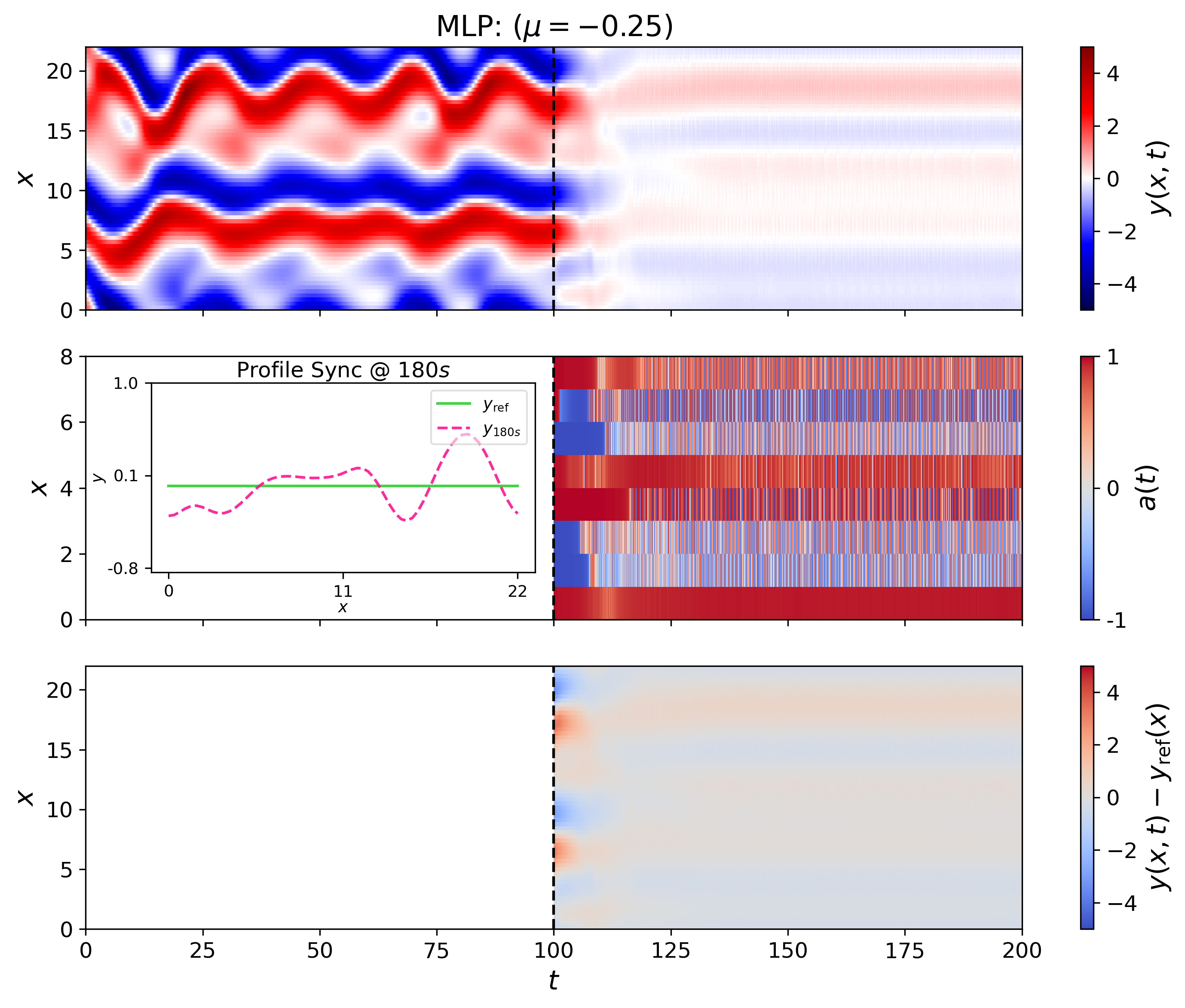}
    \end{subfigure}\hfill
    \begin{subfigure}{0.49\textwidth}
        \centering
        \includegraphics[width=0.9\linewidth]{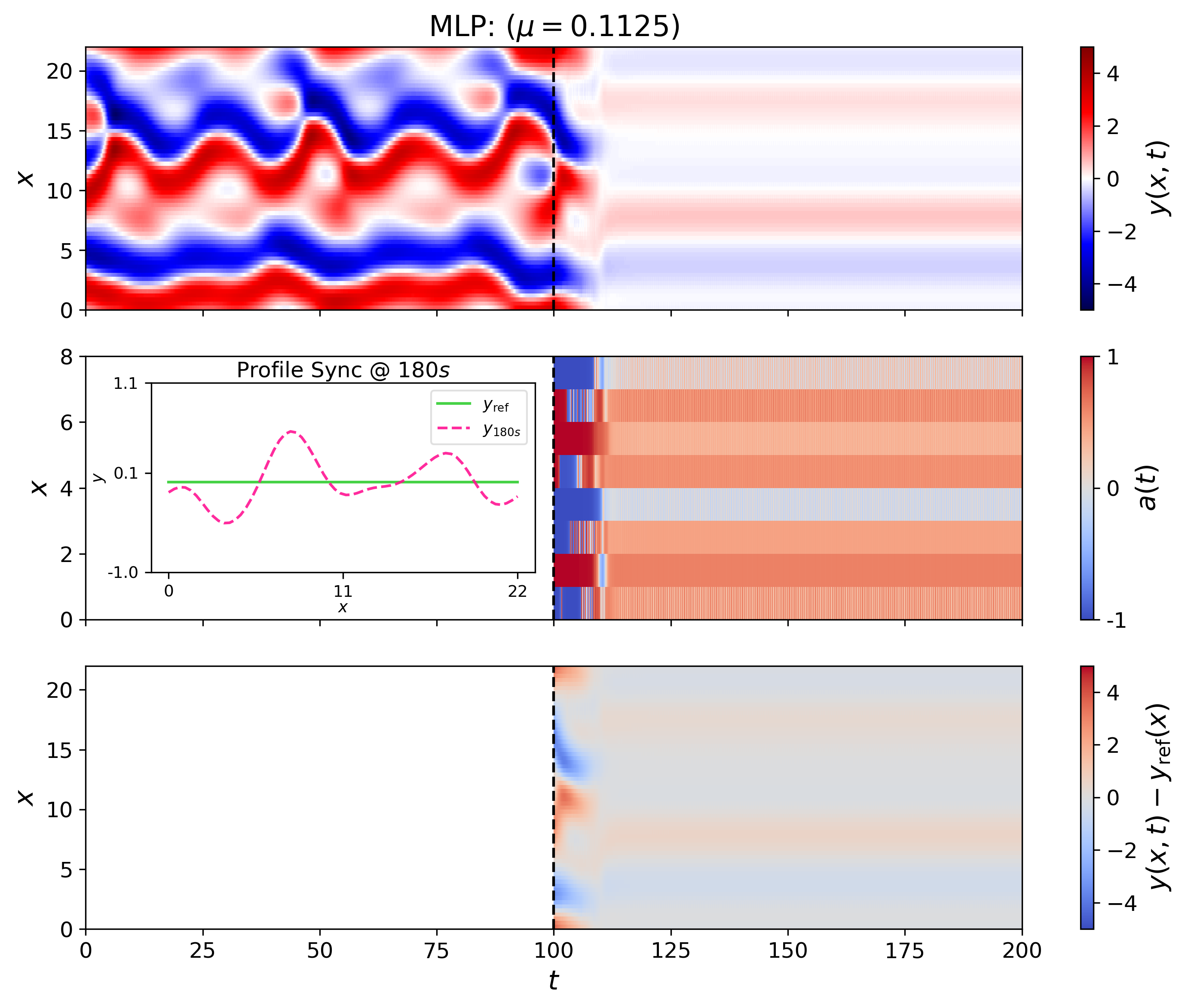}
    \end{subfigure}

    \vspace{0.4em}
    \begin{subfigure}{0.49\textwidth}
        \centering
        \includegraphics[width=0.9\linewidth]{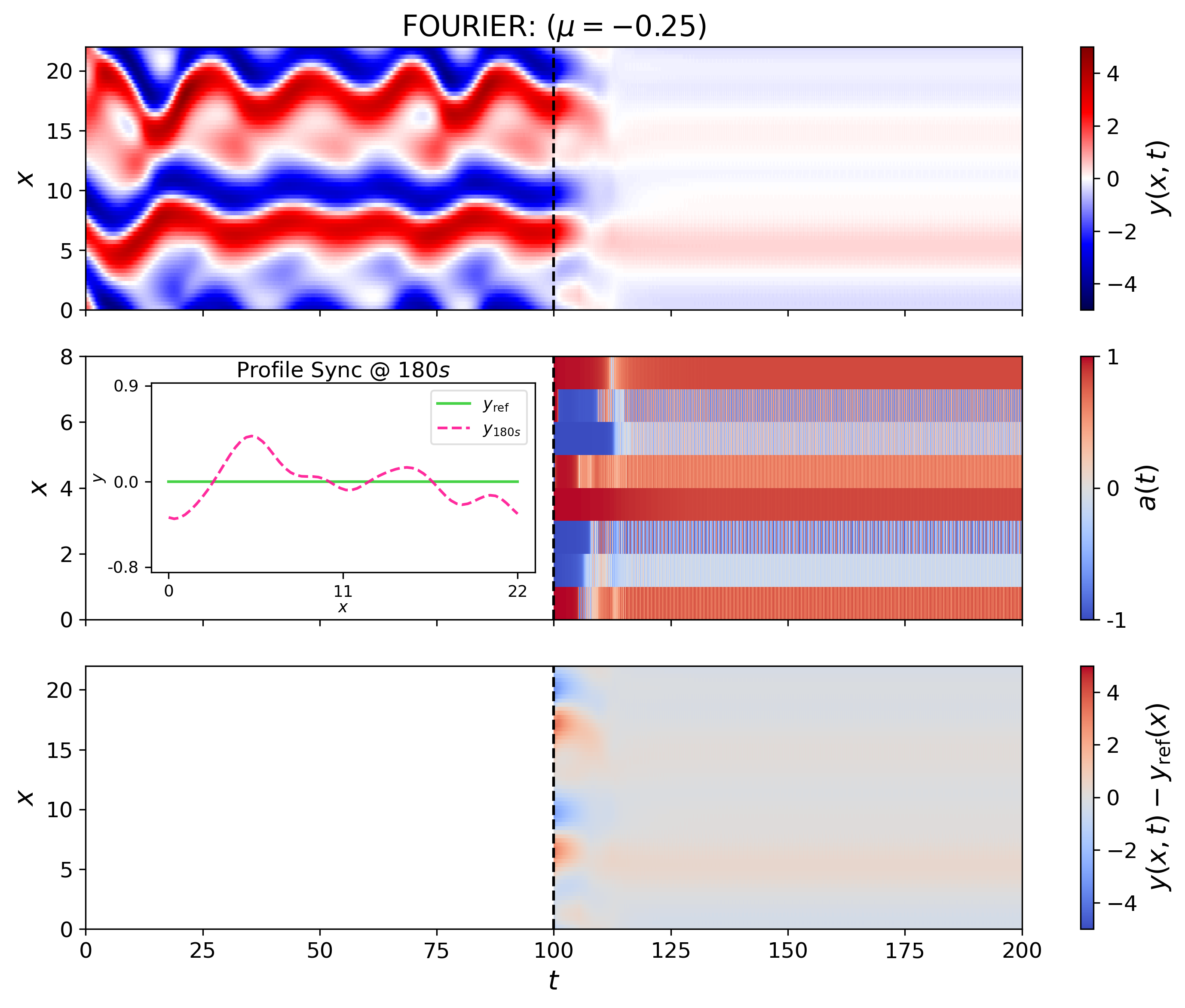}
    \end{subfigure}\hfill
    \begin{subfigure}{0.49\textwidth}
        \centering
        \includegraphics[width=0.9\linewidth]{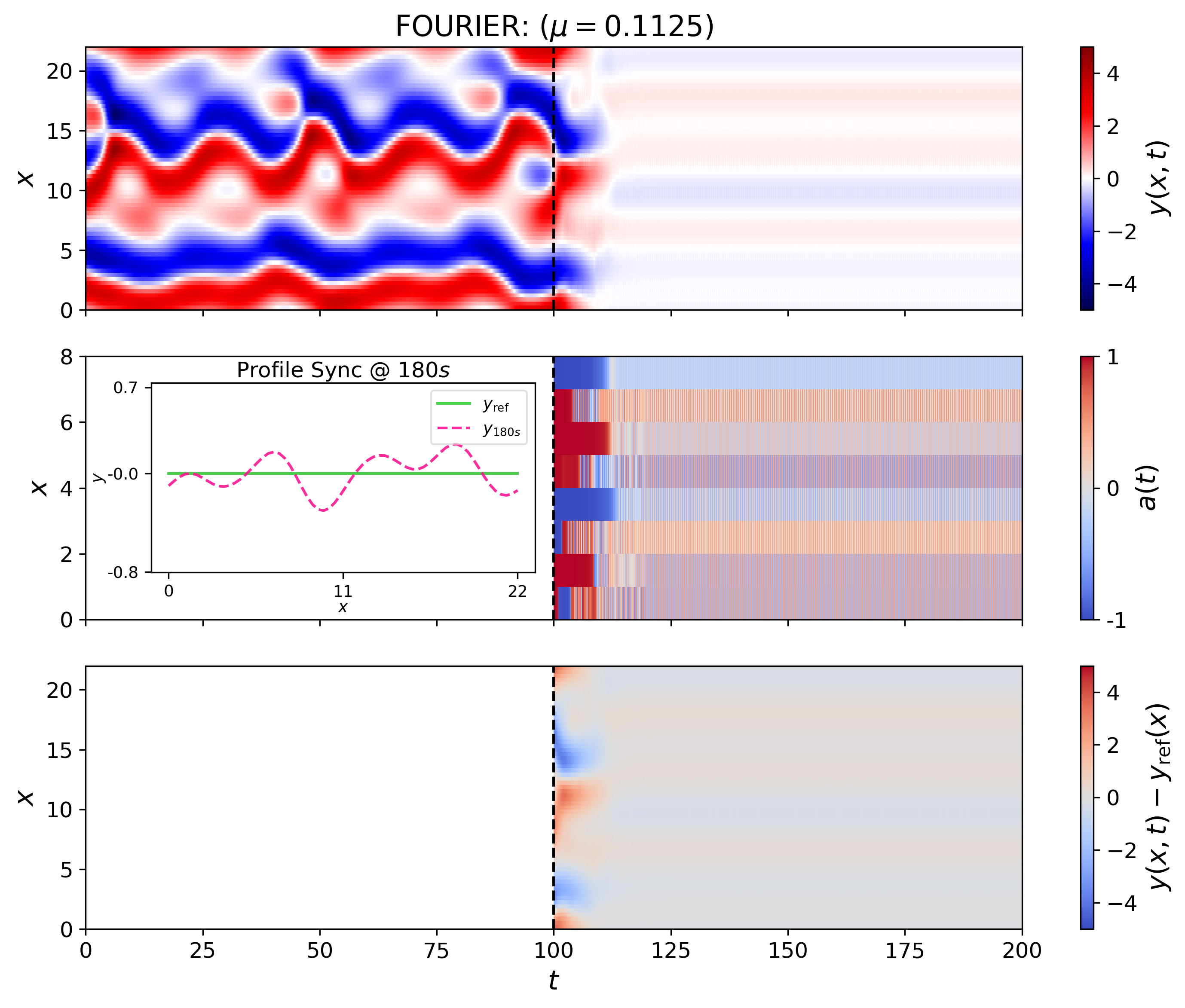}
    \end{subfigure}

    \vspace{0.4em}
    \begin{subfigure}{0.49\textwidth}
        \centering
        \includegraphics[width=0.9\linewidth]{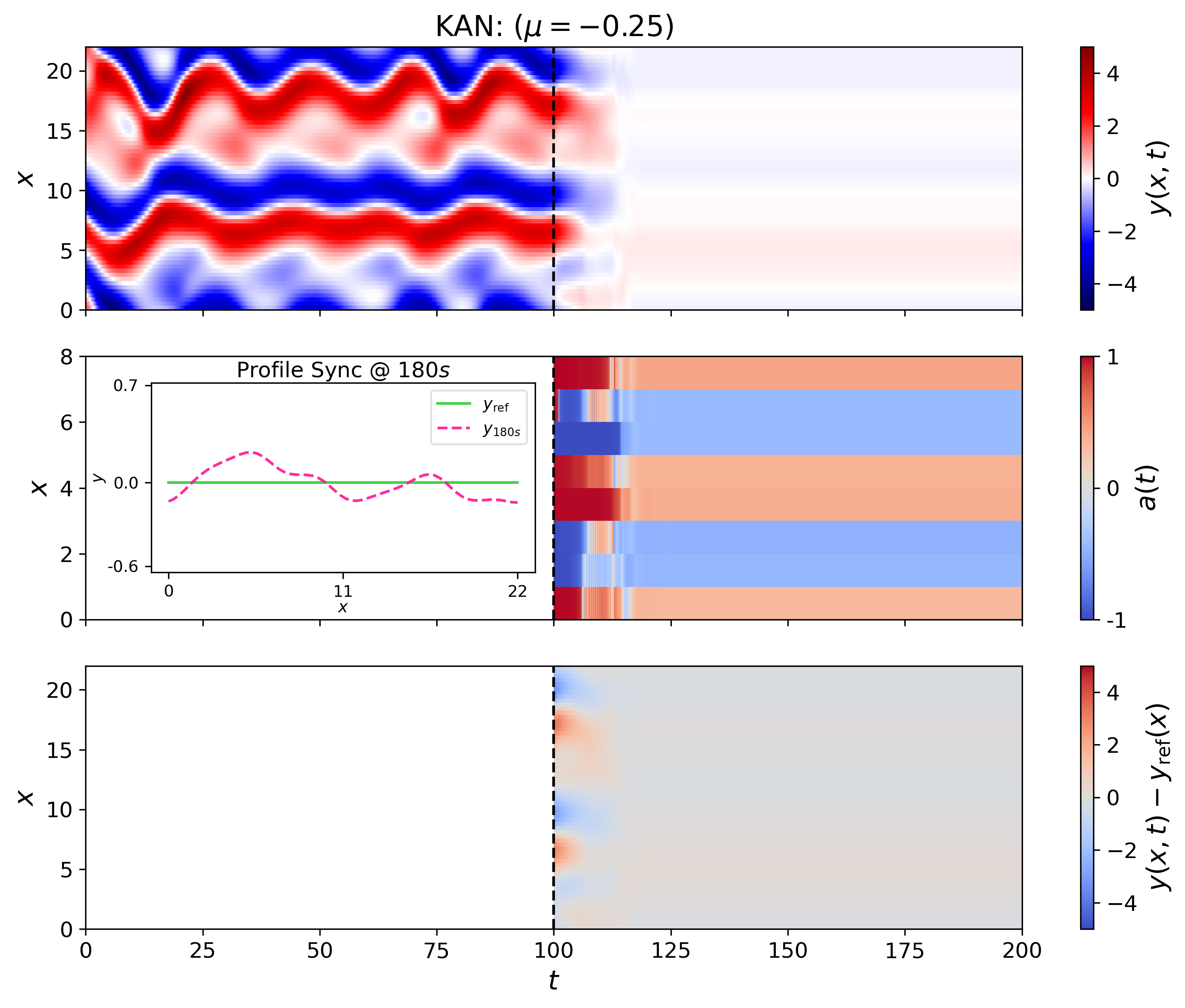}
    \end{subfigure}\hfill
    \begin{subfigure}{0.49\textwidth}
        \centering
        \includegraphics[width=0.9\linewidth]{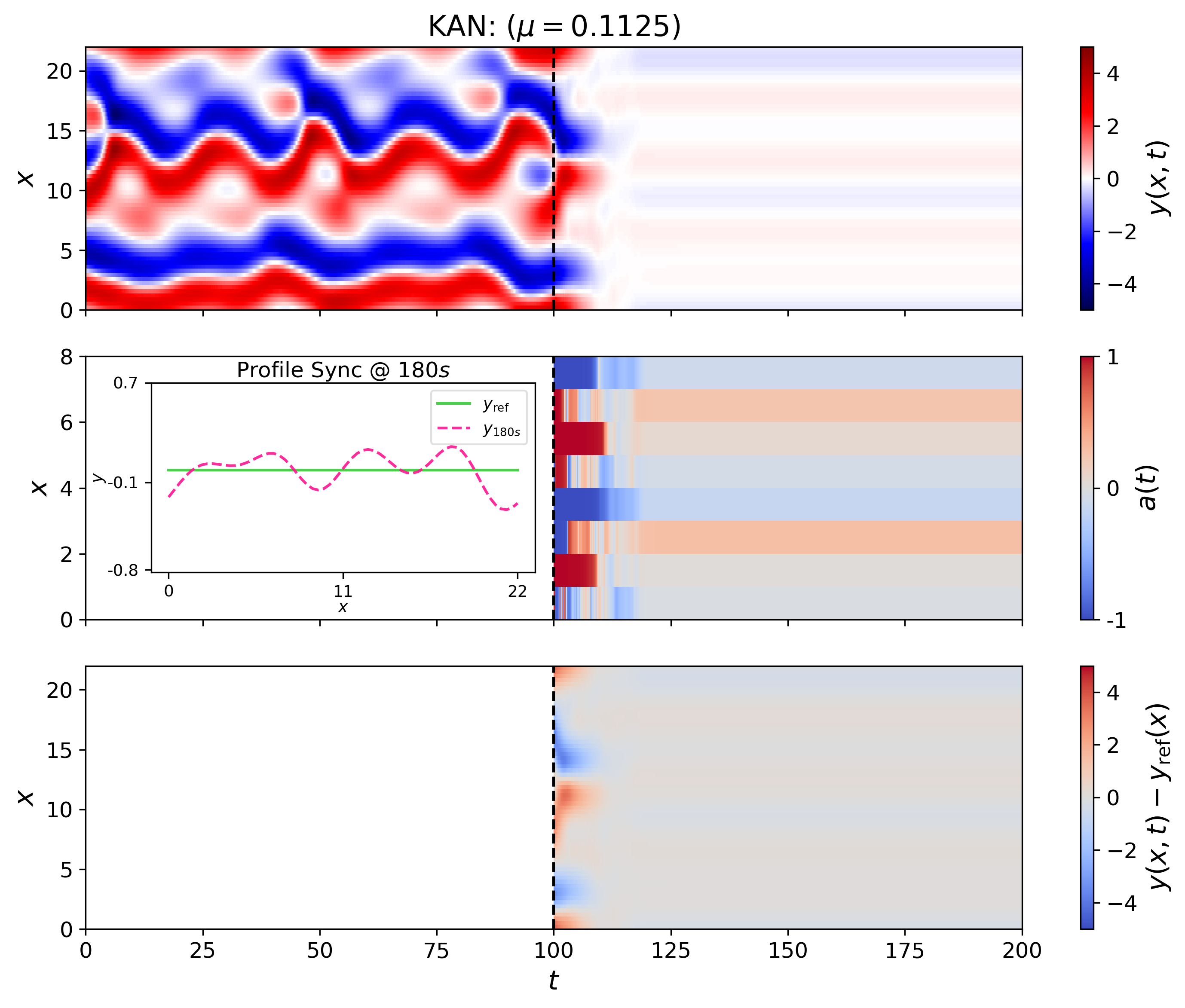}
    \end{subfigure}

    \caption{Physical representation of Case 1 (zero-reference control) via spatiotemporal contour fields $y(x,t)$. Time evolution progresses along the vertical axis while spatial domains periodically wrap across the horizontal. Results map the action of the final converged agents navigating two distinct parameter environments: an unseen interpolation ($\mu=0.1125$, right) and a mild extrapolation ($\mu=-0.25$, left). All active models decisively suppress the rapid energy cascades intrinsic to the unforced KS PDE, though the ActNet-KAN policy establishes the smoothest stabilized invariant manifold, eliminating nearly all traveling wave signatures that persist under the MLP residual.}
    \label{fig:heatmaps_tqc0}
\end{figure*}

\begin{figure*}[p]
    \centering
    \begin{subfigure}{0.49\textwidth}
        \centering
        \includegraphics[width=0.9\linewidth]{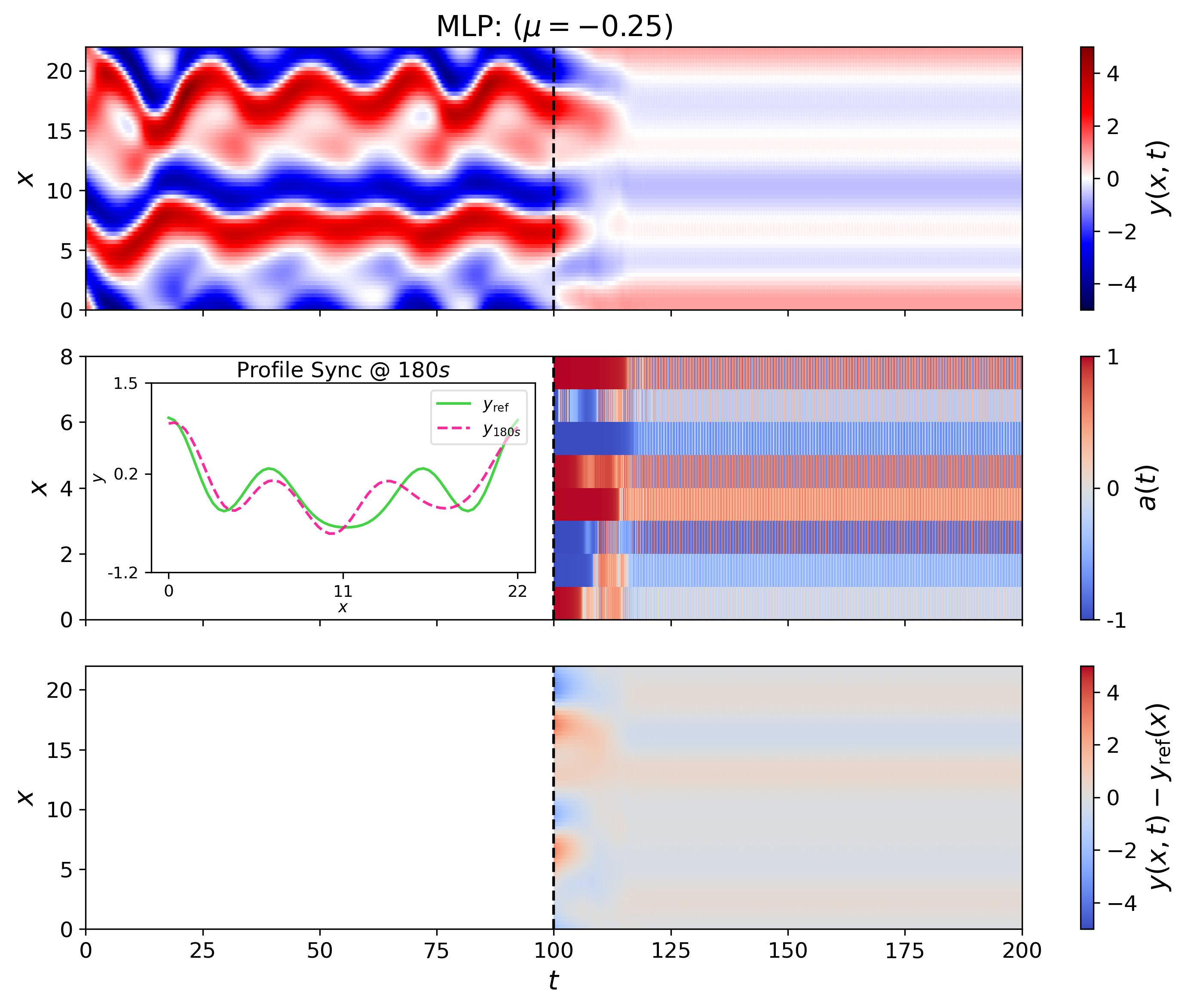}
    \end{subfigure}\hfill
    \begin{subfigure}{0.49\textwidth}
        \centering
        \includegraphics[width=0.9\linewidth]{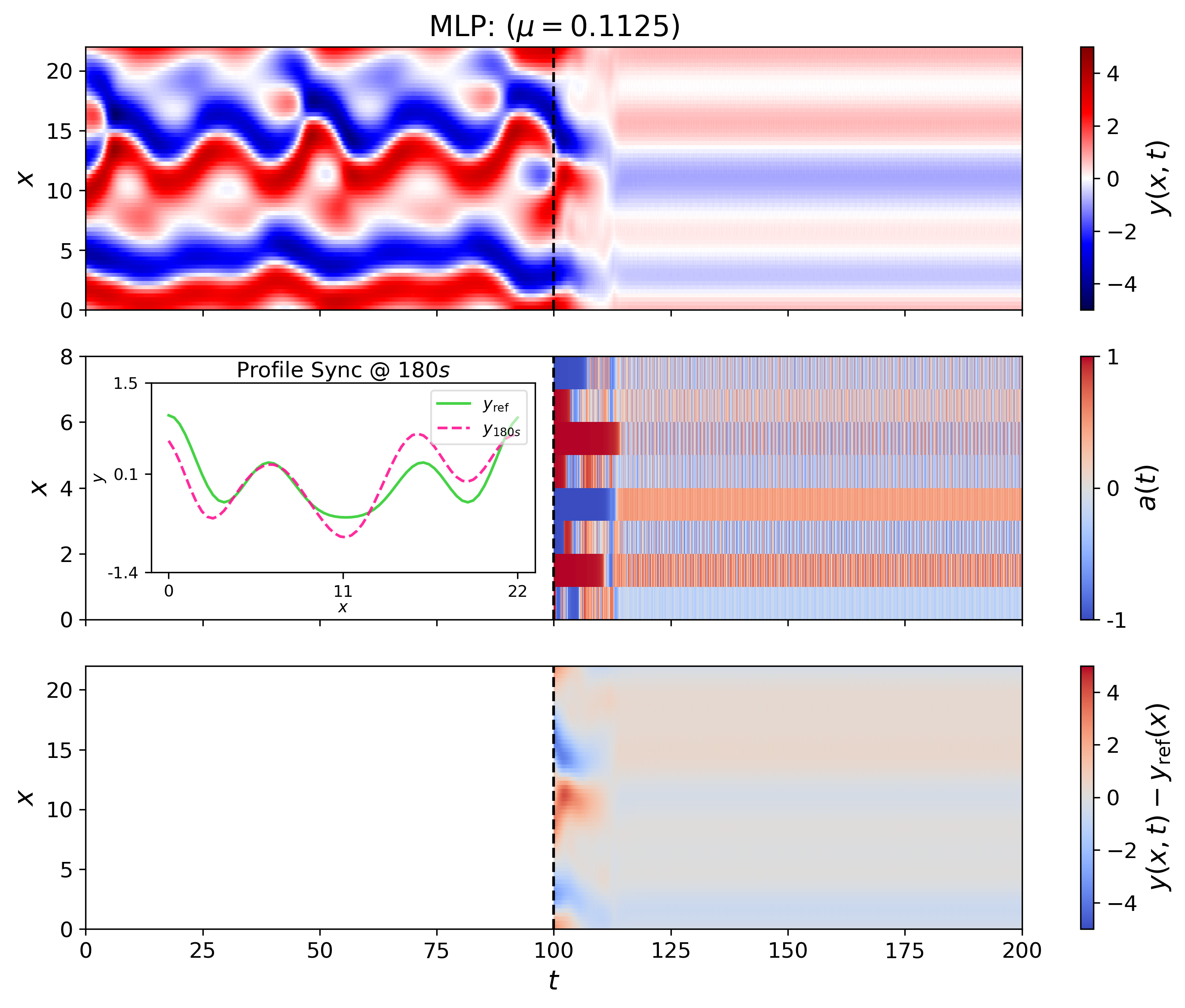}
    \end{subfigure}

    \vspace{0.4em}
    \begin{subfigure}{0.49\textwidth}
        \centering
        \includegraphics[width=0.9\linewidth]{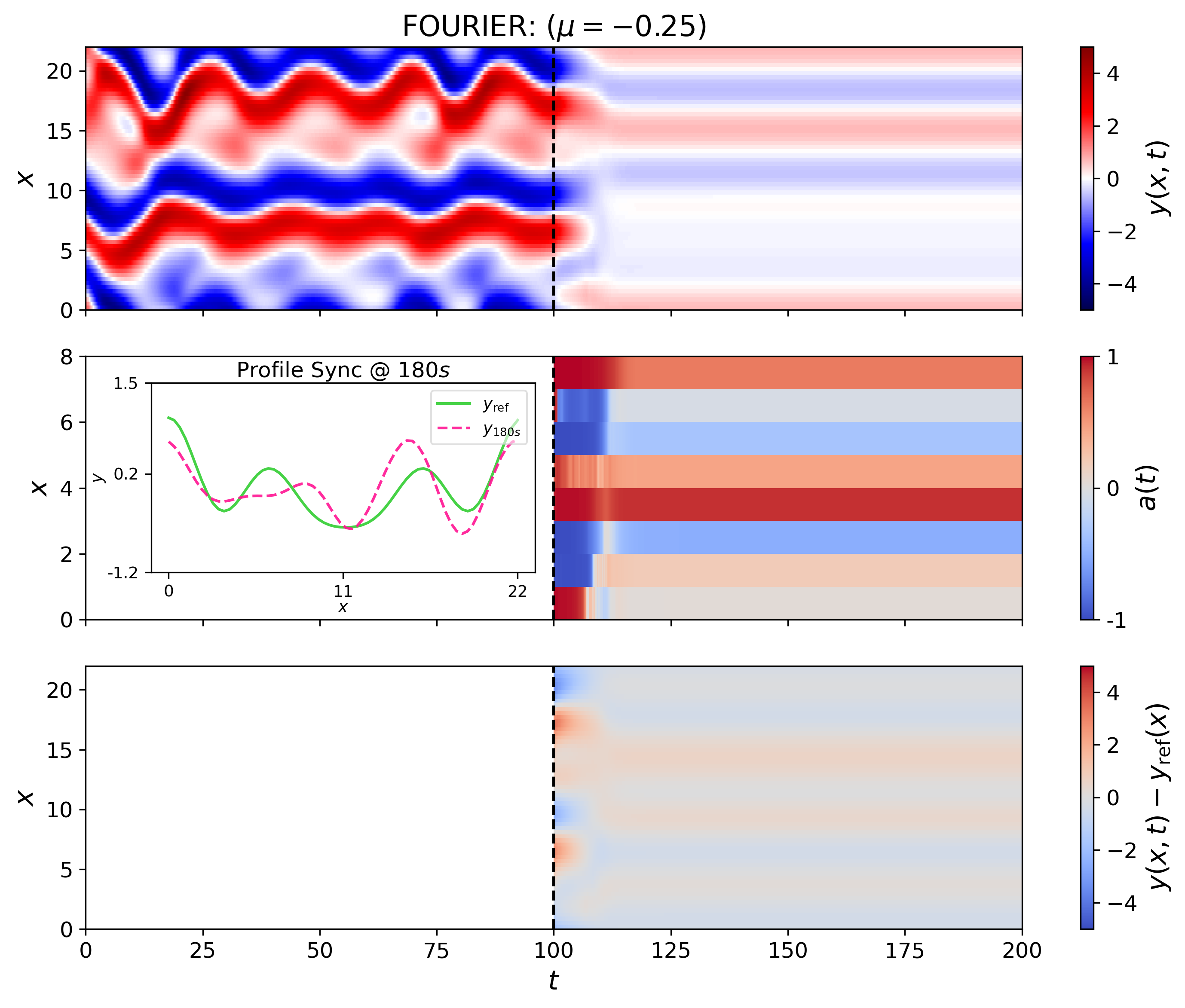}
    \end{subfigure}\hfill
    \begin{subfigure}{0.49\textwidth}
        \centering
        \includegraphics[width=0.9\linewidth]{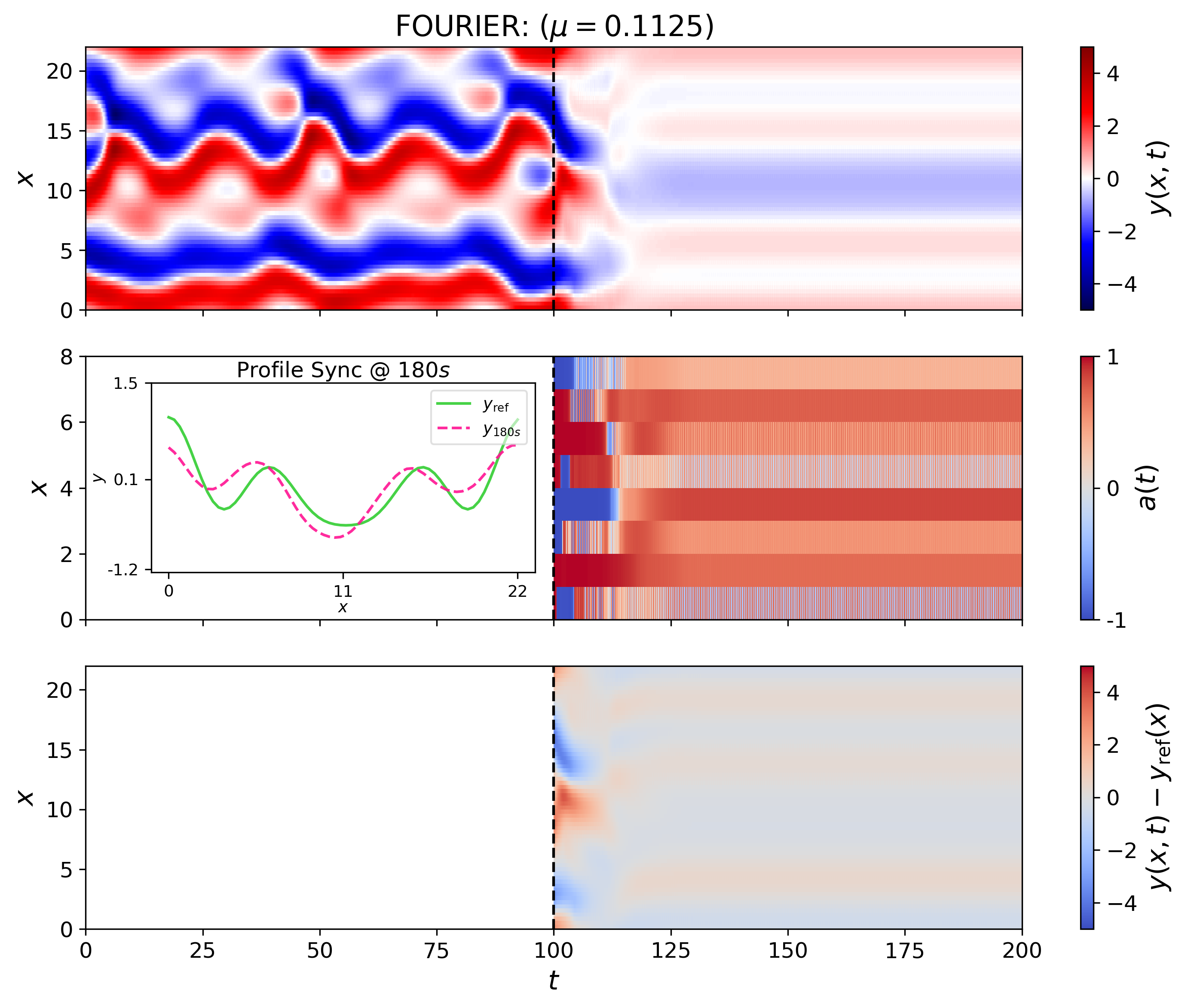}
    \end{subfigure}

    \vspace{0.4em}
    \begin{subfigure}{0.49\textwidth}
        \centering
        \includegraphics[width=0.9\linewidth]{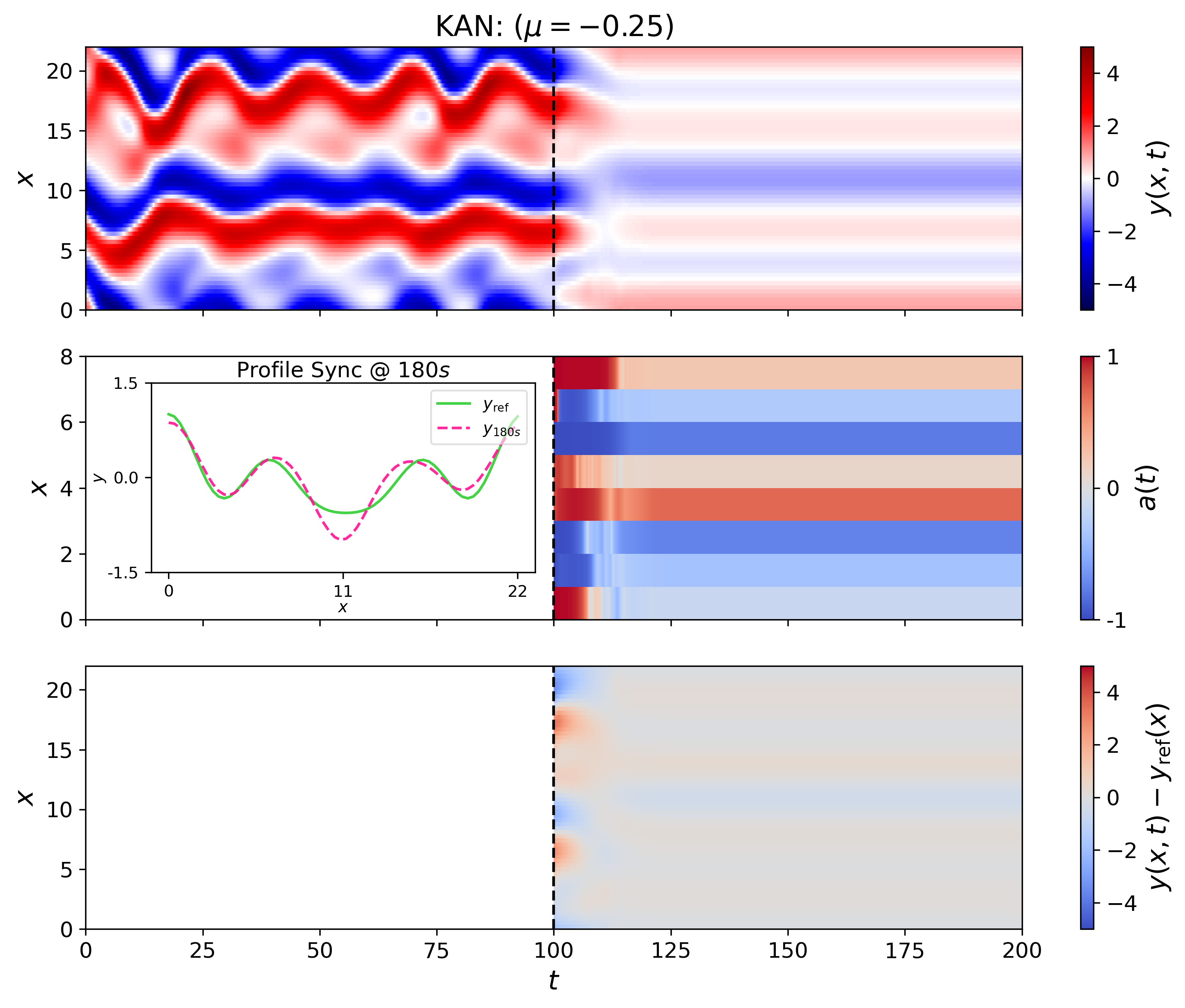}
    \end{subfigure}\hfill
    \begin{subfigure}{0.49\textwidth}
        \centering
        \includegraphics[width=0.9\linewidth]{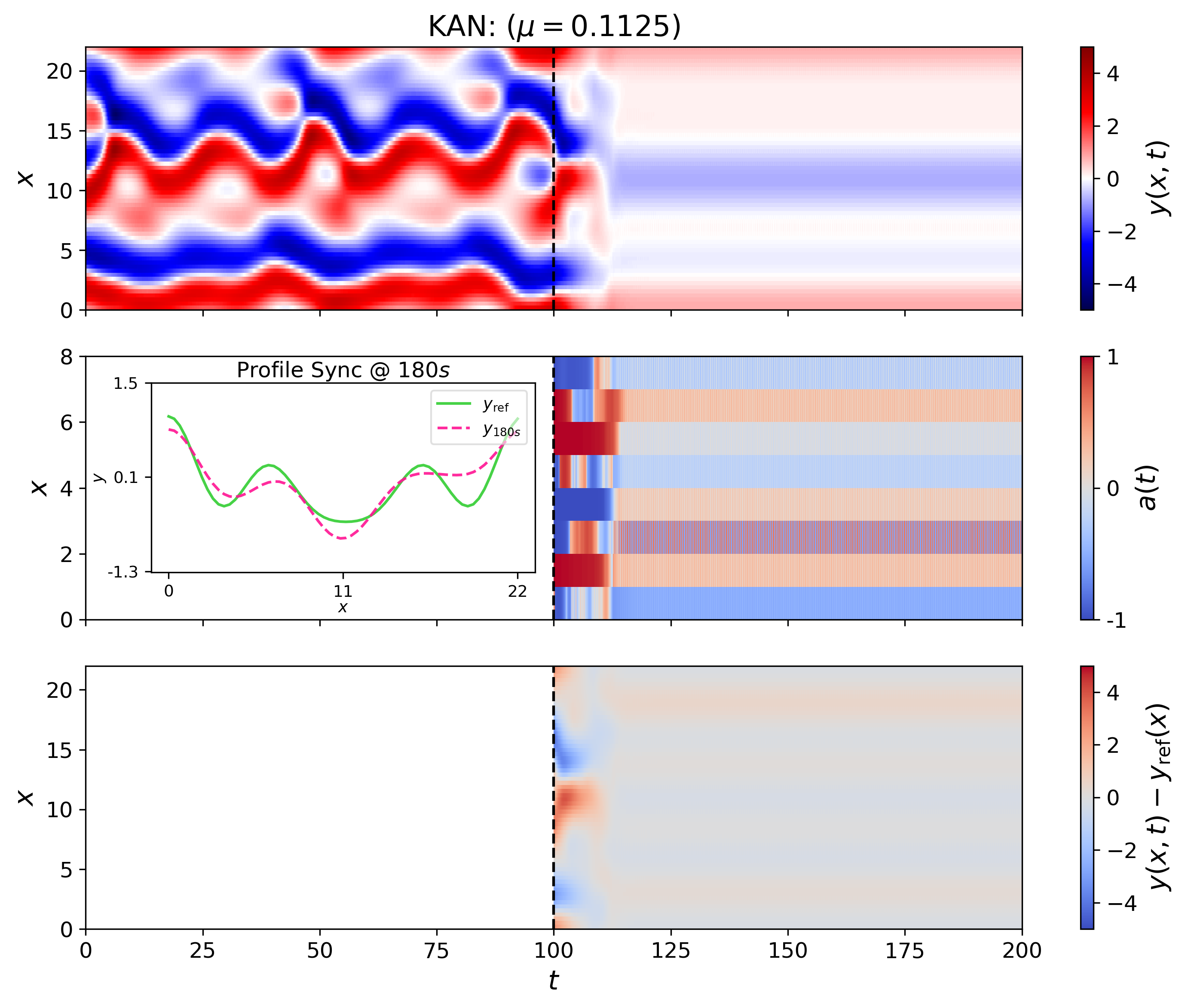}
    \end{subfigure}
    
    \caption{Spacetime evaluations $y(x,t)$ for Case 2, requiring the agent to continuously force the chaotic KS medium into a structured four-mode spatial oscillation. Columns compare an extrapolative forcing environment ($\mu=-0.25$) against an interpolative case ($\mu=0.1125$). Here, the policy must not only halt runaway turbulence but intelligently distribute specific energy profiles matching the structural geometry of the reward target. While the MLP allows significant phase distortion and temporal noise at the boundaries, ActNet-KAN retains sharp, coherent topological structures even outside the explicit training bounds.}
    \label{fig:heatmaps_tqc_cosine}
\end{figure*}

\begin{figure*}[p]
    \centering
    \begin{subfigure}{0.49\textwidth}
        \centering
        \includegraphics[width=0.9\linewidth]{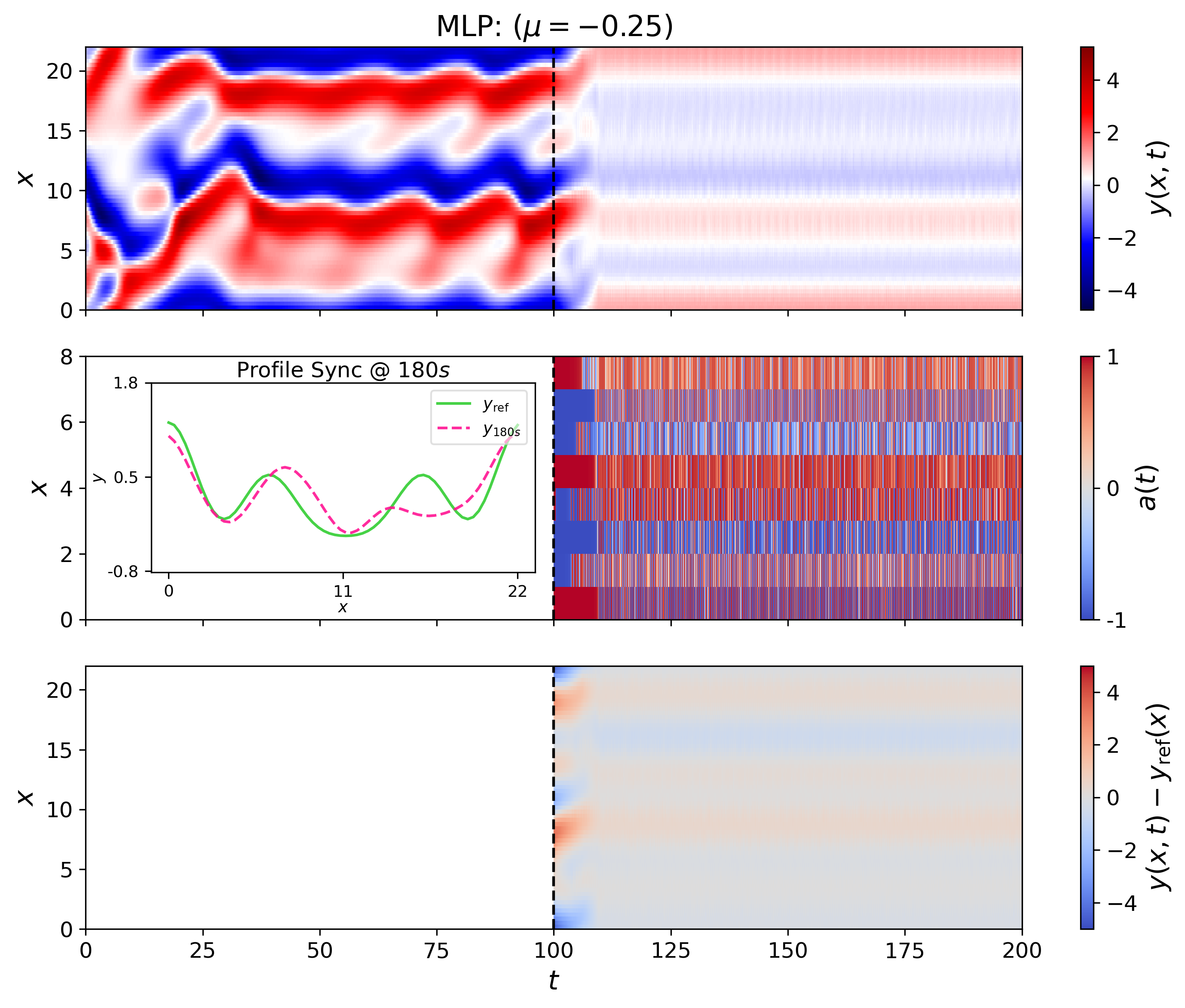}
    \end{subfigure}\hfill
    \begin{subfigure}{0.49\textwidth}
        \centering
        \includegraphics[width=0.9\linewidth]{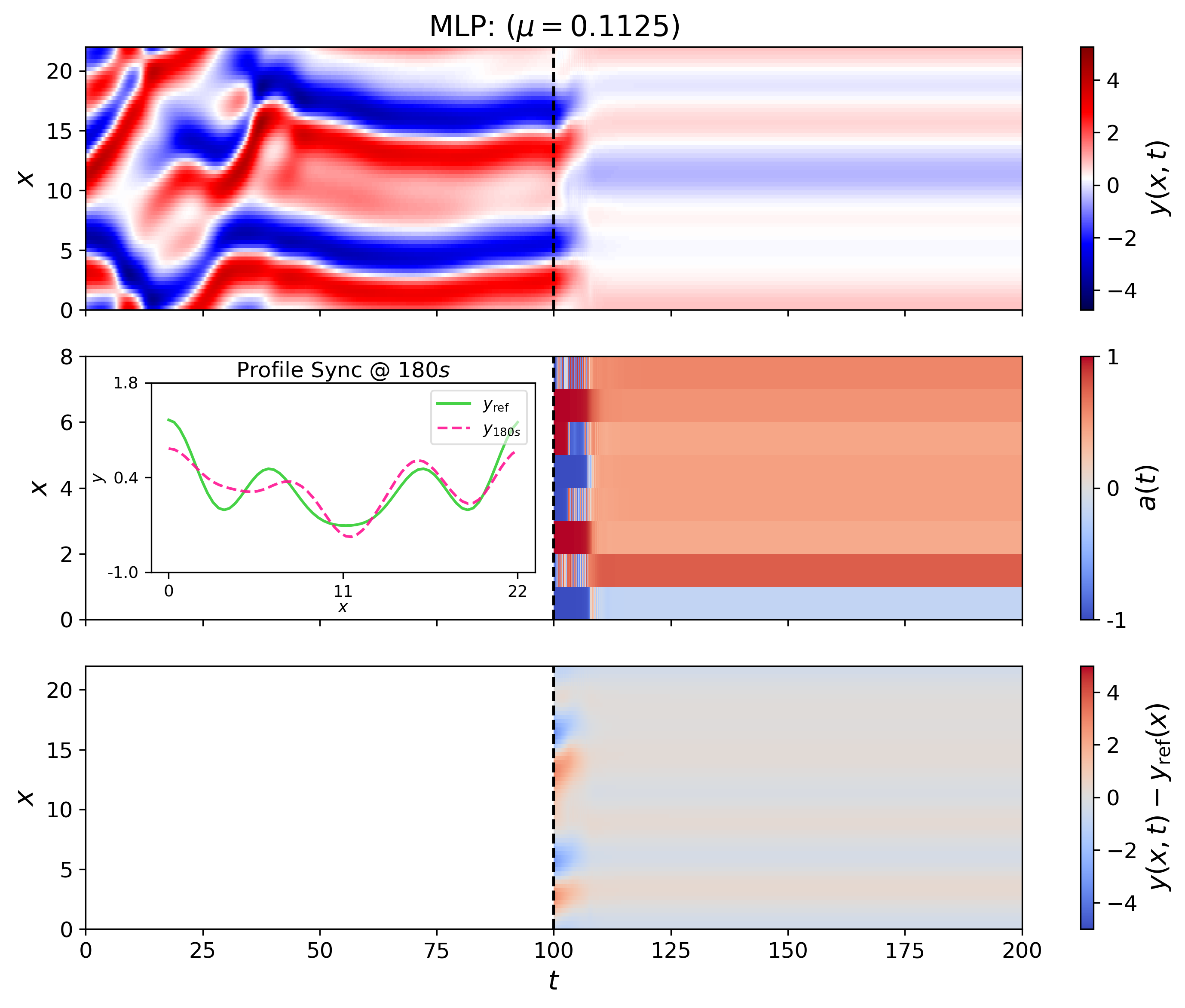}
    \end{subfigure}

    \vspace{0.4em}
    \begin{subfigure}{0.49\textwidth}
        \centering
        \includegraphics[width=0.9\linewidth]{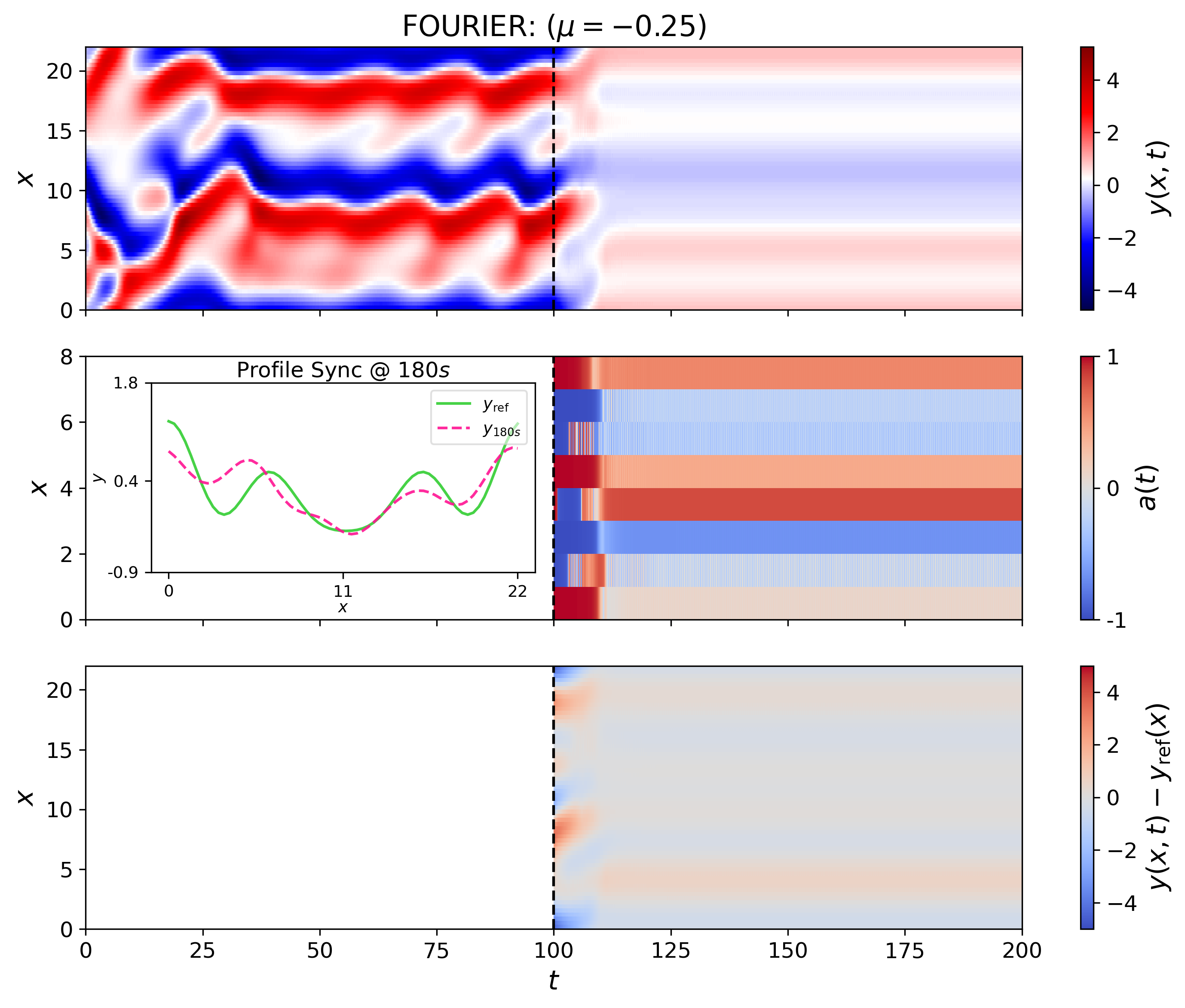}
    \end{subfigure}\hfill
    \begin{subfigure}{0.49\textwidth}
        \centering
        \includegraphics[width=0.9\linewidth]{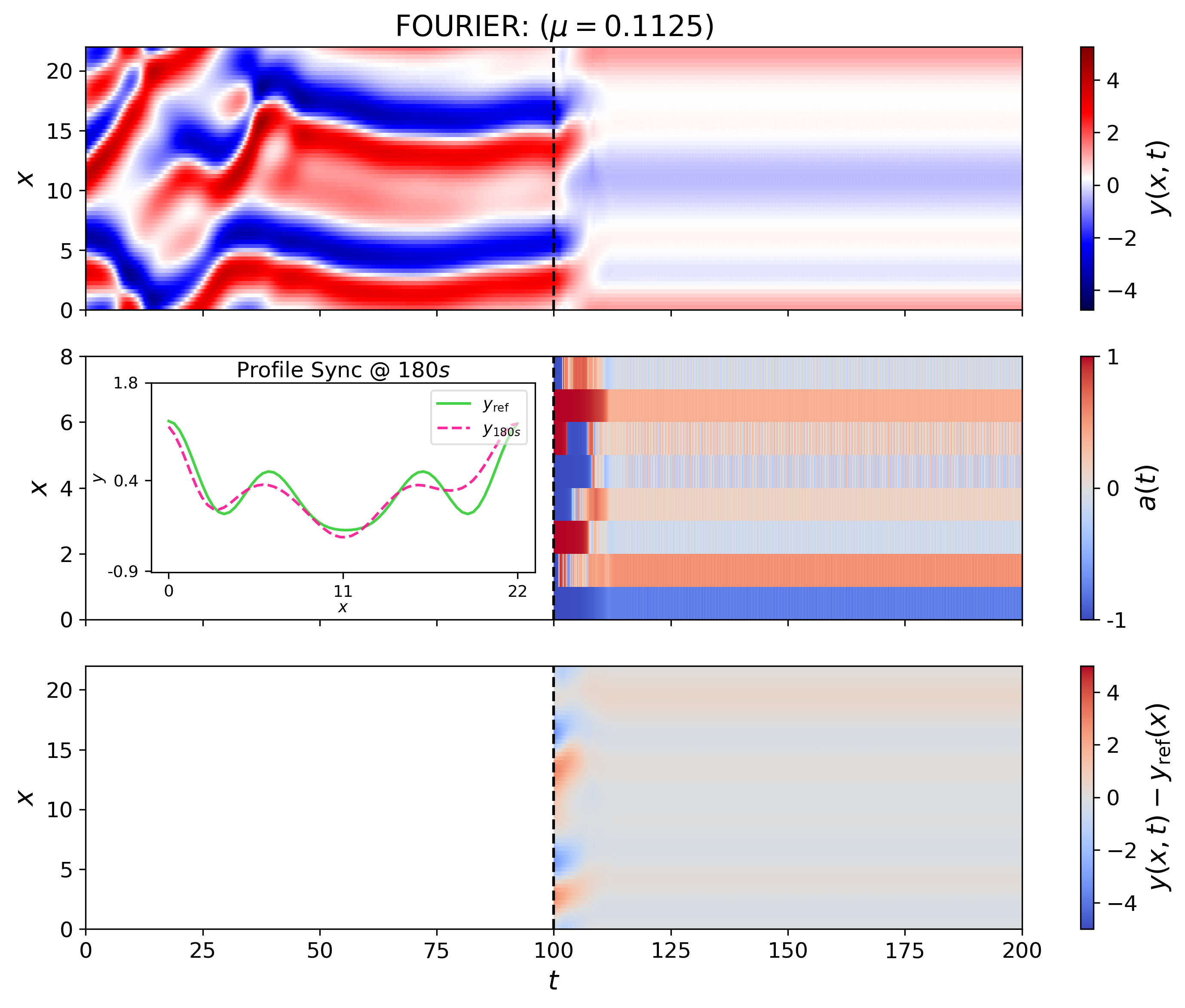}
    \end{subfigure}

    \vspace{0.4em}
    \begin{subfigure}{0.49\textwidth}
        \centering
        \includegraphics[width=0.9\linewidth]{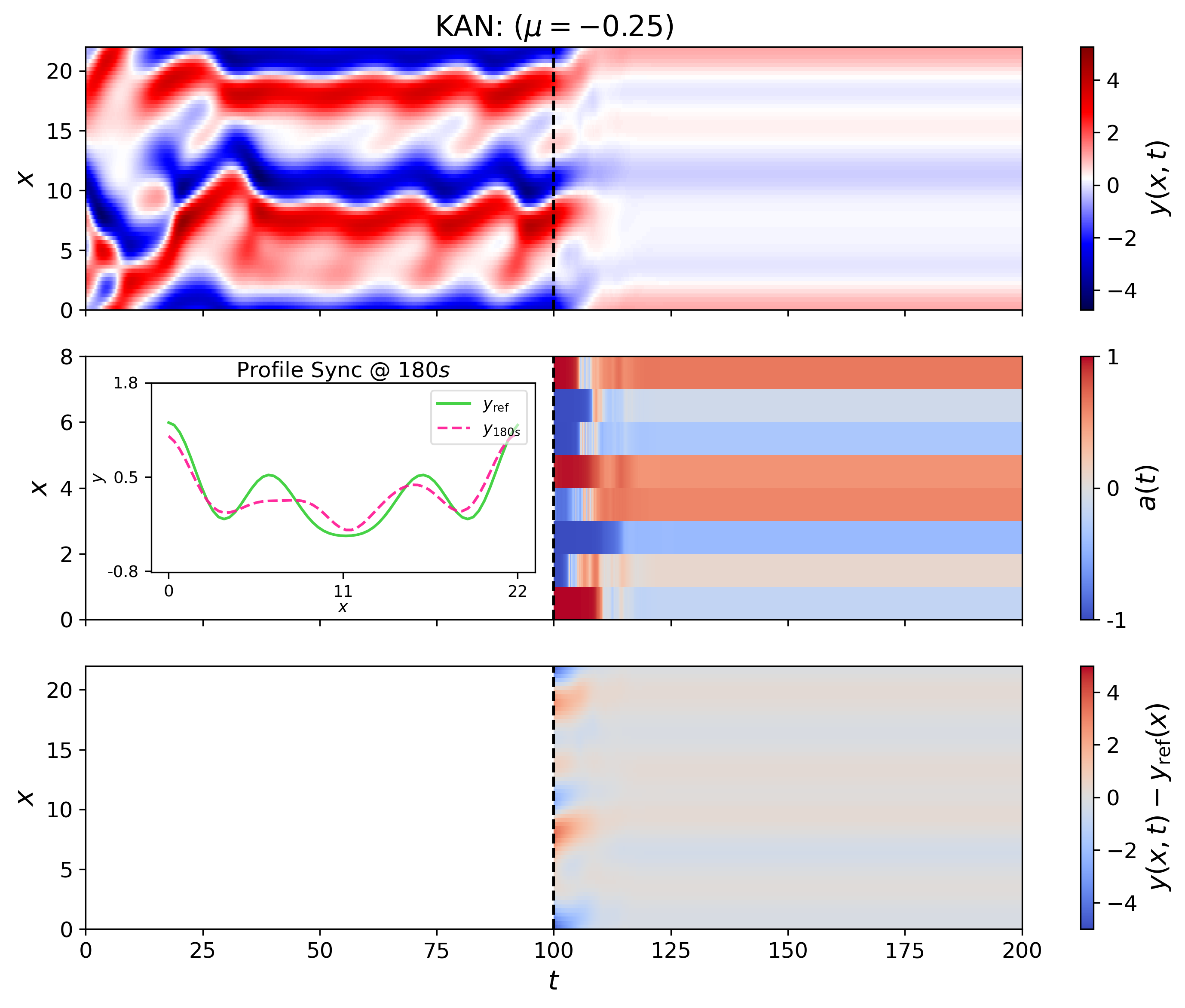}
    \end{subfigure}\hfill
    \begin{subfigure}{0.49\textwidth}
        \centering
        \includegraphics[width=0.9\linewidth]{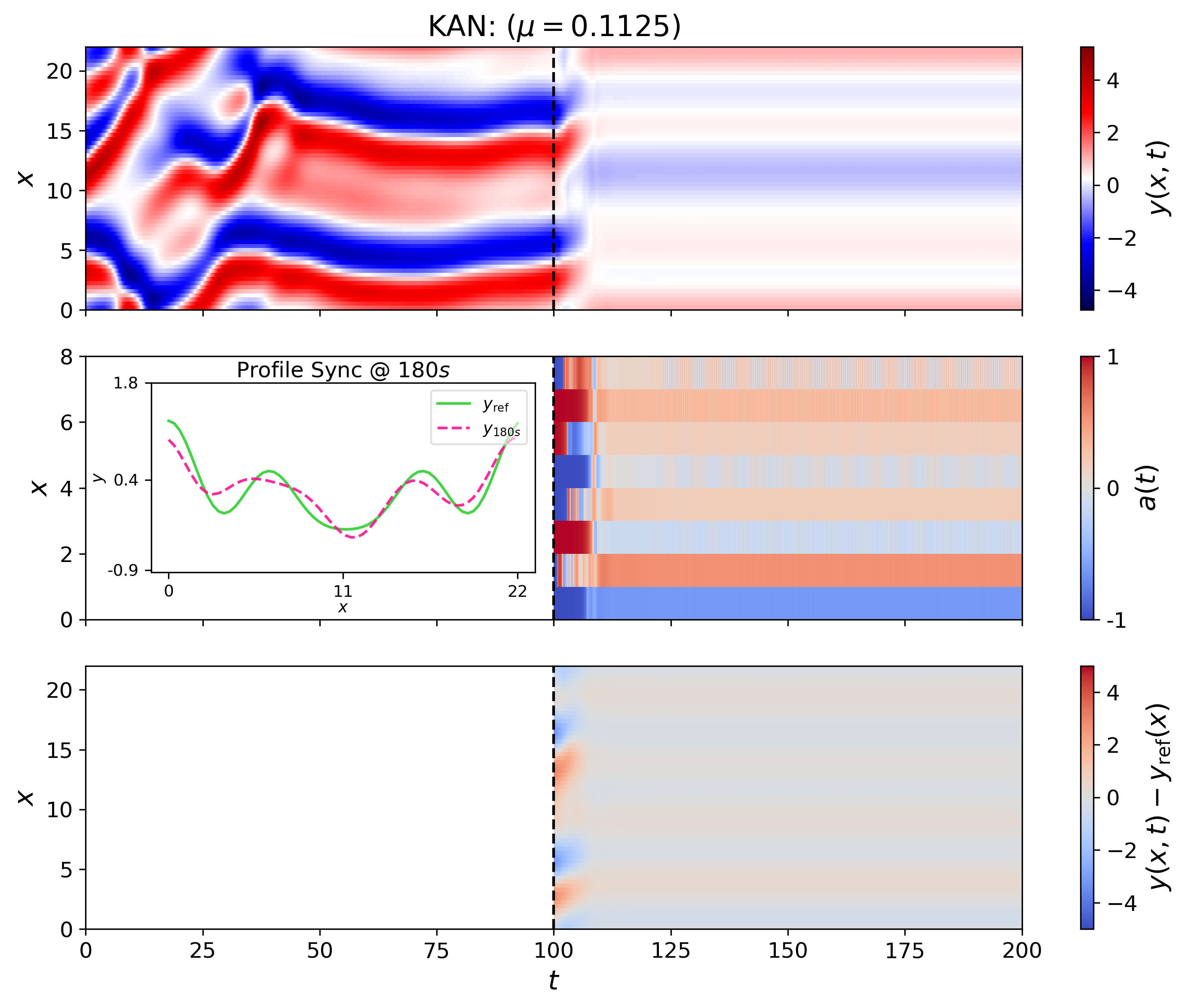}
    \end{subfigure}
    \caption{Spacetime evaluations $y(x,t)$ for Case 3, combining four-mode geometric tracking with an explicitly enforced non-zero spatial background mean. This configuration tests the agents' ability to maintain a prescribed standing wave pattern while simultaneously shifting the equilibrium state of the chaotic medium. Columns compare extrapolative ($\mu=-0.25$) and interpolative ($\mu=0.1125$) forcing levels. While all models achieve functional tracking, the KAN and Fourier representations, leveraging their intrinsic periodic basis functions, provide significantly better spatial coherence and lower boundary distortion than the MLP under these multi-objective constraints.}
    \label{fig:heatmaps_tqc_cosine_offset}
\end{figure*}

Figure~\ref{fig:rewards_all} provides quantitative reward trajectories for Cases 1--3, while Figures~\ref{fig:heatmaps_tqc0}, \ref{fig:heatmaps_tqc_cosine}, and \ref{fig:heatmaps_tqc_cosine_offset} provide the corresponding spacetime fields.

In the representative Case 1 heatmaps (Figure~\ref{fig:heatmaps_tqc0}),
we observe the physical mechanism of stabilization: the policy must suppress the high-wavenumber energy cascade typical of chaotic KS dynamics. KAN tends to produce the most uniform post-control field, effectively arresting the formation of traveling wave structures. In contrast, Fourier and MLP allow intermittent bursts of localized instability before acting, especially at the highly non-linear OOD point $\mu=-0.25$.
This dynamical interpretation aligns with the quantitative ordering in Section~\ref{sec:arch_comparison}.

Across Cases 2 and 3 (Figures~\ref{fig:heatmaps_tqc_cosine} and \ref{fig:heatmaps_tqc_cosine_offset}), all encoders achieve qualitative target tracking, actively balancing the background spatial forcing to maintain the prescribed standing wave geometries. The visualizations confirm that the policies are not merely dissipating energy indiscriminately, but rather learning to dynamically project the chaotic system onto the stabilized target manifold. KAN preserves cleaner phase-aligned boundaries and exhibits significantly lower residual distortion under OOD checks. Taken together with the five-seed quantitative study, the evidence supports a compelling physical and computational conclusion for this benchmark: a massively parallel formulation navigating at $GS=2$ provides the optimal speed/quality trade-off, while ActNet-KAN supplies the most robust parametric embeddings for maintaining precise spatial coherence under extrapolative forcing.

\section{Conclusion and Future Work}\label{sec:conclusion}

This work introduced \emph{hyperFastRL}, a unified reinforcement-learning framework for parametric control of chaotic PDE dynamics, and evaluated it on the 1D Kuramoto--Sivashinsky benchmark. The central design combines parameter-conditioned Hypernetworks with a high-throughput FastTD3/TQC training pipeline, enabling a single controller family to adapt across forcing-parameter regimes. Across the experiments reported in Section~\ref{sec:results}, the approach achieved stable training behavior and competitive generalization trends for both interpolation and mild extrapolation test settings.

A key empirical finding is computational: leveraging massively parallel environments to navigate the performance--throughput Pareto front (notably navigating to $GS=2$) provided the optimal practical operating point, intentionally trading a fraction of peak statistical reward for critical wall-clock tractability. Under a fixed high-throughput protocol, encoder choice dictated the fidelity of the learned control manifold; ActNet-KAN showed the most consistent improvement over the MLP baseline in suppressing chaotic energy cascades and traveling waves, while Fourier embeddings provided mixed extrapolation robustness.

Taken together, these results demonstrate that a single neural policy, parameterized via a Hypernetwork, can effectively track and stabilize a chaotic PDE manifold across varying forcing amplitudes without catastrophic interference. This shifts the computational paradigm from recursively tuning custom adjoint or isolated RL controllers per-regime toward learning a unified parametric control law.

However, these results must be interpreted within the study's methodological scope and empirical limits. First, the evaluation is constrained to a 1D spatial domain with a targeted parametric range ($\mu \in [-0.225, 0.225]$). Consequently, the out-of-distribution checks represent mild extrapolation (e.g., $\mu = -0.25$) rather than true zero-shot generalization to drastically distinct physics; nevertheless, this confirms the hypernetwork is successfully interpolating control manifolds rather than merely memorizing local instances. Second, because chaotic flow control is exceptionally sensitive to initialization, increasing the seed count beyond the five evaluated here would be required to establish strict statistical dominance regarding mean reward limits, though the high consistency of KAN's test-reward variance already provides strong evidence for its physical robustness. Finally, the comparisons in this work focus strictly on deep neural encoders within the hypernetwork paradigm to establish internal algorithmic hierarchy. Future extensions should benchmark this unified approach against online adaptive control or model predictive control (MPC) to fully characterize the practical utility and data-efficiency of parameter-conditioned RL in higher-dimensional fluid applications.

\noindent\textbf{Future Work.} Several extensions are natural and important:
\begin{itemize}
    \item \textbf{RL for data assimilation:} investigate how reinforcement learning can support sequential state estimation and correction under partial and noisy observations.
    \item \textbf{Different PDE settings:} evaluate transferability beyond 1D KS to additional PDE regimes and control tasks.
\end{itemize}

Overall, \emph{hyperFastRL} provides a practical foundation for learning unified controllers across parametric chaotic dynamics, and the present study motivates broader, statistically stronger evaluations toward real-world PDE-control deployment.

\bibliographystyle{unsrt}
\bibliography{references}  

\appendix
\section*{Appendix}\label{sec:appendix}

\subsection*{Appendix A: Shared Hyperparameters}\label{sec:app_hyper}

\begin{table}[!htb]
    \centering
    \caption{Key hyperparameters shared across all runs.}
    \label{tab:hyperparams}
    \small
    \begin{tabular}{lp{1.8cm}p{3.5cm}}
        \toprule
        \textbf{Parameter} & \textbf{Value} & \textbf{Rationale} \\
        \midrule
        Parallel environments & 1024 & Maximise state diversity \\
        Staggered reset & Enabled & Decorrelate initial states across parallel environments \\
        Burn-in steps & 100 & Advance KS dynamics before logged control rollout \\
        Replay buffer & $4 \times 10^6$ & Off-policy decorrelation \\
        Exploration fraction & 0.05 & Initial random-action phase (5\% of total steps) \\
        Batch size ($B$) & 32\,768 & Amortise GPU launch cost \\
        N-step returns ($n$) & 3 & Variance/bias trade-off \\
        Quantile atoms ($M$) & 25 & TQC distributional resolution \\
        Top-$d$ drop & 5 & 10\% pooled truncation \\
        Actor LR & $3 \times 10^{-4}$ & AdamW + cosine annealing \\
        Critic LR & $3 \times 10^{-4}$ & AdamW + cosine annealing \\
        Polyak coefficient ($\tau$) & 0.01 & Slow target tracking \\
        Discount ($\gamma$) & 0.99 & $\approx 100$-step effective horizon \\
        Control-cost ($\alpha$) & 0.1 & Prioritise stabilisation \\
        \bottomrule
    \end{tabular}
\end{table}

\subsection*{Appendix B: Network Details}\label{sec:app_network}

\begin{table*}[htb]
    \centering
    \caption{Network details and parameter counts for each model variant (state dimension $=65$, action dimension $=8$, target hidden width $=256$).}
    \label{tab:param_counts}
    \setlength{\tabcolsep}{5pt}
    \begin{tabular}{llp{11cm}}
        \toprule
        \textbf{Encoder} & \textbf{Field} & \textbf{Value} \\
        \midrule
        MLP & Actor layers & $64 \rightarrow 256 \rightarrow 8$ \\
        MLP & Critic layers (per head) & $72 \rightarrow 256 \rightarrow 25$ \\
        MLP & Hypernet details & ResNet backbone: $1 \rightarrow 256 \rightarrow 512 \rightarrow 1024$; affine weight heads \\
        MLP & Trainable params & 90,011,252 \\
        MLP & Non-trainable params & 0 \\
        MLP & Total params & 90,011,252 \\
        \midrule
        Fourier & Actor layers & $64 \rightarrow 256 \rightarrow 8$ \\
        Fourier & Critic layers (per head) & $72 \rightarrow 256 \rightarrow 25$ \\
        Fourier & Hypernet details & Fourier map: $1 \rightarrow 513$ (skip + sin/cos, mapping size 256), then ResNet $513 \rightarrow 256 \rightarrow 512 \rightarrow 1024$; affine heads; fixed RFF buffers ($B$, scale) \\
        Fourier & Trainable params & 90,404,468 \\
        Fourier & Non-trainable params & 771 \\
        Fourier & Total params & 90,405,239 \\
        \midrule
        KAN & Actor layers & $64 \rightarrow 256 \rightarrow 8$ \\
        KAN & Critic layers (per head) & $72 \rightarrow 256 \rightarrow 25$ \\
        KAN & Hypernet details & KAN-ResNet backbone: $1 \rightarrow 256 \rightarrow 512 \rightarrow 1024$ with ActNet residual blocks; KAN (sinusoidal) heads \\
        KAN & Trainable params & 95,975,322 \\
        KAN & Non-trainable params & 0 \\
        KAN & Total params & 95,975,322 \\
        \bottomrule
    \end{tabular}
\end{table*}

All three variants use the same parameter-conditioned Hypernetwork pipeline,
but differ in the encoder that maps the physically scaled parameter
$\tilde{\mu}$ (normalized and scaled to range approximately $[-10, 10]$ to provide highly dynamic input ranges to the encoding frequencies) to a latent feature vector. The target policy/critic layer update is

\begin{equation}
    h_\ell = \sigma_\ell\!\left(s_\ell \odot (W_\ell h_{\ell-1}) + b_\ell\right).
\end{equation}

The three encoder choices are:

\begin{equation}
    z_{\mathrm{MLP}}(\tilde{\mu})
    = \phi_{L}\!\left(\phi_{L-1}(\cdots \phi_{1}(\tilde{\mu}))\right),
\end{equation}

\begin{align}
    z_{\mathrm{Fourier}}(\tilde{\mu}) &= \phi\!\left(\gamma(\tilde{\mu})\right), \\
    \gamma(\tilde{\mu}) &= \bigl[\tilde{\mu},\,\sin(2\pi B\tilde{\mu}),\,\cos(2\pi B\tilde{\mu})\bigr], \nonumber
\end{align}

\begin{align}
    h^{(0)} &= \tilde{\mu}, \\
    h^{(\ell+1)}_i &= \sum_{j=1}^{d_\ell} a^{(\ell)}_{ij}\,\sin\!\left(\omega^{(\ell)}_{ij} h^{(\ell)}_j + b^{(\ell)}_{ij}\right), \\
    z_{\mathrm{KAN}} &= h^{(L)}, \nonumber
\end{align}

\paragraph{Weight-head mappings (key architectural difference).}
For MLP/Fourier variants, each target layer uses affine heads from the encoder feature:

\begin{align}
    \mathrm{vec}(W_\ell) &= A^{(\ell)}_{W} z + c^{(\ell)}_{W}, \\
    b_\ell &= A^{(\ell)}_{b} z + c^{(\ell)}_{b}, \\
    s_\ell &= \mathbf{1} + A^{(\ell)}_{s} z + c^{(\ell)}_{s}, \nonumber
\end{align}

with $z=z_{\mathrm{MLP}}$ for the MLP model and $z=z_{\mathrm{Fourier}}$ for the Fourier model.

For the KAN variant, each head is itself an ActNet/KAN mapping (sinusoidal edge functions):

\begin{align}
    u^{(0)} &= z_{\mathrm{KAN}}, \\
    u^{(r+1)}_i &= \sum_{j=1}^{m_r} \alpha^{(r)}_{ij}\,\sin\!\left(\Omega^{(r)}_{ij} u^{(r)}_j + \beta^{(r)}_{ij}\right), \nonumber
\end{align}

\begin{align}
    \mathrm{vec}(W_\ell) &= u^{(R_{W,\ell})}_{W,\ell}, \\
    b_\ell &= u^{(R_{b,\ell})}_{b,\ell}, \\
    s_\ell &= \mathbf{1} + u^{(R_{s,\ell})}_{s,\ell}, \nonumber
\end{align}

This makes the distinction explicit: MLP/Fourier heads are linear projections of encoder features, while KAN heads are nonlinear sinusoidal function expansions.

\end{document}